\documentclass[11pt,a4paper]{article}
\usepackage{jcappub}
\usepackage{bm}
\usepackage{times}

\newcommand{\rbr}[1]{\left(#1\right)}
\newcommand{\sbr}[1]{\left[#1\right]}
\newcommand{\dr}{\mathrm d\bar r\,}
\newcommand{\rz}{{\bar r_z}}
\newcommand{\etaz}{\bar\eta_z}
\newcommand{\etao}{\bar\eta_o}
\newcommand{\Dkk}{\frac{\mathrm d^3k}{(2\pi)^3}}
\newcommand{\Dkone}{\frac{\mathrm d^3k_1}{(2\pi)^3}}
\newcommand{\Dktwo}{\frac{\mathrm d^3k_2}{(2\pi)^3}}
\newcommand{\Dkthr}{\frac{\mathrm d^3k_3}{(2\pi)^3}}
\newcommand{\Hz}{\mathcal H_z}
\newcommand{\habla}{\hat\nabla}
\newcommand{\kone}{\mathbf{k}_1}
\newcommand{\ktwo}{\mathbf{k}_2}
\newcommand{\kthree}{\mathbf{k}_3}
\newcommand{\klong}{\mathbf{k}_l}
\newcommand{\xone}{\mathbf{x}_1}
\newcommand{\xtwo}{\mathbf{x}_2}
\newcommand{\xthree}{\mathbf{x}_3}
\newcommand{\q}{\mathbf q}
\newcommand{\Dq}{\frac{\mathrm d^3q}{(2\pi)^3}}
\newcommand{\FF}{{\cal F}}
\newcommand{\deobs}{\de_m^\up{obs}}
\newcommand{\RR}{{\cal R}}
\newcommand{\IIetc}[1]{\noindent \textit{\textbf{#1}}:}
\newcommand{\xisqz}{\xi_{\rm sqz}}
\newcommand{\xisqzlim}{\xi_{\rm sqz}^\up{lim}}
\newcommand{\avg}{\sigma}
\newcommand{\ens}{\Xi}
\newcommand{\up}[1]{{\rm #1}}
\newcommand{\etc}{\noindent $\bullet$~}
\newcommand{\Ietc}[1]{\etc \textit{\textbf{#1}}.---}
\newcommand{\bdv}[1]{{\bf #1}}
\newcommand{\beeq}{\begin{equation}}
\newcommand{\eneq}{\end{equation}}
\newcommand{\bear}{\begin{eqnarray}}
\newcommand{\enar}{\end{eqnarray}}
\newcommand{\nnn}{\nonumber \\}
\newcommand{\nn}{\nonumber}
\newcommand{\AVE}[1]{\left\langle#1\right\rangle}
\newcommand{\RA}{\rightarrow}
\newcommand{\pa}{\partial}
\newcommand{\Dquad}{\qquad\qquad}
\newcommand{\fnl}{f_\up{NL}}
\newcommand{\OO}{\mathcal{O}}
\newcommand{\kvec}{\bdv{k}}
\newcommand{\xvec}{\bdv{x}}
\newcommand{\Zang}{\bdv{\hat z}}
\newcommand{\Vang}{\bdv{\hat n}}
\newcommand{\al}{\alpha}
\newcommand{\be}{\beta}
\newcommand{\ga}{\gamma}
\newcommand{\de}{\delta}
\newcommand{\HH}{\mathcal{H}}   
\newcommand{\rbar}{\bar r}      
\newcommand{\dz}{\delta z}      
\newcommand{\drr}{\delta r}     
\newcommand{\dtt}{\delta\theta}   
\newcommand{\dpp}{\delta\phi}    
\newcommand{\gbar}{\bar g}      
\newcommand{\px}{\varphi_{\chi}}
\newcommand{\n}{\Vang}

\newcommand{\DD}{{\mathfrak D}}

\begin{document}

\begin{titlepage}

\setcounter{page}{1} \baselineskip=15.5pt \thispagestyle{empty}
\pagenumbering{roman}

\bigskip

\vspace{1cm}
\begin{center}
{\fontsize{20}{28}\selectfont \bfseries Non-Gaussianity in the Squeezed
Three-Point Correlation from the Relativistic Effects}
\end{center}

\vspace{0.2cm}

\begin{center}
{\fontsize{13}{30}\selectfont Jaiyul Yoo,$^{a,b}$
Nastassia Grimm,$^{a,c}$  and Ermis Mitsou$^a$}
\end{center}

\begin{center}
\vskip 8pt
\textsl{$^a$ Center for Theoretical Astrophysics and Cosmology,
Institute for Computational Science}\\
\textsl{University of Z\"urich, Winterthurerstrasse 190,
CH-8057, Z\"urich, Switzerland}

\vskip 7pt

\textsl{$^b$Physics Institute, University of Z\"urich,
Winterthurerstrasse 190, CH-8057, Z\"urich, Switzerland}

\vskip 7pt

\textsl{$^c$D\'epartement de Physique Th{\'e}orique \& Center
for Astroparticle Physics, Universit\'e de Gen\`eve,\\
Quai E. Ansermet 24, CH-1211 Gen\`eve 4, Switzerland}

\vskip 7pt

\today

\end{center}

\note{jyoo@physik.uzh.ch,~~~~ nastassia.grimm@unige.ch,~~~~ 
ermitsou@physik.uzh.ch}

\vspace{1.2cm}
\hrule \vspace{0.3cm}
\noindent {\sffamily \bfseries Abstract} \\[0.1cm]
Assuming a $\Lambda$CDM universe in a single-field inflationary scenario,
we compute the three-point correlation function of the observed matter
density fluctuation in the squeezed triangular configuration, accounting
for all the relativistic effects at the second order in perturbations.
This squeezed three-point correlation function characterizes the local-type
primordial non-Gaussianity, and it has been extensively debated in literature  
whether there exists a prominent feature in galaxy clustering on large scales
in a single-field inflationary scenario
either from the primordial origin or the intrinsic nonlinearity in general
relativity. First, we show that theoretical descriptions of galaxy bias are
incomplete in general relativity due to ambiguities in spatial gauge choice, 
while those of cosmological observables are independent of spatial gauge 
choice. Hence a proper relativistic description of galaxy bias is needed
to reach a definitive conclusion in galaxy clustering.  Second, we demonstrate
that the gauge-invariant calculations of the cosmological
observables remain unaffected by extra coordinate transformations like CFC or 
large diffeomorphism like dilatation. Finally, we show that
the relativistic effects associated with light propagation in
observations cancel each other, and hence there exists {\it no} 
non-Gaussian contribution from the so-called projection effects
in the squeezed three-point correlation function.
\vskip 10pt
\hrule

\vspace{0.6cm}
\end{titlepage}

\clearpage

\noindent\hrulefill \tableofcontents \noindent\hrulefill

\pagenumbering{arabic}

\section{Introduction}

In the standard model of cosmology, cold dark matter is the majority of
the matter content of the Universe, and a cosmological constant dominates
the energy content today. The initial condition is set by a slow-rolling 
single scalar field during the inflationary expansion in the early Universe,
and the primordial fluctuations are highly Gaussian. Despite several
deficiencies, the standard model of cosmology has been extremely successful 
in explaining cosmological observations on a wide range of scales, 
spanning galactic scale to horizon scale (see, e.g., 
\cite{TEEIET06,PLANCKfnl13,PLANCKcos18}). Among
the issues in the standard model, the origin of the Universe is particularly
interesting and puzzling. The standard single-field inflationary model
is so generic that very little is known about the scalar field and its
potential, except that
it was slow-rolling and its energy density is dominant in the early Universe
(see, e.g., \cite{LYRI99,LINDE00} 
for a review). A further investigation of the standard
inflationary model revealed \cite{MALDA03} that the primordial fluctuations
deviate slightly from a perfect Gaussianity, and the deviation is characterized
by a parameter~$\fnl$ in the squeezed limit bispectrum. The standard
single-field inflationary model predicts negligibly small non-Gaussianity
$\fnl\sim(n_s-1)\sim\varepsilon$ \cite{MALDA03},
where $n_s$ is the spectral index
and~$\varepsilon\sim0.01$ is the slow-roll parameter.
In contrast, other non-standard inflationary models such as multi-field
models and models with non-canonical kinetic term predict non-Gaussianity
much larger than the standard inflationary model (see, e.g., \cite{BAKOET04}).
Robust detection of the primordial non-Gaussianity is, 
therefore, one of the important
targets in the current and the upcoming large-scale surveys that can
reveal the nature of the early Universe.

Recently it was shown \cite{DADOET08} that in the presence of a local-type
primordial non-Gaussianity the galaxy bias exhibits a strong scale-dependence
in the power spectrum on very large scales, where the bias factor is expected
to be a constant and the galaxy number density fluctuation is in proportion
to the matter density fluctuation. Further investigations show 
\cite{VEMA09,CACAET15,CAMASA15,DIDV18} that measurements of the
galaxy power spectrum in the upcoming surveys
would provide constraints on the primordial 
non-Gaussianity tighter than the constraints obtained in measurements
of the cosmic microwave background (CMB) anisotropies, as the upcoming galaxy
surveys will have larger three-dimensional volumes, compared to the
two-dimensional snapshot from the CMB measurements. Certainly, this would be
a promising avenue in cosmology, given the recent developments in large-scale
galaxy surveys such as the Dark Energy Spectroscopic Instrument \cite{DESI13} 
and the Vera C. Rubin Observatory (formerly LSST)
\cite{LSST04} and two space missions,
Euclid \cite{EUCLID11} and the Nancy Grace Roman Space Telescope (formerly
WFIRST) \cite{WFIRST12}. 

The question then arises naturally for the standard inflationary model:
Since the primordial non-Gaussianity is small but non-zero, 
will this induce a scale-dependent bias in the galaxy power spectrum
in the standard model? Moreover, the relativistic computation of the
matter density fluctuation beyond the linear order in perturbations shows
\cite{BRHIET14,BRHIWA14,CAMASA15,VIVEMA14,YOGO16,MAPIRO21} that there exist
nonlinear relativistic effects in the initial condition, arising from
the nonlinear Hamiltonian constraint in general relativity, even when
the primordial fluctuation is linear and Gaussian. Will this non-Gaussianity
again induce a scale-dependent bias in the galaxy power spectrum in
the standard model? It has been argued 
\cite{PASCZA13,DEDOGR15,DEDOET17} against this implication
that for a single-field inflationary model there exists only one
degree of freedom and a long-mode fluctuation
by this degree can be absorbed
into a coordinate transformation, such that the local two-point correlation
is {\it not} affected by the presence of a long mode fluctuation.
This argument can be equally applied to the nonlinear relativistic effects
from the Hamiltonian constraint, and  no scale-dependent
galaxy bias is predicted in the case of a single-field inflationary model.

Given that the expectation for the standard inflationary model is
either small ($\sim\varepsilon$) or zero if absorbed by a coordinate
transformation, an order unity correction
from the nonlinear relativistic effects is a significant contamination, if
real, for future observations. Furthermore, the power spectrum on large
scales or the squeezed limit bispectrum in an infinite hypersurface
are {\it not} a direct observable we can measure from large-scale surveys; 
observations are made on a light cone
volume in terms of the observed redshift and the observed angular position
on our sky, which gives rise to additional relativistic effects
in cosmological observations \cite{YOFIZA09,YOO10,BODU11,CHLE11,JESCHI12}.
It was argued 
\cite{VEMA09,BRCRET12,PASCZA13,
CAMASA15,KEDIET15,DIPEET17,UMJOET17,JOUMET17,KOUMET18}
that the relativistic effects associated
with the light propagation and observations add extra contributions
to measurements of the primordial non-Gaussianity 
at the level similar to the intrinsic nonlinear relativistic effects,
and this contamination is {\it always present}, 
no matter what inflationary models are considered. Therefore, it is of
significant interest that we obtain accurate predictions for the
primordial non-Gaussianity for each cosmological model
that can be measured from large-scale surveys.
Here we examine the issue critically, accounting for all the relativistic
effects from the Hamiltonian constraint and the light propagation.
The past work for or against the extra contributions to the primordial
non-Gaussianity is an important first step, but the final answer to the
level of observable non-Gaussianities is yet to be derived:
While a proper second-order theory is needed to compute the bispectrum or
the three-point correlation, most computations 
\cite{PASCZA13,DEDOGR15,DEDOET17} are based on the linear-order 
calculations, supplemented by coordinate transformations to absorb long-mode
fluctuations. In particular, these calculations are performed 
in a coordinate system with a finite range of validity, 
and a Fourier transformation is made to compute
the squeezed limit bispectrum, presumably outside the validity range.
In other works 
\cite{KEDIET15,DIPEET17,UMJOET17,JOUMET17,BERAET18,KOUMET18,UMKOET19,CLDEET19,
UMKO19,MAJOET20b},
the relativistic effects from the light propagation and observations are
also considered, but {\it not} in full entirety. 
For example, the relativistic contributions
at the observer position or along the line-of-sight direction are often
neglected in the past work, while those individual terms could potentially
add an order unity correction to the primordial non-Gaussianity.
Here we improve the previous work and provide a complete second-order
gauge-invariant calculation of the three-point correlation function.

In fact, we find that there exist {\it several flaws} 
in theoretical descriptions
that need to be addressed in full but are {\it less} known in literature, 
before one can reach a definitive answer. These issues are discussed 
in detail in Section~\ref{issues}. In particular, any computations beyond
the linear order in perturbations require a choice of spatial gauge condition,
as they change with spatial gauge transformation and so do the bispectrum
and the three-point correlation function. We show that when observable
quantities are computed, their theoretical descriptions are independent of
a choice of spatial gauge. However, 
the relation between the galaxy and the matter
distributions is independent of observations, and it requires a physical
explanation for a specific choice of spatial gauge.
Moreover, we show that the gauge-invariant calculations of cosmological
observables leave {\it no} residual degrees of freedom, often employed
in a coordinate transformation like dilatation or conformal Fermi coordinates.
Adopting a $\Lambda$CDM model in the standard inflationary model and
using the Einstein equation, we derive exact analytical solutions for
the second-order perturbation variables and compute the observed three-point
correlation function. We demonstrate that the relativistic effects
associated with light propagation and observations generate extra
contributions to the primordial non-Gaussianities but {\it they all add up to
cancel each other, if all the relativistic effects are properly considered.}

The organization of the paper is as follows: In Section~\ref{smp},
we compute the bispectrum of the matter density fluctuation
in the standard perturbation theory and the relativistic perturbation
theory, and we discuss the implications for the scale-dependent
galaxy bias. Theoretical flaws in the previous
work are extensively discussed in Section~\ref{issues}: Convention
for the primordial non-Gaussianity in Section~\ref{convention},
ambiguity in spatial gauge choices in Section~\ref{spatial},
coordinate dependence of the ensemble average in Section~\ref{ensavg},
possibility of 
extra symmetry in gauge-invariant calculations in Section~\ref{extra}, and
gauge choice for galaxy biasing in Section~\ref{sec:gauge}.
In Section~\ref{sec:matt},
we present how such theoretical issues are resolved in the 
observed matter density fluctuation and compute the observed
three-point correlation function in the squeezed limit.
Our main findings are summarized in Section~\ref{findings},
and the implication of our work is discussed in Section~\ref{sec:discuss}.
The details of the second-order relativistic calculations are 
presented in two Appendices~\ref{append} and~\ref{details}.

\section{Second-order matter density fluctuation}
\label{smp}
In this section we briefly review the theoretical descriptions of
the second-order matter density fluctuation in the standard Newtonian
perturbation theory and the relativistic perturbation theory. We then
compute the matter bispectra in the squeezed limit for both cases
and compare them to the previous work.

\subsection{Standard and relativistic perturbation theories}
\label{sptrpt}
In the standard Newtonian perturbation theory (SPT), the governing equations
of the Poisson equation, the Euler equation, and the continuity equation
are often solved for the pressureless medium to yield recurrent solutions for
the matter density fluctuation (see, e.g., 
\cite{GOGRET86,MASASU92,JABE94,BECOET02}). In particular, a simple exact
perturbative solution can be derived in a $\Lambda$CDM universe \cite{YOGO16},
and the matter density fluctuation up to the second order in perturbations is
\beeq
\label{newt}
\delta_m(x^\mu)=-D_1\Delta \RR+\frac 57D_{A}\nabla_\al\left(\RR^{,\al}\Delta \RR
\right)+\frac 17D_{B}\Delta\left(\RR^{,\al}\RR_{,\al}\right)~,
\eneq
where commas represent spatial derivatives, $\Delta=\gbar^{\al\be}\nabla_\al
\nabla_\be$ is a Laplacian operator,
three time-dependent growth functions $D_1(t)$, $D_A(t)$, and $D_B(t)$
are given in Eqs.~\eqref{growth} and~\eqref{growth2}, and the comoving-gauge
curvature perturbation~$\RR(\bm{x})$ in a hypersurface at early time~$t_i$
specifies the initial condition (see Appendix~\ref{setup}). The first
growth function~$D_1(t)$ corresponds to the standard linear-order
growth factor~$D(t)$
in literature, if normalized to unity at present time. So, the first
term is just the linear-order matter density fluctuation:
\beeq
\de_m^{(1)}(\xvec,t_i)=-D_1(t_i)\Delta\RR(\xvec)~.
\eneq
The other two
growth functions are often approximated as $D_A=D_B\approx D_1^2$
in literature, while this equality is true only in the Einstein-de~Sitter
universe. Two quadratic terms in Eq.~\eqref{newt}
comprise the second-order matter density fluctuation often characterized
by the Fourier kernel~$F_2$ in the standard perturbation theory (SPT),
when expressed in terms of the initial density fluctuation~$\de_m(\xvec,t_i)$.

In the relativistic perturbation theory, there exist more degrees of freedom
due to the diffeomorphism symmetry in general relativity. In particular,
the matter density fluctuation, defined as
\beeq
\delta(x^\mu):={\rho_m(x^\mu)\over\bar\rho_m(t)}-1~,
\eneq
is gauge-dependent, as it changes depending on our choice of hypersurface,
where $\rho_m(x^\mu)$ is the matter density and the background matter 
density~$\bar\rho_m(t)$ is just a function of time.
Even at the linear order in perturbations, many different matter density
fluctuations exist as the solutions of the Einstein equation in different
gauge conditions (see, e.g., \cite{BARDE80,YOO14b}). 
Consequently, we need to choose
one gauge condition and its matter density fluctuation among many other
choices, when we want 
to relate it to the galaxy number density fluctuation, which
is known as galaxy bias \cite{KAISE84,POWI84,BCEK91,FRY96} (see also
\cite{DEJESC18} for a recent review). Naturally, we demand that
the matter density fluctuation at the linear order in relativistic perturbation
theory reduces to one in the
standard Newtonian perturbation theory. This condition leads to a choice of
hypersurface described by
the synchronous gauge or the comoving gauge, in which the matter density
fluctuations are indeed identical at the linear order
to each other and the Newtonian one ({\it not} in Newtonian gauge).
The hypersurface in those gauge choices
represents the proper-time hypersurface of the matter fluid 
\cite{CHLE11,YOHAET12,JESCHI12,YOO14b}.

The situation beyond the linear order in perturbations is somewhat ambiguous,
as the matter density fluctuations in the synchronous gauge
and the comoving gauge are {\it different} \cite{YOO14b}.
Furthermore, the spatial gauge condition,
which is of {\it no} relevance for scalar fluctuations at the linear order, 
starts to affect the matter
density fluctuation beyond the linear order.
The matter density fluctuation up to the second order in perturbations is then
\cite{YOGO16}
\beeq
\label{propt}
\delta_m(x^\mu)=D_1\left(-\Delta \RR+\frac 32 \RR^{,\al}\RR_{,\al}
+4\RR\Delta \RR
  \right)+\frac 57D_{A}\nabla_\al\left(\RR^{,\al}\Delta \RR
\right)+\frac 17D_{B}\Delta\left(\RR^{,\al}\RR_{,\al}\right)~,
\eneq
and it is almost identical to the matter density fluctuation in
Eq.~\eqref{newt}, except two extra terms in the round bracket, arising from
the nonlinear constraint equation of general relativity (see also
\cite{BRHIET14,BRHIWA14}
for other derivations in the synchronous gauge). 
These relativistic contributions are generic
in the proper-time hypersurface, and independent of spatial gauge choice,
which only affect the Newtonian contributions in Eq.~\eqref{propt}.

In Section~\ref{issues} further discussion is presented in regard to
the gauge choice and its consequence.

\subsection{Bispectrum of the matter density fluctuation in the squeezed limit}
\label{MVformalism}

Using the matter density fluctuation at the second order in perturbations,
we compute its three-point correlation function and the bispectrum. In 
particular, our primary interest is the three-point correlation function 
$\xi_m(\xone,\xtwo,\xthree)$  in the squeezed triangular configuration,
in which two points $\xone=\xtwo$ are identical (or close enough)
and a third point~$\xthree$ is far away from the two points:
\beeq
\label{xisqz}
\xisqz:=\xi_m(\xone,\xtwo,\xthree)~,\Dquad |\xone-\xtwo|\ll |\xone-\xthree|~.
\eneq
The correlation function in this special triangle is useful in the limit
(also known as the squeezed limit):
\beeq
\label{xisqzlim}
\xisqzlim:=\lim_{|\xone-\xthree|\RA\infty}\xisqz~,
\eneq
in which the correlation function~$\xisqzlim$
is subject to various consistency relations
\cite{MALDA03,CRZA04,CRNOET13,KERI13,PEPI13,HUJOWO19,MIYO22a},
providing clues for
the nature of the initial conditions at the early Universe. Moreover,
the galaxy two-point correlation function or the power spectrum
receives corrections from the squeezed three-point correlation~$\xisqz$
in the presence of non-Gaussianity \cite{MALUBO86,MAVE08}, 
which is evident for non-vanishing
three-point correlation function on large scales.
While the non-Gaussian contribution from the Newtonian
non-linear evolution in the matter density fluctuation in Eq.~\eqref{newt}
is negligible on large scales, the non-Gaussian contribution from the
initial conditions characterized by
\beeq
\label{ini}
\RR(\xvec):=\RR_g(\xvec)+\frac35\fnl\RR_g^2(\xvec)~ \Dquad
\up{at}\quad t=t_i~,
\eneq
shows prominent features in the galaxy power spectrum on large scales
\cite{DADOET08,MAVE08,SLHIET08}, where~$\RR_g$ represents the linear-order
Gaussian fluctuation and $\fnl$ is often assumed to be 
constant.\footnote{Though we already used the subscript~$g$ to represent 
galaxies in the number density fluctuation~$n_g$, we use the same subscript~$g$
here to indicate the Gaussian field, following the convention.}
Note that the numerical factor 3/5 in the convention reflects the linear-order
time-evolution of the Newtonian gauge potential~$\px$ in Eq.~\eqref{Newtg}
from the radiation dominated era to the matter dominated era.

In the simplest model of galaxy formation,
galaxies (or dark matter halos) form in an over-dense region, where the
matter density fluctuation is above a threshold~$\de_m\geq\delta_c$,
characterized by a critical density contrast~$\de_c$ 
\cite{PRSC74,KAISE84,BCEK91}.
For the Gaussian distribution of the matter density fluctuation, the galaxy
number density is a biased tracer of the matter density fluctuation, and its
two-point correlation function can be analytically computed 
\cite{KAISE84,POWI84,BBKS86,BCEK91} as
\beeq
\xi_g(\xone,\xtwo)={\nu^2\over\sigma^2_R}~\xi_m(\xone,\xtwo)~,
\eneq
where $\sigma_R^2$ is the rms matter density fluctuation with smoothing
length~$R$ and $\nu:=\delta_c/\sigma_R$. In the presence of non-Gaussianity
due to the non-Gaussian initial conditions or the non-linear evolution,
the probability distribution of the
matter density fluctuation is altered, affecting the number density of the
peaks above the threshold. The leading correction to the galaxy two-point
correlation function arises from the three-point correlation function~$\xisqz$
in the squeezed triangular configuration  \cite{MALUBO86,MAVE08}
\beeq
\label{MVcorr}
\Delta \xi_g(\xone,\xtwo)={\nu^3\over\sigma^3_R}~\xisqz(\xone,\xone,\xtwo)~.
\eneq
In this context, it is evident that 
the primordial non-Gaussianity in the initial hypersurface in Eq.~\eqref{ini}
induces a non-vanishing three-point correlation function, in particular
$\xisqz$ in the squeezed configuration, 
and in turn generates the correction to
the galaxy two-point correlation function~$\xi_g$ or the power spectrum~$P_g$.
Similarly, the non-linear relativistic terms in Eq.~\eqref{propt}
also contribute to the three-point correlation function~$\xisqz$ and hence
the galaxy two-point correlation function~$\Delta\xi_g$, according
to Eq.~\eqref{MVcorr}.

Here we present the computation of the three-point correlation function~$\xisqz$
in the squeezed triangular configuration,
using the matter density fluctuation in Eq.~\eqref{propt}.
For the leading contribution in the bispectrum or the three-point correlation
function, we need to contract one second-order contribution~$\de_m^{(2)}$
and two linear-order contributions~$\de_m^{(1)}$
to the matter density fluctuation.
The expression of~$\xisqz$ is derived in Eq.~\eqref{sqz} as
\beeq
\label{sqzhere}
\xi_{\rm sqz}(\xone,\xone,\xtwo)
=\int\Dkone\int\Dktwo~ e^{i\kvec_{12}\cdot(\xone-\xtwo)}
\bigg[B_{112}+B_{211}+B_{121}\bigg]
(\mathbf k_1,\mathbf k_2,-\mathbf k_{12})~,
\eneq
where $\kvec_{12}:=\kvec_1+\kvec_2=-\kvec_3$ and 
three bispectrum contributions $B_{112}$, $B_{211}$, and $B_{121}$ defined
in Eqs.~\eqref{BBBA}$-$\eqref{BBBB} are  expressed in terms
of the Fourier kernels~$\FF(\kone,\ktwo)$ of the individual
components in Eq.~\eqref{propt}. The detailed computation of the individual
Fourier kernels is presented in Appendix~\ref{details}, and
all the components in Eq.~\eqref{propt} being at the source position
are categorized as the contributions at the source position in 
Appendix~\ref{srccont}. Furthermore, if the correlation function~$\xisqz$
in the squeezed triangle is treated as a two-point correlation function 
as in Eq.~\eqref{MVcorr} and hence just as a function of
its separation, its Fourier transformation yields the power spectrum
\beeq
\label{pokmain}
\Delta P(k)=\int d^3L~e^{-i\mathbf k\cdot\mathbf L}~\xi_{\rm sqz}=
\int\Dkone \bigg[B_{112}+B_{211}+B_{121}\bigg]
(\mathbf k_1,\mathbf k-\kvec_1,-\mathbf k)~,
\eneq
where we defined the separation vector
\beeq
\bdv{L}:=\xone-\xtwo~.
\eneq
Note that such power spectrum obtained in a hypersurface is {\it not} a
direct observable, in particular, on large scales, 
and we present further discussion about this issue later.

To compute~$\xisqz$ in Eq.~\eqref{sqzhere},
consider two second-order relativistic contributions to the
matter density fluctuation in Eq.~\eqref{propt}. Two Fourier kernels
for the relativistic terms are computed in Appendix~\ref{srccont} as
\bear
&&\bullet~~
\frac 32 D_1(z)\RR^{,\al} \RR_{,\al}:\quad
\FF_s(\kone,\ktwo)=-\frac 32 D_1(z)\kone\cdot \ktwo ~,\\
&&\bullet~~4D_1(z)\RR\Delta \RR:\qquad
\FF_s(\kone,\ktwo)=-2D_1 \left(k_1^2+k_2^2\right) ~,
\enar
and their one-point ensemble averages are  
\beeq
\label{cancelfnl}
\left\langle\frac 32 D_1\RR^{,\al} \RR_{,\al}\right\rangle=
\frac 32 D_1\avg_2\,,\Dquad 
\bigg\langle4D_1\RR\Delta \RR\bigg\rangle=-4D_1\avg_2~.
\eneq
where we defined a dimensionful constant
\beeq
\label{Deltan}
\avg_n:=\int\Dkk ~k^n P_\RR(k)\,,\Dquad [\avg_n]=L^{-n}~.
\eneq
Though not diverging in the infrared,
the ensemble averages of the relativistic contributions
are {\it non-vanishing} ($\avg_2\neq0$), and so is the ensemble average of the
matter density fluctuation~$\de_m$ in Eq.~\eqref{propt}.
We discuss this point in detail
in Section~\ref{ensavg}. After subtracting these non-vanishing constants,
their contributions to the bispectrum~$B_{112}$ are
\beeq
B_{112}=-\frac32D_1^3k_1^2k_2^2k_3^2P_\RR(k_1)P_\RR(k_2)\left[1+\frac53
  \left({k_1^2\over k_3^2}+{k_2^2\over k_3^2}\right)\right]~,
\eneq
where $k^3P_\RR(k)\propto k^{n_s-1}$ is the primordial curvature
power spectrum with the spectral index $n_s\simeq1$ and we used
\beeq
\kone\cdot\ktwo=\frac12\left(k_3^2-k_1^2-k_2^2\right)~.
\eneq
In the squeezed limit, in which the separation $L=|\xone-\xtwo|$ becomes
infinite, the exponential factor in Eq.~\eqref{sqzhere} imposes 
\beeq
\mathbf k_l:=\mathbf k_{12}=\mathbf k_1+\mathbf k_2\RA0~,\Dquad
\kvec_s:=\mathbf k_1\simeq -\mathbf k_2~,
\eneq
and the bispectrum~$B_{112}$ behaves in this limit as
\beeq
B_{112}=-\frac32D_1^3k_s^4k_l^2P^2_\RR(k_s)\left(1+{10\over3}
{k_s^2\over k_l^2}\right)\propto -\frac32 
{1\over k_s^2}P_m^2(k_s)\left({k_l^2\over k_s^2}+{10\over3}\right)~,
\eneq
where the matter density power spectrum scales as
$P_m(k)\propto k^4P_\RR(k)\propto k^{n_s}$.
The integration over the short mode of the bispectrum~$B_{112}$ yields
its contribution to $\xisqzlim$ in the squeezed limit, or its contribution
to $P(k)$ in the squeezed limit, both of which vanish as $k_l\propto
1/L\RA0$. The contribution of the term with constant 10/3
is removed from the tadpole contribution discussed in Appendix~\ref{FDEV}.

The other two bispectra can be computed in the same way, and they are
identical
\beeq
B_{211}=B_{121}=-\frac32D_1^3k_1^2k_2^2k_3^2P_\RR(k_2)P_\RR(k_3)\left[1+\frac53
  \left({k_2^2\over k_1^2}+{k_3^2\over k_1^2}\right)\right]~,
\eneq
and in the squeezed limit they become
\beeq
\label{eftfnl}
B_{211}=B_{121}\propto -\frac32{1\over k_s^2}P_m(k_s)P_m(k_l)
\left[\left(1+\frac53\right){k_s^2\over k_l^2}+{5\over3}\right]~.
\eneq
The long-mode contribution in the round bracket diverges in the limit 
$k_l\RA0$, and  this contribution is known as the relativistic correction
to the primordial non-Gaussianity signal. With the volume factor $d^3k_l$
in Fourier space, this contribution to~$\xisqzlim$ vanishes in fact in the
squeezed limit. However, the power spectrum in the hypersurface scales
as
\beeq
\Delta P(k_l)\propto {1\over k_l^2}P_m(k_l)~,
\eneq
and hence the correction to the galaxy power spectrum
would scale in the same way, according to Eq.~\eqref{MVcorr}:
\beeq
\Delta P_g(k_l)\propto {1\over k_l^2}P_m(k_l)~,
\eneq
which should give rise to a prominent feature on large scales, 
similar to that by the presence of~$\fnl$ (see, e.g., 
\cite{VEMA09,BRHIET14,BRHIWA14}).

In essence, the same computation has been performed \cite{MAVE08}
in the presence of the local-type primordial non-Gaussianity in
Eq.~\eqref{ini}, but without the second-order relativistic contributions.
The presence of $\fnl\RR^2$ term in the initial condition
would yield
\beeq
\label{fnleq}
-D_1\Delta\RR(\xvec)=-D_1\left[\Delta\RR_g+\frac65\fnl\left(
\RR_g^{,\al}\RR_{g,\al}+\RR_g\Delta\RR_g\right)\right]~,
\eneq
whose second-order Fourier kernels are similar to those for the relativistic
contributions as 
\beeq
\frac65\fnl D_1~\kvec_1\cdot\kvec_2~,\Dquad \frac35\fnl D_1(k_1^2+k_2^2)~,
\eneq
and their contribution to the bispectrum is
\beeq
B_{112}=\frac65\fnl D_1^3k_1^2k_2^2k_3^2P_\RR(k_1)P_\RR(k_2)~.
\eneq
In the squeezed limit, the bispectrum $B_{112}$ vanishes, but 
the other two bispectra scale with~$k_l$ as
\beeq
B_{211}=B_{121}\propto-\frac32{1\over k_s^2}P_m(k_s)P_m(k_l)\left[-\frac45
\fnl{k_s^2\over k_l^2}\right]~.
\eneq
In comparison to Eq.~\eqref{eftfnl}, one reaches the conclusion 
\cite{BRHIET14,BRHIWA14,CAMASA15,VIVEMA14,MAPIRO21} 
that the relativistic contributions in the matter density fluctuation 
(or nonlinearity in general relativity) generate the {\it effective}
non-Gaussianity
\beeq
\Delta\fnl=-{10\over3}~,
\eneq
or $\Delta F_\up{nl}=-5/3$ for different notation convention in Eq.~\eqref{FNL}.
A few remarks are in order, regarding the extra terms in Eq.~\eqref{fnleq}
with~$\fnl$, compared to the relativistic corrections in Eq.~\eqref{propt}.
With the same coefficients with~$\fnl$ for both terms, their one-point
ensemble averages cancel each other, as can be 
inferred in Eq.~\eqref{cancelfnl}. The peculiar scale-dependence
$P_m(k_l)/k_l^2$ arises solely from the contribution $\RR\Delta\RR$, which
can originate from the presence of the primordial non-Gaussianity or 
the Hamiltonian constraint equation in general relativity. 
While $\fnl$ can be zero in the initial condition, the Hamiltonian constraint
is satisfied all the time (hence the constraint).
The effective non-Gaussianity
can therefore be read off from the coefficient of $\RR\Delta\RR$ without
computing the three-point correlation function or the bispectrum, as
shown in Eq.~\eqref{comp}.

We continue with two Newtonian contributions in Eq.~\eqref{propt}.
Two Fourier kernels of the Newtonian contributions
are computed in Eqs.~\eqref{sptone} and~\eqref{spttwo}:
\bear
\FF_s(\kone,\ktwo)&=&\frac{5}{14} D_A(z) \left(\kone+\ktwo\right)\cdot
\left(k_2^2\kone+k_1^2\ktwo\right)~,\\
\FF_s(\kone,\ktwo)&=&\frac 17 D_B(z) \left(\kone+\ktwo\right)^2
\kone\cdot \ktwo ~.
\enar
In the Einstein-de~Sitter Universe, where two growth factors become equivalent,
i.e., 
$D_A=D_B=D_1^2$, two Fourier kernels add up to be the SPT kernel derived
in Eq.~\eqref{stdf2}:
\beeq
\label{sptf2}
\FF_2(\kone,\ktwo)=D_1^2(t)\left[\frac57+\frac12{\kone\cdot\ktwo\over k_1k_2}
\left({k_1\over k_2}+{k_2\over k_1}\right)+\frac27
\left({\kone\cdot\ktwo\over k_1k_2}\right)^2\right]~.
\eneq
Naturally, the ensemble averages of two Newtonian
contributions vanish. Their contributions to the bispectrum are
then computed as 
\beeq
B_{112}=D_1^2k_1^2k_2^2P_\RR(k_1)P_\RR(k_2)\left\{{5\over14}D_A\left[k_3^2
(k_1^2+k_2^2)-(k_1^2-k_2^2)^2\right]+\frac17D_B\left[k_3^4-k_3^2(k_1^2+k_2^2)
\right]\right\}~,
\eneq
and two other bispectra can be readily obtained by permutation of its
arguments. With $k_3\RA0$ and $k_1^2=k_2^2$, 
all three bispectra naturally vanish in the squeezed limit. 
It is well established \cite{BECOET02} that the nonlinearity 
in the Newtonian perturbation theory vanishes in the infrared.

\subsection{Setting the stage:
Relativistic corrections to the primordial non-Gaussian signal?}
\label{stage}
Up to this point, we have performed a straightforward computation of the
three-point correlation function or the bispectrum, given the expression
of the matter density fluctuation in Eq.~\eqref{propt}. It appears that
the relativistic contributions inherent in general relativity give rise
to correction terms to the standard second-order expression for the matter
density fluctuation and the particular term~$\RR\Delta\RR$ yields
unique behavior  on large scales in the power spectrum, 
which is then also related to the galaxy power spectrum. 
While there is no doubt
in the sanity for the matter density fluctuation in Eq.~\eqref{propt},
it has been intensively debated 
\cite{PASCZA13,BRHIET14,BRHIWA14,DEDOGR15,DEDOET17,MAPIRO21}
whether the relativistic contributions
give rise to a correction in the galaxy power spectrum on large scales
in the same way the presence of the primordial non-Gaussianity affects
the galaxy power spectrum. For instance, it has been argued 
\cite{PASCZA13,DEDOGR15,YOGO15,DEDOET17,KOUMET18,UMKOET19} 
that
an extra coordinate transformation like the dilatation or the conformal
Fermi coordinate can remove such contribution, at least, for the single-field
inflationary model, and hence no relativistic correction to the primordial
non-Gaussian signal on large scales in the galaxy power spectrum.

In Section~\ref{issues} we present several issues associated with the 
computation in this section, some of which are largely {\it unknown} 
in literature and some of which are {\it debated} in the community.
We present solutions to some of the issues, but {\it not} all the issues
are resolved. In short, we show that the relativistic contributions in the 
matter density fluctuation {\it cannot} be removed by 
any coordinate transformation.  However, there exist {\it gauge ambiguities}
in relating the matter density fluctuation to the galaxy number density
fluctuation.
The theoretical description of the observed galaxy
number density should be gauge-invariant, and it is proved at the linear
order in perturbations \cite{YOFIZA09,YOO10,BODU11,CHLE11,JESCHI12}.
However, it becomes tricky \cite{YOO14b,KOUMET18,UMKOET19,UMKO19}
in general relativity 
with many subtle and unresolved issues beyond the linear order 
in perturbations to relate the matter density fluctuation to 
the (intrinsic) galaxy number density fluctuation (or galaxy bias), 
which is not yet an observable. 

Furthermore, it is important that in the end
we need to provide theoretical predictions for observable quantities
such as the observed galaxy correlation function. While the power spectrum
or the bispectrum is a useful statistics, these statistics are defined
in a hypersurface outside the observed light-cone volume, particularly
when our primary interest lies in their signals on large scales.
While the (theory) power spectrum in a hypersurface is related to the
observed power spectrum \cite{GRSCET20}, a simple way of measuring the
power spectrum in observations
yields signals very different from the (theory) power spectrum in a hypersurface
on large scales due to the wide angle effect, the time-evolution along
the line-of-sight direction, the curvature of the sky, and so on
(see, e.g., \cite{FEKAPE94,TEHAET98,RASAPE10,YODE13,YOSE12} 
for detailed discussion of those
complications). For example, note that the correlation function
$\xisqz(\xone,\xone,\xtwo)$ in Eq.~\eqref{sqzhere} is {\it not} just a function
of its separation $\bdv{L}=\xone-\xtwo$; it depends on two observed
angles~$\Vang_1$ and~$\Vang_2$ for two positions~$\xone$ and~$\xtwo$.
Without proper consideration of those issues,
a simple computation of the (theory) power spectrum in a hypersurface 
would lead to conclusions that are highly biased when compared to observations,
in particular for the primordial non-Gaussian signature.
On the other hand, the correlation function involves no such complication,
and we compute the observed matter density correlation function 
in Section~\ref{sec:matt}.

\section{Theoretical issues in the previous calculations}
\label{issues}

\subsection{Convention for the primordial non-Gaussianity: Gaussian vs
non-Gaussian?}
\label{convention}
We first 
clarify different notation conventions in literature, regarding the
primordial non-Gaussianity. While they 
are just a matter of notational preference,
we show that the separation into Gaussian and non-Gaussian fields
becomes {\it ambiguous} beyond the linear order in perturbations, where things
are generally non-Gaussian. Given the general metric representation in 
Eq.~\eqref{metric}, the initial condition~$\RR(\xvec)$
is set in terms of the comoving-gauge curvature perturbation~$\varphi_v$
in Eq.~\eqref{cv}:
\beeq
\RR(\xvec):=\varphi_v(\xvec,t_i)~,
\eneq
where $t_i$ represents some early time~$t_i$ for the initial conditions.
The comoving-gauge curvature perturbation~$\varphi_v$ is conserved in time
outside the horizon to all orders in perturbations,
and it is conserved on all scales at the linear order in a universe with
pressureless medium. The other popular choice to set up the initial condition 
is to adopt a different representation of the spatial metric:
\beeq
\label{expo}
ds^2=-a^2\left(1+2\alpha\right)d\eta^2-2a^2\beta_{,\al}dx^\al d\eta+
a^2e^{2\zeta}\de_{\al\be}dx^\al dx^\be~,
\eneq
where a rectangular coordinate~$\de_{\al\be}$ is chosen in the three metric 
and the spatial C-gauge ($\ga\equiv0$)
in Eq.~\eqref{CCC} is adopted. By expanding the 
exponential factor, the relation between two notation conventions is
\beeq
\RR=\zeta+\zeta^2+\frac23\zeta^3+\frac13\zeta^4+\cdots~.
\eneq
Assuming that the same comoving gauge condition in Eq.~\eqref{gauge} is 
adopted, both notation conventions yields the same quantity at the linear
order:
\beeq
\RR^{(1)}=\zeta^{(1)}~,
\eneq
but there exist obvious differences beyond the linear order in perturbations
\beeq
\RR^{(2)}=\zeta^{(2)}+\big[\zeta^{(1)}\big]^2~.
\eneq
Furthermore, while the spatial gauge condition is not relevant at the linear
order, it becomes so beyond the linear order, where scalar, vector, and
tensor components start to mix with each other. In particular, the presence of 
traceless transverse component~$\ga_{\al\be}$ (gravitational waves)
in the exponential representation
\beeq
a^2\exp\left[2(\zeta~\de_{\al\be}+\ga_{\al\be})\right]dx^\al dx^\be
=a^2\left[(1+2\zeta+2\zeta^2+\cdots)\de_{\al\be}+2(\ga_{\al\be}+\ga_{\al\ga}
\ga_{\ga\be}+\cdots)\right]dx^\al dx^\be~,
\eneq
generates a non-vanishing off-diagonal scalar component~$\ga$ beyond the 
linear order (for instance, from $\ga_{\al\ga}\ga_{\ga\be}$),
changing the spatial gauge condition adopted in the exponential
metric representation, i.e., it is {\it no longer} equivalent to the spatial
C-gauge.
Therefore, care must be taken in interpreting the calculations in two
different notation conventions, in particular for the initial conditions
beyond the linear order in perturbations.

Given our convention for the initial conditions in Eq.~\eqref{ini},
the other convention for the initial conditions in literature is
\beeq
\label{FNL}
\zeta(\xvec):=\zeta_g(\xvec)+\frac35F_\up{NL}~\zeta_g^2(\xvec)~,
\eneq
which implies
\beeq
\RR^{(1)}=\zeta^{(1)}=\RR_g=\zeta_g~,\Dquad
\RR^{(2)}=\frac35\fnl~\RR_g^2=\frac35F_\up{NL}~\zeta_g^2+\zeta_g^2~,
\eneq
and
\beeq
\label{fnlFNL}
\fnl=F_\up{NL}+\frac53~.
\eneq
It is evident that a vanishing non-Gaussianity in one convention
($\fnl=0$ or $F_\up{NL}=0$) means a non-vanishing non-Gaussianity
in the other convention. Note that the parametrization of~$\fnl$ or $F_\up{NL}$
is a matter of notational preference or representation.
This shows that it makes {\it less} sense to distinguish Gaussian and
non-Gaussian fields
beyond the linear order in perturbations, as they are all generically
non-Gaussian. Explicitly, the matter density fluctuation
in Eq.~\eqref{propt} in two different conventions is
\bear
\label{comp}
\delta_m(x^\mu)&=&D_1\left[-\Delta \RR_g+\frac65\left(\frac54-\fnl\right)
\RR^{,\al}_g\RR_{g,\al}+\frac65\left(\frac{10}3-\fnl\right)\RR_g\Delta\RR_g
\right]+\de_\up{Newt.}^{(2)} \nnn
&=&
D_1\left[-\Delta \zeta_g-\frac65\left(\frac5{12}+F_\up{nl}\right)\zeta_g^{,\al}
\zeta_{g,\al}+\frac65\left(\frac53-F_\up{nl}\right)\zeta_g\Delta\zeta_g
\right] +\de_\up{Newt.}^{(2)} ~,
\enar
where we defined the second-order Newtonian contribution to the matter
density fluctuation
\beeq
\label{newt2}
\de_\up{Newt.}^{(2)}:=\frac 57D_{A}\nabla_\al\left(\RR_g^{,\al}
\Delta \RR_g\right)+\frac 17D_{B}\Delta\left(\RR_g^{,\al}\RR_{g,\al}\right)~.
\eneq
Equation~\eqref{comp} is often expressed in terms of a Newtonian 
potential~$\phi_g=(3/5)\RR_g$ in the initial condition as
\bear
\delta_m(x^\mu)&=&{5D_1\over3}\left[-\Delta \phi_g+
2\left(\frac54-\fnl\right)\phi_g^{,\al}\phi_{g,\al}
+2\left(\frac{10}3-\fnl\right)\phi_g\Delta\phi_g
\right]+\de_\up{Newt.}^{(2)} \nnn
&=&
{5D_1\over3}\left[-\Delta \phi_g-2\left(\frac5{12}
+F_\up{nl}\right)\phi_g^{,\al}
\phi_{g,\al}+2\left(\frac53-F_\up{nl}\right)\phi_g\Delta\phi_g
\right] +\de_\up{Newt.}^{(2)} ~.
\enar
The matter density fluctuation is generically non-Gaussian beyond the
linear order, even in the limit $t\RA0$. This does not imply that any of the
non-Gaussian parametrization is useless.
Given a theory for the initial conditions, 
accurate predictions for~$\fnl$ or $F_\up{NL}$ can be made, and these
numbers can be observationally tested.

\subsection{Spatial gauge transformation: pure gauge mode or not?}
\label{spatial}
Diffeomorphism is a symmetry of general relativity, and this symmetry in
cosmology is often expressed as gauge freedom in a general coordinate
transformation:
\beeq
\label{ct}
\tilde x^\mu=x^\mu+\xi^\mu~,
\eneq
i.e., the same physical point is described by two different coordinate
values in two different coordinate systems. A gauge choice then  amounts to
completely fixing four degrees of freedom in Eq.~\eqref{ct}, which are
temporal and spatial \cite{BARDE80}. A temporal gauge choice or time-slicing
implies a choice of three-dimensional hypersurface of simultaneity in such
a coordinate system, and various temporal gauge choices are
discussed in literature (see, e.g., \cite{ELBR89}),
in conjunction with their physical meaning. However, a spatial gauge
choice has received {\it little} attention in literature. The main reason is 
due to the
symmetry in a homogeneous and isotropic universe, for which the background
quantities are just a function of time only. Since any linear-order
perturbation~$\de T$ of a tensorial quantity~$T$ gauge transforms
in terms of Lie derivative~$\cal L$ as
\beeq
\de_\xi~\de T=-{\cal L}_\xi\bar T+\OO(2)~,
\eneq
the symmetry in the background universe renders the spatial gauge
transformation irrelevant at the linear order for scalar fluctuations
({\it not} for general tensors). However, beyond the linear
order, perturbations gauge transform as 
\beeq
\label{GT1}
\de_\zeta~\de T=-{\cal L}_\zeta\bar T+\frac12{\cal L}_\zeta{\cal L}_\zeta\bar T
-{\cal L}_\zeta\de T+\OO(3)~,
\eneq
and a spatial gauge choice plays a role in determining the perturbation
variables uniquely \cite{BRSO99,YODU17}, even for the scalar fluctuations.
Note that we have used the exponential parametrization of a general 
coordinate transformation
\beeq
\tilde x^\mu=e^{\zeta^\nu\pa_\nu}x^\mu=x^\mu+\zeta^\mu+\frac12\zeta^\mu{}_{,\nu}
\zeta^\nu+\OO(3)~,
\eneq
where two parametrizations are related as
\beeq
\xi^\mu=\zeta^\mu+\frac12\zeta^\mu{}_{,\nu}\zeta^\nu+\OO(3)~.
\eneq

As a simple illustration, consider the matter density fluctuation~$\de_m$
in a given hypersurface, which is fixed by a temporal gauge choice
($\xi^\eta=0$).
The matter density fluctuation is uniquely fixed at the linear order,
but ambiguities appear beyond the linear order in perturbations as
\beeq
\label{dmgt}
\delta_m(x^\mu)=\tilde \delta_m(\tilde x^\mu)=\tilde \delta_m(x^\mu)+
\xi^\al~\pa_\al\tilde \delta_m\Big|_{x^\mu}~,
\eneq
where the background matter density~$\bar\rho_m(t)$
is identical in the same hypersurface.
In other words, though the matter density fluctuation at a given physical point
is invariant, its functional form in a given hypersurface changes,
according to a spatial gauge choice.  It was shown \cite{YOO14b} that
two different spatial gauge choices in a proper-time hypersurface result
in different second-order matter density fluctuations.
Consequently, the three-point correlation
function, whose leading order term depends on the second-order expression,
is affected by a spatial gauge choice:
\beeq
\label{bischange}
\AVE{\tilde\de(\xvec_1)\tilde\de(\xvec_2)\tilde\de(\xvec_3)}=
\Big\langle\de(\xvec_1)\de(\xvec_2)\de(\xvec_3)\Big\rangle-\left[\Big\langle
\xi^\al(\xvec_1)\pa_\al\de(\xvec_1)~\de(\xvec_2)\de(\xvec_3)\Big\rangle+
\up{cycl.}\right]+\OO(5)~.
\eneq
Unless the gauge choice is fully specified, the three-point
correlation function remains ambiguous and gauge-dependent.
This result is generic and applicable
to any three-point correlation function, e.g., the consistency relation
for a single-field inflation in the squeezed limit.
In \cite{MIYO22a}, the full three-point correlation function 
in a single-field inflationary model was computed, and the dependence of
a spatial gauge choice was investigated.

We close this section by concluding that spatial gauge choices are {\it not}
pure gauge modes.
In Section~\ref{theory} we show that the arbitrariness of a spatial gauge
choice is lifted in describing the observable quantities.

\subsection{Ensemble average: coordinate-dependent?}
\label{ensavg}
The ensemble average $\AVE{\OO}$ of a field~$\OO(x)$ is widely used in
cosmology, but often in a way its exact definition or meaning is left
ambiguous or implicit. Such ambiguity can then cause inconsistencies in
perturbation theory, in particular, beyond the linear order.
A field~$\OO(x)$ at a given spacetime position~$x^\mu$,
such as the matter density fluctuation
$\rho_m(x^\mu)$, can be split into the background $\bar\OO(t)$ and
the perturbation $\de\OO$ around it as
\beeq
\OO(x^\mu)=\bar\OO(t)\big[1+\de\OO(x^\mu)\big]~,
\eneq
and the dimensionless fluctuation~$\de\OO$ 
is often assumed to be Gaussian distributed
with zero mean at the linear order. In fact, the correct statement is
that at each wave number~$\kvec$, the linear-order fluctuation~$\de\OO(\kvec;t)$
in Fourier space defined in a given hypersurface set by~$t$ is independent
with different Fourier modes and Gaussian distributed with its variance
specified by the power spectrum $P_{\de\OO}(k;t)$. This applies to any
fields~$\OO$ at the linear order in perturbation theory,
because they are all linearly related.
To be more specific, consider a Gaussian probability functional 
$\mathcal{P}[\de\OO]$,
\bear
\left\langle\de\OO^{(1)}_{\kvec;t}\right\rangle&:=&
\int{\mathcal D}_{\de\OO}~\mathcal{P}[\de\OO]~\de\OO^{(1)}_{\kvec,t}=0~,\\
\left\langle\de\OO^{(1)}_{\kvec;t}\de\OO^{(1)}_{\kvec';t}\right\rangle&:=&
\int{\mathcal D}_{\de\OO}~\mathcal{P}[\de\OO]~\de\OO^{(1)}_{\kvec,t}
\de\OO^{(1)}_{\kvec',t}=(2\pi)^3\de^D(\kvec+\kvec')P_{\de\OO}(k;t)~, \nn
\enar
where we defined the ensemble average,
the integral is over all values of $\de\OO$, and the isotropy
of the power spectrum over the wave vector is assumed, though not needed.
Hence, the ensemble average of~$\OO(x)$ at the {\it linear order} yields
\beeq
\left\langle\OO(x^\mu)\right\rangle=\bar\OO(t)~,
\eneq
where the average over PDF and the Fourier transformation commute.
The ensemble average is over multiple realizations
of the universe, in which different values of~$\de\OO(\kvec;t)$ are given
for a fixed~$\kvec$ in a $t$-hypersurface (or the same spacetime position).
In other words, it is a local process, but with homogeneity it is identical
everywhere.

So far, we summarized explicitly the standard procedure for treating
the random fluctuations and the ensemble average. It is also common in the
standard procedure to assume the Ergodicity of the system ---
Once the random fluctuations are averaged
over a sufficiently large volume, the resulting average is ought to be
equivalent to the ensemble average over multiple realizations of the universe:
\beeq
\left\langle\OO(\xvec;t)\right\rangle\underset{\up{Erg.}}{\equiv}
\lim_{V\RA\infty}{1\over V}\int_Vd^3x'~\OO(\xvec';t)~,
\eneq
where we made it clear that the $t$-hypersurface is fixed in the local
ensemble average or the volume average over the $t$-hypersurface.
It is obvious from the
definitions that the ensemble average depends on a hypersurface set by
a time-coordinate~$t$, which changes depending
on our choice of gauge conditions. This point is often left
implicit or less emphasized in literature. At the linear order
in perturbations, however, this gauge dependence matters only for the
power spectrum, not for the mean, as the gauge transformation~$\mathcal{G}(x)$
at the linear
order is also a Gaussian random variable with zero mean, for instance,
\bear
\widetilde{\de\OO}{}^{(1)}_x&=&\de\OO^{(1)}_{x}+\mathcal{G}^{(1)}_{x}~,\Dquad
\left\langle\mathcal{G}^{(1)}_{\kvec;t}\right\rangle=0~,\Dquad
\left\langle\widetilde{\de\OO}^{(1)}_{\kvec;t}\right\rangle=0~,\\
\label{pokavg}
P_{\widetilde{\de\OO}}(k;t)&=&P_{\de\OO}(k;t)+P_{\mathcal{G}}(k;t)
+2P_{\de\OO,\mathcal{G}}(k;t)\neq P_{\de\OO}(k;t)~,
\enar
where $P_{\de\OO,\mathcal{G}}$ is the cross-power spectrum. It is important
to note that the same coordinate values in terms of~$x^\mu$ (or wave
vector~$\kvec$ in a $t$-hypersurface) correspond to
different physical spacetime points
(or different hypersurfaces) under the gauge transformation.

Beyond the linear order in perturbations, however, things are not as simple
as in the linear-order calculations, and one consequence is that
the ensemble average of~$\OO(x)$ {\it beyond the linear order}
quite often yields
\beeq
\label{holy}
\left\langle\OO(x^\mu)\right\rangle\neq\bar\OO(t)~.
\eneq
Before we proceed, we stress that there exist coordinate systems,
in which the relation $\AVE{\OO}=\bar\OO(t)$ holds beyond the linear order,
but there exist a lot more coordinate systems, in which the relation does
not hold beyond the linear order. This is a natural consequence, since the
ensemble average in LHS depends on a coordinate system, while the background
value~$\bar\OO(t)$ in RHS is just one number at a fixed value of~$t$.
For instance, the matter density fluctuation at the second order in
perturbations is given in Eq.~\eqref{newt}, where the first term is
the linear-order contribution and the remaining two terms
are the Newtonian second-order contributions. The Newtonian second-order
contributions are often expressed in terms of their Fourier kernel
$F_2(\kvec_1,\kvec_2)$ in Eq.~\eqref{sptf2}.
It is well-known that this second-order
Newtonian matter density fluctuation vanishes, when the ensemble
average is taken. However, there exist extra relativistic contributions at the
second order, or the quadratic terms in proportion to~$D_1(t)$
in Eq.~\eqref{propt}, where their ensemble averages
are non-vanishing as computed in Eqs.~\eqref{grd1} and~\eqref{grd2}.
Furthermore, the spatial gauge-transformation changes the second-order
Newtonian contributions in Eq.~\eqref{propt}, according to Eq.~\eqref{dmgt}.
One of the  popular choices is the synchronous gauge at the second order,
in which the hypersurface is identical to one for Eq.~\eqref{propt}, but
the spatial gauge is different. Consequently, the second-order Newtonian
terms are different from the standard contributions with~$F_2(\kvec_1,\kvec_2)$,
while the second-order relativistic contributions are identical.
It was shown \cite{YOO14b} that the ensemble average of the second-order
Newtonian contributions in the synchronous gauge is also non-vanishing.

So, in summary, we have demonstrated that the ensemble average depends
on a choice of hypersurface set by time coordinate~$t$ and
it also depends on a choice of spatial gridding, i.e., it depends on
a choice of full gauge condition. The time-slicing
matters already at the linear order, as borne out in Eq.~\eqref{pokavg},
and the spatial-gridding becomes relevant beyond the linear order, as shown
in Section~\ref{spatial}. Furthermore, the background
solution~$\bar\OO(t)$ in Eq.~\eqref{holy}, for example the background matter
density~$\bar\rho_m(t)$, is obtained by solving the Einstein equation
under the assumption that the solution is {\it just a function of time}
due to homogeneity and isotropy, which involves {\it no} average of any sort.
Hence the equality between $\AVE{\OO}$ and~$\bar\OO(t)$ is {\it not} expected to
be valid in general. This issue has been extensively discussed
in literature (see, e.g., 
\cite{BUEH97,MUABBR97,HISE05,LAREN09,ISWA06,CLELET11})
under the name of back-reaction, in which
it was shown that the background solutions obtained by ignoring any spatial
derivatives in the governing equation are different from those obtained
by averaging over a spatial
volume in a given hypersurface due to the non-commutativity
of two different procedures.

We conclude this subsection with a brief remark about the
``average in observation.'' It is quite common that the observers measure
average quantities of, for example, the luminosity distances,
the galaxy number density, the cosmic microwave
background temperature, and so on. These observational averages are certainly
independent of any coordinate choice we assume to describe observations.
Furthermore, these quantities are in fact the average over the observed
angle~$\Vang$
in the sky at a fixed (observed) hypersurface. For instance, the average
of the galaxy number density or the luminosity distance can be obtained
by averaging those quantities over the sky at a fixed observed redshift~$z$.
The average CMB temperature is also obtained by averaging the
observed CMB temperature $T(\Vang)$
over the sky at the Earth. These observational averages are
naturally different from the ensemble average or the background quantity
(see \cite{YOMIET19,GRSCET20,YOMIET19T,BAYO21}).
While the observer coordinate is fixed up to a trivial rotation
and is independent of coordinate systems in our theoretical calculations,
the ensemble average involves a full Euclidean average over the hypersurface.
 Since the observers can only perform angle average over the
sky at one position (or at the Earth), the lack of translation in the observer
position results in the cosmic variance or the discrepancy between the
observational average and the ensemble average over the observer hypersurface
\cite{MIYOET20}, i.e., the ensemble average is {\it not} a direct
observable.

\subsection{CFC, dilatations, special conformal transformations: 
extra gauge freedom?}
\label{extra}

In the context of cosmological perturbation theory the 
presentation of general covariance is through the corresponding gauge 
transformations of the fluctuations and in particular their 
scalar-vector-tensor decomposition \cite{BARDE80}.
It is then often left implicit that, 
for this transformation to be unique and well-defined, the corresponding 
generating vector field must decrease sufficiently fast at spatial infinity, 
or at least be bounded in space. Indeed, this is required for preserving
 the global validity of perturbation theory, but also by the Fourier 
transformation that underlies the scalar-vector-tensor decomposition.
Indeed, this is required for
preserving the global validity of perturbation theory, but also by the
Fourier transformation that underlies the scalar-vector-tensor decomposition.
It was shown in \cite{MAPIRO21} 
that this decomposition is {\it not} uniquely defined 
over a finite range of validity employed in coordinates
such as CFC (or even dilatation, but expanded over a limited range).

In the last decade two new types of transformations appeared in the
 field which do not satisfy the aforementioned condition. The first set
of transformations includes 
the dilatations \cite{WEINB03} and special conformal 
transformations (including tensor analogues) \cite{CRNOSI12,HIHUKH12} which 
lie at the heart of the consistency relations of inflationary correlation 
functions \cite{CRZA04,CHFIET08,CRNOSI12,CRNOET13,HIHUKH14}.
 These transformations arise as a global residual freedom of a
 fully gauge-fixed metric and do not reduce to the identity at spatial 
infinity, thus corresponding to so-called ``large gauge transformations.''
The other type includes the transformations that relate a typical coordinate 
system of cosmological perturbation theory (e.g., conformal Newtonian gauge) 
to the conformal Fermi coordinates (CFC) associated with some geodesic
 world-line \cite{PASCZA13,DAPASC15a,CAPASC17}. The CFC construction
 essentially corresponds to a deformed exponential map associated with 
a particular tetrad along the world-line and is therefore uniquely 
determined in some surrounding finite world-tube. However, since it is 
built order by order in a spatial expansion, its asymptotic behavior is 
obscure, so the corresponding transformation might or might not be a 
large gauge transformation and it is usually not even defined at infinity. 
Moreover, since one always stops at finite order, 
in practice its implementation takes the form of a large gauge transformation.

The utility of these extra types of transformations in cosmology 
lies in the fact that they affect the first two terms in a spatial expansion 
of the metric fluctuations and are therefore always used to simplify the
 metric representation {\it within  a finite patch} of space. 
This is in contrast to the standard 
practice of gauge transformations, in which the scalar-vector-tensor 
description are globally defined  in real space, 
privileging in particular {\it no point} or {\it direction} therein. 
Given these aspects and especially the fact that large gauge transformations
 arise {\it on top} of the standard gauge transformations, 
there seems to be some confusion
 in the literature about how the former affect cosmological observables.
 Indeed, while it is obvious that any physical quantity must be invariant
 under standard gauge transformations, the use of large gauge transformations 
is sometimes worded as ``going to the physical frame of the observer", i.e. 
the coordinates that an observer at $x^i = 0$ would use in practice to map 
their local patch. Attributing a physical meaning to the transformation might
 then seem to distinguish the available frames into ``right" and ``wrong" 
ones, which leads to the question: what happens to a cosmological observable 
if one picks a ``wrong" frame?

Here we wish to stress that the cosmological observables and their theoretical
descriptions are {\it invariant under all of these transformations},
simply because they are all just different types of
{\it coordinate transformations} (diffeomorphism).
Indeed, observables can be defined in a 
completely coordinate-independent way and at the fully non-linear level, i.e. 
without any reference to the background solution 
(see, e.g., \cite{MIYO19}), so they do not see the difference between 
standard and large gauge transformations. The fact that some coordinate system
 can be given a physical interpretation does not mean that one has to commit 
to that system to obtain the correct answer to a physical question ---
 the essence of general covariance. 
The content of this subsection is discussed in more detail and depth 
in a dedicated paper \cite{MIYO22b}, which addresses in particular some 
more subtle aspects. There we also show that the relevant property 
of CFC in cosmology, i.e. that the metric satisfies
$g_{\mu\nu} = a^2 \left( \eta_{\mu\nu} + {\cal O}(\bdv{x}^2) \right)$, 
can be achieved with a standard gauge transformation, which 
therefore avoids the ambiguity at spatial infinity.

\subsection{Proper-time hypersurface: unique choice for galaxy bias?}
\label{sec:gauge}
The observed galaxy clustering is described by two physically distinct 
effects: the volume effect and the source effect \cite{YOO09,YOFIZA09}.
The volume effect can be uniquely determined by solving the geodesic equation
for the mismatch between the observed and the physical source positions and 
volumes occupied by the source galaxies.
The source effect deals with the intrinsic
properties of the source galaxy sample and its mismatch, compared to the
observed properties of the source galaxy sample.
More importantly, the most dominant contribution to the source effect and also
to galaxy clustering overall is the matter density fluctuation that drives 
the fluctuation in the observed galaxy number density.

The relation between the galaxy and the matter density distributions, known
as galaxy bias \cite{KAISE84},
is an area of intense research in literature. In the simplest form discussed in 
Section~\ref{MVformalism}, galaxies form in an over-dense region with 
$\de_m\geq\de_c$, and this simple model yields that the galaxy
number density fluctuation~$\de_g$ is linearly proportional to the
matter density fluctuation, i.e.,  
\beeq
\label{simplebias}
n_g(\xvec)=\bar n_g(t)[1+\de_g(\xvec)]~,\Dquad \de_g(\xvec)=b~\de_m(\xvec)~,
\eneq
or the linear bias relation \cite{KAISE84,POWI84,BBKS86}, where the bias 
factor~$b$ is a constant. Beyond the linear order,
the bias relation can be further extended to
incorporate the higher-order perturbation contributions by introducing
nonlinear bias factors such~$b_n$ with $n\geq2$ \cite{FRGA93,FRY96,SCWE98,
SHTO99}, the tidal gravitational bias factor~$b_t$ \cite{BASEET12}, 
the relative velocity bias~$b_r$ between the 
baryon and the matter distributions 
\cite{DAPESE10,YODASE11,YOSE13b,SCHMI16}, or
effective field descriptions 
\cite{MCDON08,MCRO09,POSEZA14,SENAT15,VLWHAV15,SCELET19,VLCHSC20}
(see, e.g., \cite{DEJESC18} for recent review of galaxy bias on large scales).
However, put in the context of general relativity, this linear bias relation is
ambiguous and ill-defined. Under a change of coordinates in Eq.~\eqref{ct},
the galaxy number density fluctuation in Eq.~\eqref{simplebias} transforms 
at the linear order as
\beeq
\tilde\de_g(x^\mu)=\de_g(x^\mu)-{\bar n_g'\over\bar n_g}~T~,
\eneq
and for a matter distribution we recover the gauge-transformation relation
for the matter density fluctuation
\beeq
\tilde \de_m(x^\mu)=\de_m(x^\mu)+3\HH T~,\Dquad \bar\rho_m\propto a^{-3}~.
\eneq
To maintain both the linear bias relation to the matter distribution and the
gauge-transformation relation, a strict condition for the galaxy number
density is imposed:
\beeq
\label{strict}
b=-{1\over3\HH}{d\ln\bar n_g\over d\eta}={d\ln\bar n_g\over d\ln\bar\rho_m}~,
\eneq
such that the galaxy bias factor has to be related to the number density
evolution.
Again, for a matter distribution we recover the consistency relation $b=1$,
but for a general galaxy distribution in observations we already know that
this relation is {\it not} valid (see, e.g., \cite{EIANET01,WHBLET11}). 

Several attempts have been made in literature to generalize the bias
relation in general relativity.  Noting that 
the bias relation should reduce to the Newtonian description on small scales,
the matter density fluctuation~$\de_v$ in the comoving-synchronous gauge
was advocated in \cite{CHLE11} for the linear bias relation:
\beeq
\de_g(x^\mu)=b~\de_v(x^\mu)~,
\eneq
where the gauge-invariant matter fluctuation~$\de_v$ 
in both the synchronous gauge and the comoving gauge is
equivalent to the Newtonian matter density fluctuation at the linear order.
In this bias model, a gauge choice is made by hand, but with 
the Newtonian correspondence.
It was argued \cite{JESCHI12} that galaxies can only measure their own local
time, such that the bias model should be in a hypersurface of a constant-age:
\beeq
n_g(x^\mu)=\bar n_g(t_p)(1+b~\de_v)~,
\eneq
in support of the bias model in \cite{CHLE11}, where $t_p$ represents the
proper-time coordinate (or constant-age) in the rest frame of the source
galaxies. Since the proper time of local galaxies is {\it not} observable,
there exists an extra contribution, when the proper time is expressed
in terms of the observed redshift. Hence the most general linear-order
expression for galaxy bias in general relativity is then
\cite{YOHAET12}
\beeq
n_g(x^\mu)=\bar n_g(z)[1+b~\de_v-e~\dz_v]~,
\eneq
where two bias parameters are
\beeq
b:={d\ln\bar n_g\over d \ln\bar \rho_m}\bigg|_{t}~,\Dquad
e:={d\ln\bar n_g\over d\ln(1+z)}~,
\eneq
and the observed redshift~$z$ is related to the redshift~$z_p$ at the proper 
time~$t_p$ as $1+z=(1+z_p)(1+\dz_v)$ at the linear order with~$\dz_v$
in the comoving gauge.

Beyond the linear order in perturbations, the galaxy bias relation poses
more challenges in general relativity.
With the arguments for a proper-time hypersurface,
the galaxy number density can be generically written \cite{YOZA14} as
\beeq
\label{intflc}
n_g(x^\mu_s)=\bar n_g(t_p)\left[1+\de_g^\up{int}(x^\mu)\right]~,
\eneq
where $\de_g^\up{int}$ represents the intrinsic (nonlinear)
galaxy fluctuation in a proper-time hypersurface. 
It was argued \cite{YOZA14} that the intrinsic 
fluctuation should vanish upon average over the proper-time hypersurface:
\beeq
\label{vanish}
\bar n_g(t_p)=\AVE{n_g}_{t_p}~,\Dquad \AVE{\de_g^\up{int}}_{t_p}=0~.
\eneq
It was shown \cite{YOO14b} that both the synchronous gauge in 
Eqs.~\eqref{sync} and~\eqref{BBB} (denoted as gauge-II)
\beeq
\alpha\equiv0~,\Dquad v\equiv0~,\Dquad \beta\equiv0~,
\eneq
and the comoving gauge in Eqs.~\eqref{gauge} and~\eqref{CCC} (denoted as
gauge-I) 
\beeq
v\equiv0~,\Dquad \ga\equiv0~,
\eneq
describe the same proper-time hypersurface, but the matter density
fluctuations in two gauges are different, due to the difference in the
spatial gauge condition. The difference arises only at the second order.
Since only the matter density fluctuation 
in gauge-I satisfies the condition in Eq.~\eqref{vanish} and
the one in gauge-II
has non-vanishing one-point average, the gauge-I was favored for galaxy 
bias in \cite{YOO14b}. However, the matter density fluctuation in gauge-I 
in fact has non-vanishing one-point average shown in Eq.~\eqref{cancelfnl}, 
due to the intrinsic relativistic effects
in Eq.~\eqref{propt}, which was missing in \cite{YOO14b}.
Beyond the linear order in literature, many different attempts have been
made. With the focus on the second-order volume effect, a simple model $b=1$
was assumed in \cite{DIDUET14}. The matter density fluctuation
in the synchronous gauge was chosen \cite{BEMACL14,BEMACL14b}, while
gauge-I was favored in \cite{UMJOET17,JOUMET17}. In both cases, 
no intrinsic relativistic effects were considered.

In studying the relativistic contributions to the primordial non-Gaussianity,
a simple Newtonian biasing $b_1$ and $b_2$ was used \cite{KOUMET18} 
in terms of the matter density fluctuation in gauge-II, but without the
intrinsic relativistic contribution. Identifying the synchronous gauge 
(gauge-II) as the Lagrangian frame, the Lagrangian description of
galaxy bias was used \cite{UMKOET19,UMKO19}:
\beeq
1+\de_g^\up{int}=(1+\de_g^L)(1+\de_m)~,\Dquad 
\de_g^L:=b_1^L\de_l^{(1)}+\frac12b_2^L [\de_l^{(1)}]^2+\cdots~,
\eneq
where the bias factors~$b_n^L$ are defined in Lagrangian space,
$\de_l^{(1)}$ is the linear-order matter density in Lagrangian space,
and~$\de_m$ is the nonlinear matter density. Note that the biasing prescription
is based on the linear-order matter density fluctuation. 
The intrinsic relativistic
effects like $\RR\Delta\RR$ arises in this model from the volume 
fluctuation ($1+\de_m$), while no such terms are multiplied by the bias
factors up to the second order.

In summary, when the galaxy bias model is considered,
the proper-time hypersurface appears to be the right choice
for time slicing, as the proper time is the only clock available 
in the local galaxy and matter distribution and it provides the right
Newtonian correspondence at the linear order. However, 
{\it beyond the linear order, the spatial gauge choice matters} and 
there is no physical argument to prefer one spatial gauge  over the others.
For example, while two different gauge choices (gauge-I and-II) can describe the
same proper-time hypersurface, the matter density fluctuations in each
choice are different.\footnote{Often in literature, the matter density
fluctuations in gauge-I and gauge-II are referred to as the matter fluctuation
in the Eulerian and the Lagrangian frames. However, this is a {\it misnomer.}
Given the exact definition of the
Eulerian and the Lagrangian frames in the standard Newtonian perturbation
theory, the analogy and correspondence of each gauge choice in general
relativity to the Newtonian frames
are {\it not} exact. For a pressureless medium in a flat Universe,
the equations of motion for the matter density fluctuation in two gauges
are identical to the Newtonian equations in the Eulerian and the Lagrangian
frames, only up to the second order in perturbations  
\cite{HWNO05,HWNO05b,HWNO06}. Furthermore, the nonlinear constraint equations
in general relativity impose extra conditions that are absent in the Newtonian
dynamics, which are the origin of the intrinsic relativistic effects in the
matter density fluctuation in Eq.~\eqref{propt}. 

The Lagrangian dynamics reproduces the Eulerian dynamics \cite{BOCOET95},
and this is valid even in GR \cite{BEBAET15}. The situation of our
interest is, however, different. The matter density 
fluctuation~$\de_m(x^\mu)$ in each gauge choice in the same proper-time
hypersurface is described by different functional forms, and in GR there
is no preference for one coordinate choice to another.} 
Only the matter density fluctuation
in Eq.~\eqref{propt} in gauge-I has the correct correspondence to the
Newtonian second-order contributions, though it has extra second-order
relativistic contributions. 

Therefore, the proper-time hypersurface with spatial C-gauge appears as
the best choice for describing galaxy bias, but the choice of spatial gauge
remains to be explained in a successful galaxy bias model in general 
relativity.  
Keep in mind that the spatial gauge ambiguities are removed,
when we compute the observable quantities in terms of the observed
redshift and angle, i.e., any choice of spatial gauge would yield the same
answer to the observers. However, the relation between the galaxy
and the matter distributions (or galaxy bias) should be independent
of whether any observers exist, i.e., we need a certain choice of gauge
condition (temporal and spatial) with physical explanations.
If we take this choice and apply the bias relation developed
in \cite{MAVE08}, we arrive at the conclusion that the intrinsic nonlinear
relativistic effects in general relativity generate corrections to
the primordial non-Gaussianity $\Delta\fnl=-10/3$ (or $\Delta F_\up{nl}=-5/3$).
Furthermore, our calculations in Section~\ref{sec:matt} show that
extra relativistic effects associated with the light propagation and 
observations also generate corrections to the primordial non-Gaussianity, but
they cancel together, leaving only the correction from the intrinsic 
relativistic effects. 

In contrast, one can also argue \cite{PASCZA13} that the galaxy
bias models are based on Newtonian descriptions, so that they should be
considered only in local coordinates such as CFC, in which the intrinsic
relativistic effects are absorbed into the local coordinates in a single-field
inflationary scenario. While it provides a useful framework for interpreting
Newtonian descriptions in general relativity, it is not clear whether this is
enough. With all the relativistic corrections, CFC is {\it not a Newtonian 
coordinate} either. Moreover, such local coordinates
as CFC have a finite range of validity, in which Fourier transformation is
ill-defined, in particular for long wavelength modes in the squeezed limit. 
For example, the non-Gaussian correction in Eq.~\eqref{MVcorr} needs to be
computed in the squeezed limit. Certainly, 
we need a better and consistent description of 
galaxy bias in general relativity.

Here we focused on the
so-called Eulerian bias, because our goal is to describe
the cosmological observables, such that the linear bias factor~$b$ is always
multiplied by the nonlinear matter density fluctuation.

\section{Contribution of the matter density fluctuation to the observed
  galaxy bispectrum}
\label{sec:matt}

\subsection{Theoretical considerations}
\label{theory}
Here we compute the major contribution to the observed galaxy three-point
correlation function in the squeezed limit, resolving all the issues
discussed in Section~\ref{issues}. In particular, we consider the contribution
of the matter density fluctuation to the observed galaxy clustering.
Given the number counts~$dN_g^\up{obs}$
of the observed galaxies in a unit solid angle~$d\Omega$ and a
unit redshift bin~$dz$, the observed galaxy number density is constructed
in terms of observable quantities as 
\beeq
\label{ngobs}
n_g^\up{obs}(\Vang,z):={dN_g^\up{obs}(\Vang,z)\over d\bar V_\up{obs}(\Vang,z)}=
n_g^\up{phy}(x^\mu_s)~{dV_\up{phy}(x^\mu_s)\over
  d\bar V_\up{obs}(\Vang,z)}=n_g^\up{phy}(1+\de V)~,
\eneq
where the observed volume is
\beeq
d\bar V_\up{obs}(\Vang,z):={\rz^2~d\Omega ~dz\over H(z)(1+z)^3}~,
\eneq
the volume element in the background universe,  $n_g^\up{phy}(x^\mu_s)$
is the physical galaxy number density at the source position,
and
\beeq
\label{volf}
dV_\up{phy}(x_s^\mu)=:\left(1+\de V\right)d\bar V_\up{obs}(\Vang,z)
\eneq
is the 3D physical volume in 4D spacetime that appears
subtended by the observed redshift bin~$dz$ and solid angle~$d\Omega$.
Note that we used the superscript ``phy'' for the galaxy number density
to contrast with the observed galaxy number density.
In Eq.~\eqref{volf} we defined the dimensionless and gauge-invariant
volume fluctuation~$\de V$.
Evident in Eq.~\eqref{ngobs}, all the contributions to the observed galaxy
clustering can be split into two physically distinct effects
\cite{YOO09,YOFIZA09}: the volume effect~$\de V$ associated with
$dV_\up{phy}$ and the source effect associated with~$n_g^\up{phy}$.

The former is the ratio of two
volume elements in Eq.~\eqref{ngobs}, and it includes the redshift-space
distortion and the gravitational lensing 
in addition to other relativistic effects associated with the light propagation
(e.g., \cite{GUNN67,SAWO67,KAISE87,MIRAL91b,KAISE92}).
For instance, the mismatch between the observed angular position and the
real position of the source galaxies gives rise to the gravitational lensing 
effect \cite{GUNN67,MIRAL91b,KAISE92}, 
and the mismatch in volume due to the observed redshift and the real
positions gives rise to the redshift-space distortion \cite{KAISE87}. 
These two well-known
effects belong to the volume effect. The latter (or the source effect)
comes from the physical galaxy number density~$n_g^\up{phy}$ (or the
source) discussed in Section~\ref{sec:gauge}, and
its main contribution is the matter density fluctuation as galaxies are
a biased tracer \cite{KAISE84} 
of the matter density. Additional contributions in the
source effect arise from the fact that the physical number density is
expressed in terms of the observed redshift and angle, given the
observational constraint on the galaxy sample such as the luminosity threshold.
For instance, the magnification bias \cite{NARAY89,BARTE95,JAIN02}
belongs to the source effect, and it
arises from the imposed threshold in 
observation in terms of the inferred luminosity for the galaxy sample.
A complete treatment of observed galaxy clustering with full relativistic
treatment is first given in \cite{YOFIZA09,YOO10,BODU11,CHLE11,JESCHI12}
(see \cite{YOO14a} for review), and the formalism was extended to
the second order in perturbations \cite{YOZA14,DIDUET14,BEMACL14}.

As discussed in Section~\ref{sec:gauge}, the matter density contribution
with galaxy bias factor in the source effect is the dominant contribution
to the observed galaxy clustering, and more importantly it is a distinct
effect that is separable from other contributions due to its unique
combination of galaxy bias factor~$b$. The intrinsic galaxy fluctuation
at the source position in Eq.~\eqref{intflc} can be written as
\beeq
\label{intde}
\de_g^\up{int}(x^\mu_s)=b~\de_m(x^\mu_s)+\cdots~,
\eneq
where the galaxy bias factor~$b$ is also called the linear bias
(sometimes denoted as~$b_1$) and
we omitted other contributions such as the nonlinear bias factors,
the tidal tensor bias, and so on (see, e.g., 
\cite{FRY96,MOWH96,CALUET98,CAPOKA00,SMT01,
YODASE11,BASEET12,SCJEDE13,DEJESC18}). According to the discussion
in Section~\ref{sec:gauge}, the intrinsic galaxy fluctuation must contain
at least the linear bias factor~$b$
to reproduce galaxy clustering in the Newtonian limit, and
the matter density~$\de_m$ multiplied by~$b$ is the nonlinear matter
density fluctuation. At the leading order in the bispectrum,
this contribution~$b\de_m$ in~$\de_g^\up{int}$ is of our primary interest
here for the contributions to the observed galaxy bispectrum on large scales.
First, the other nonlinear contributions in Eq.~\eqref{intde}
such as $b_2\de_m^2$ provide Newtonian contributions to the observed
galaxy bispectrum, which is negligible on large scales or in the squeezed
triangular configuration \cite{SCSEZA04,SEKO07,YODASE11}.
 Hence, the relativistic effects
from the second-order in the intrinsic fluctuation, i.e.,
\beeq
\label{prim}
\big\langle b\de_m^{(2)}~b\de_m^{(1)}~b\de_m^{(1)}\big\rangle~,
\eneq
provide the dominant contribution to the observed galaxy bispectrum on
large scales. Second, the other important contributions
to the observed galaxy bispectrum on large scales arise from the relativistic
effects in the volume effect~$\de V$. These contributions at the leading
order in bispectrum are
\beeq
\label{others}
\big\langle b\de_m^{(1)}~b\de_m^{(1)}~\de V^{(2)}\big\rangle~,\qquad
\big\langle b\de_m^{(1)}~\de V^{(2)}~\de V^{(1)}\big\rangle~,\qquad
\big\langle b\de_m^{(2)}~\de V^{(1)}~\de V^{(1)}\big\rangle~,\qquad
\big\langle \de V^{(2)}~\de V^{(1)}~\de V^{(1)}\big\rangle~,
\eneq
and they are as important on large scales as
the contribution in Eq.~\eqref{prim}. However, as discussed in
Section~\ref{sec:gauge} these individual contributions are separately
gauge-invariant, and they are distinct in terms of scaling with
galaxy bias factor~$b$. Given the level of difficulties in computing
the second-order relativistic effect in~$\de V$, here we focus on the
dominant contribution in Eq.~\eqref{prim}
from the intrinsic matter density fluctuation and call it the {\it observed
 matter density} contribution to the observed galaxy bispectrum.
In this way, our calculations are {\it not} affected by 
the uncertainty in galaxy formation theory in general relativity.
Calculations of the other contributions in Eq.~\eqref{others}
will be performed in future work.

\subsection{Observed matter density fluctuation}
\label{obsmat}
While the proper-time hypersurface is our best physical choice for relating the
matter density fluctuation to the intrinsic galaxy fluctuation, the spatial
gauge choice remains undetermined as discussed in Section~\ref{spatial}.
Any change in spatial gauge choice alters
 the prediction of the bispectrum, as illustrated in Eq.~\eqref{bischange}.
A natural question arises:
What would be the best physical choice for spatial gauge?
In fact, there is {\it no convincing physical preference} for any spatial gauge,
as discussed in Section~\ref{sec:gauge}. However, this arbitrariness is
completely lifted in observable quantities such as galaxy clustering, once
the source position is expressed in terms of the observed redshift and angle.
To the second order in perturbations,
the matter density fluctuation at the observed position is
\beeq
\label{delta2}
\delta_m(x^\mu_s)=\delta_m(\bar x^\mu_z)+\Delta x^\mu~\partial_\mu \delta_m
\Big|_{\bar x^\mu_z}~,
\eneq
where $\Delta x^\mu$ is the spacetime distortion of the source position
\beeq
x^\mu_s=:\bar x^\mu_z+\Delta x^\mu~,
\eneq
relative to the observed position
\beeq
\label{obspos}
\bar x^\mu_z:=\left(\bar\eta_z,~\rz\Vang\right)~.
\eneq
For a general coordinate transformation in Eq.~\eqref{ct}, the observed
position~$\bar x_z^\mu$ in Eq.~\eqref{obspos} remains unaffected, and hence
the distortion of the source position transforms as
\beeq
\widetilde{\Delta x}{}^\mu=\Delta x^\mu+\xi^\mu~.
\eneq
As seen in Eq.~\eqref{dmgt}, the matter density fluctuation gauge transforms
as
\beeq
\tilde \delta_m(x^\mu_s)=\delta_m(\bar x^\mu_z)-\xi^\mu~\pa_\mu\delta_m
\Big|_{\bar x^\mu_z}~,
\eneq
and we readily prove that the expression in Eq.~\eqref{delta2}
at the observed position is {\it fully} gauge-invariant 
 (temporal and spatial) up to the second order,
which states nothing more than the invariance under diffeomorphism
of physical quantities \cite{YODU17}. 
From now on, we call the combination,
\beeq
\label{deobs}
\deobs:=\de_m(\bar x_z^\mu)+\Delta x^\mu{\pa\over\pa x^\mu}\de_m\bigg|_{\bar
x^\mu_z}~,
\eneq
the ``observed'' matter density fluctuation
and compute its contribution to the observed galaxy bispectrum.
Let us emphasize that  the observed matter density fluctuation
is  independent of arbitrariness in choosing a
spatial gauge condition, naturally resolving the issue of spatial gauge 
choice. Furthermore, with galaxy bias factor~$b$, this contribution
is separable from other relativistic contributions in Eq.~\eqref{others}.

Splitting the distortion of the source position in terms of spherical
components set by the observed angular direction,
the observed matter density fluctuation is written explicitly as
\beeq
\label{main}
\delta(x^\mu_s)=\delta(\bar x^\mu_z)+\left(\delta \eta
 {\pa\over\pa\eta}+\delta r{\pa\over\partial r}+
\delta \theta{\pa\over\partial \theta} +\delta\phi{\pa\over\partial
  \phi}\right)\delta_m\bigg|_{\bar x^\mu_z}~,
\eneq
where the first term is the matter density fluctuation in Eq.~\eqref{propt}
at the observed position~$\bar x_z$. It is clear that we only need the
linear-order expressions of~$\Delta x^\mu$ for our calculations.
The distortion of the source position can be obtained
by solving the geodesic equation (see \cite{YOGRET18,YOO14a} 
for detailed derivations).
The radial distortion of the source position is
\beeq
\label{drr}
\delta r=(\chi_o+\delta\eta_o)-\frac{\delta z_\chi}{\mathcal H_z}
+\int_0^{\bar r_z}\mathrm d\bar r\, (\alpha_\chi-\varphi_\chi)
+n_\alpha(\delta x^\alpha+\mathcal G)_o-n_\alpha\mathcal G^\alpha\,, 
\eneq
and the angular distortion is
(similarly, $\bar r_z\sin\theta\delta\phi$ for the azimuthal distortion)
\beeq
\label{dtt}
\bar r_z\delta\theta=\bar r_z\theta_\alpha\sbr{-V^\alpha-
  \epsilon^\alpha_{ij}n^j\Omega^j}_o-\int_0^\rz\dr \rbr{\bar r_z-\bar r}
\theta_\alpha (\alpha_\chi-\varphi_\chi)^{,\alpha}+\theta_\alpha
(\delta x^\alpha+\mathcal G^\alpha)_o-\theta_\alpha\mathcal G^\alpha\,,
\eneq
where various perturbation quantities are defined in Appendix~\ref{setup}
and $\delta z_\chi:=\delta z+H\chi$. The distortion in the observed redshift is
\beeq
\label{dzz}
\delta z=-H\chi+(\mathcal H\delta\eta+H\chi)_o+\left(V_\parallel
-\alpha_\chi\right)_o^z-\int_0^{\bar r_z}\dr (\alpha_\chi-\varphi_\chi)'~,
\eneq
and it is related to the distortion in the time coordinate of the source as
\beeq
\label{deta}
\delta \eta={\dz\over \HH}~.
\eneq
Mind that in previous work the distortion in time coordinate of the source
was denoted as $\Delta\eta$, but here we use $\delta \eta$ for the 
notational consistency with $\drr$, $\dtt$, and $\dpp$.

For concreteness, we choose the temporal comoving gauge and the spatial C-gauge
described in Eq.~\eqref{gauge} and compute the individual components.
It is important to note that our choice of such gauge conditions is driven 
merely by convenience and the result is {\it independent} of our gauge choice.
While the expressions in Eqs.~\eqref{drr}$-$\eqref{deta} are general,
our interest is the specific prediction in the standard inflationary
model. Using the Einstein equation, 
the observed matter density fluctuation at the second order in perturbations
is expressed as
\bear
\label{mainRRmain}
\deobs&=&D_1\left(-\Delta \RR+\frac 32 \RR^{,\al}\RR_{,\al}
+4\RR\Delta \RR
  \right)+\frac 57D_{A}\nabla_\al\left(\RR^{,\al}\Delta \RR
\right)+\frac 17D_{B}\Delta\left(\RR^{,\al}\RR_{,\al}\right) \\
&&
+\frac{1}{\Hz}D_V \Delta \RR\left[\RR-\RR_o
-n^\alpha\left(D_V\RR_{,\alpha}-D_V\RR_{,\alpha}\big|_o\right)
+2\int_0^\rz\dr D_\Psi(\bar r)~ \partial_r \RR(\bar r\hat n)\right] \nnn
&&
-D_1{\pa\over\partial r}\Delta\RR
\bigg\{D_V(x_o)\RR_o-\frac{1}{\Hz}\bigg[D_\Psi\RR+\RR_o+n^\alpha
\left(D_V\RR_{,\alpha}-D_V\RR_{,\alpha}\big|_o\right)   \nnn
&&
\Dquad
-2\int_0^\rz\dr D_\Psi(\bar r) ~\partial_r \RR(\bar r\hat n)\bigg]
+2\int_0^\rz\dr D_\Psi(\bar r)\RR(\bar r\hat n)\bigg\}
+D_1 \left(D_V\nabla^\alpha\RR\right)_o\habla_\alpha\left(\Delta \RR\right) \nnn
&&
+2D_1\habla_\alpha\left(\Delta\RR\right)\int_0^\rz\dr
\left(\frac{\rz-\bar r}{\rz\bar r}\right)D_\Psi(\bar r)\habla^\alpha \RR 
-D_1\nabla^\al\left(\Delta\RR\right)
\int_0^{\bar\eta_o}\mathrm d\eta\,D_V\nabla_\al \RR\bigg|_{\bm{x}=0}
  \nn~,
\enar
in terms of the initial condition~$\RR$, where the detailed derivations
for the expression and the definition of the growth factors~$D_V$ and~$D_\Psi$
are presented in Appendix~\ref{details}. The first line is~$\de_m^{(2)}$
in Eq.~\eqref{propt}, and the remaining terms originate from the
coupling of~$\Delta x^\mu$ and~$\de_m$.

\subsection{One-point ensemble average}
\label{opt}
Here we present the ensemble average $\AVE{\deobs}$
of the observed matter density fluctuation
in Eq.~\eqref{mainRRmain}, and the detailed calculations can be found
in Appendix~\ref{details}. As discussed in Section~\ref{ensavg}, the ensemble
average depends on a coordinate system, while the observed
matter density fluctuation is independent of any gauge choice. This might
appear inconsistent at first glance. However, the observed matter density
fluctuation is in fact expressed in the hypersurface of the observed redshift
with spatial gridding in terms of the observed angle,
both of which are independent
of our choice of coordinates in theoretical descriptions.
The ensemble average of the observed matter density fluctuation is therefore
the average over such hypersurface with spatial gridding, and it is also
independent of our gauge choice. We emphasize again that such
hypersurface is not fully accessible to the observer at one point, as the
light-cone surface is limited to a two-dimensional intersection with the
hypersurface. However, as discussed in Appendix~\ref{FDEV}, fictitious
observers besides our vantage point can have access to the full hypersurface.
Therefore,
while the ensemble average of the observed matter density fluctuation
is {\it not} a direct observable, it is one
of the important quantities in theoretical calculations, in a way that
any power spectrum in a hypersurface is {\it not} a direct observable,
but an important statistics in theoretical calculations.

The ensemble average of the observed matter density fluctuation is derived
as 
\bear
\big\langle\deobs\big\rangle&=&-\left(\frac{5D_1}{2}
+\frac{D_V}{\Hz}\right)\avg_2+\frac{D_V}{\Hz}\avg_{2,0}+
\rbr{-\frac{D_VD_V^o}{\Hz}-3D_1 D_V^o+\frac{D_1}{\Hz}}\avg_{3,1}
-\frac{D_1D_V}{3\Hz}\avg_4 \nnn
&&+\rbr{\frac{D_1D_V^o}{3\Hz}+\int_0^{\bar \eta_o}\mathrm d\bar\eta\, 
D_V(\bar \eta)}\avg_{4,0}-\frac{2D_1D_V^o}{3\Hz}\avg_{4,2}\nnn
&&
-2 \rbr{\frac{D_V}{\Hz}+3D_1}\int \Dkk \int_0^\rz\dr D_\Psi(\bar r)
 j_1(k\Delta r) k^3 P_\RR(k) \nnn
&&+\frac{2D_1}{3\Hz}\int \Dkk \int_0^\rz\dr D_\Psi(\bar r) 
\Big[j_0(k\Delta r)-2j_2(k\Delta r)\Big] k^4 P_\RR(k)~,
\enar
where $\Delta r=\rz-\rbar$ and we defined the dimensionful quantity
\beeq
\label{avgnm}
\avg_{n,m}(z):=\int{d^3k\over(2\pi)^3}~k^nj_m(k\rz)P_\RR(k)~,
\Dquad [\avg_{n,m}]=L^{-n}~.
\eneq
Given the spectral index $n_s-1\simeq0$, none of the variances
 $\avg_n$ or $\avg_{n,m}$ diverges in the infrared
if $n>0$, and the ensemble average of the observed matter density
fluctuation is devoid of any infrared divergences.
In an Einstein-de Sitter universe, this expression is greatly simplified
as 
\bear
\label{edsvar}
\big\langle\deobs\big\rangle&=&\etaz^2\rbr{-\frac{1}{4}-\frac{1}{10}
    +\frac{3}{25}+\frac{3}{25}{+}\frac{6}{25}}\avg_2
+\etaz^2\rbr{-\frac{3}{25}+\frac{1}{10}-\frac{3}{25}{-}\frac{6}{25}}
\avg_{2,0} \\
&&
+\etaz^2\rbr{-\frac{\etao}{50}-\frac{3\etaz}{50}-\frac{\etao}{50}
+\frac{\etaz}{20}-\frac{\etao}{25}}\avg_{3,1}
-\frac{\etaz^4}{300}\avg_4+\etaz^2\etao\rbr{\frac{\etaz}{300}
+\frac{\etao}{100}}\avg_{4,0}-\frac{\etaz^3\etao}{150}\avg_{4,2}  \nnn
&=&\frac{13\etaz^2}{100}\avg_2-\frac{19\etaz^2}{50}\avg_{2,0}
-\etaz^2\rbr{\frac{\etaz}{100}+\frac{2\etao}{25}}\avg_{3,1}
-\frac{\etaz^4}{300}\avg_4+\etaz^2\etao
\rbr{\frac{\etaz}{300}+\frac{\etao}{100}}\avg_{4,0}
-\frac{\etaz^3\etao}{150}\avg_{4,2} ~.\nn
\enar

\subsection{Three-point correlation function in the squeezed limit}
\label{sqztpt}
Now we are in a position to compute the observed matter density contribution
to the galaxy bispectrum in the squeezed limit. Since Fourier transformation
involves integration over an infinite hypersurface of simultaneity, Fourier
quantities like the power spectrum and the bispectrum are {\it less suited} for
a direct comparison to observations on large scales, in which the geometry
of the sky is non-flat and the time evolution along the line-of-sight direction
becomes significant. In contrast, being a function of observations at the
observed positions only,
the correlation function is well defined within the survey region,
regardless of the geometry of a survey or the scale of our interest.
In particular, we are interested in the three-point correlation function
of the observed galaxy fluctuation in the squeezed triangle, in which the
separation of two observed positions is negligible compared to the separation
to the third position. In the limit the separation goes to infinity, also
known as the squeezed limit, this special triangular configuration for the
three-point correlation encodes
critical information about the primordial non-Gaussianity.

In general, the primordial fluctuation is constrained to be highly Gaussian.
However, a slight deviation from the Gaussianity (or the primordial 
non-Gaussianity) is expected in any inflationary models, and it is often
parametrized in terms of $\fnl$ (see, e.g., \cite{KOSP01,BAKOET04})
in the bispectrum as
\beeq
B_\RR(\kvec_1,\kvec_2,\kvec_3)=\frac65\fnl\bigg[P_\RR(\kvec_1)P_\RR(\kvec_2)
  +P_\RR(\kvec_2)P_\RR(\kvec_3)+P_\RR(\kvec_3)P_\RR(\kvec_1)  \bigg]~,
\eneq
where $\fnl$ can be a function of scale and the numerical factor is
present for the convention in literature with the bispectrum in terms of
Newtonian gauge potential~$\px$. A similar relation is also defined
for~$\zeta$ and~$F_\up{NL}$.
For the single-field inflationary models,
the prediction for the primordial non-Gaussianity 
in the squeezed limit is slow-roll suppressed \cite{MALDA03,CRZA04} as
\beeq
\label{consi}
\lim_{k_3\RA0}B_\zeta(\kvec_1,\kvec_2,\kvec_3)=-(n_s-1)P_\zeta(\kvec_1)
P_\zeta(\kvec_3)~, \Dquad F_\up{NL}=-{5\over12}(n_s-1)~,
\eneq
and this prediction in the squeezed limit is generic for any scalar-field 
potential, such that if a non-negligible~$F_\up{NL}$ 
is observed in the squeezed
limit, all classes of single-field inflationary models will be ruled out.
As pointed out in Section~\ref{spatial}, the bispectrum depends on the spatial
gauge choice \cite{MIYO22a}, and the single-field consistency relation in 
Eq.~\eqref{consi} is derived with the spatial gauge choice in 
Eq.~\eqref{expo}.

Here we compute the three-point correlation function of the observed
matter density fluctuation
$\DD:=\deobs-\langle\deobs\rangle$, given in Eq.~\eqref{DDD}, in which the 
non-vanishing ensemble average was subtracted 
to remove the tadpole contributions to the three-point correlation function.
The details of the computation are presented in Appendix~\ref{thrpt}.
In particular, we consider the squeezed triangular configuration discussed
in Section~\ref{MVformalism} and also given
in Eq.~\eqref{sqzcond} and~\eqref{sqzcond2}, in which two observed positions are
identical and the third position is in the opposite side of the sky, but 
all three positions are at the same redshift:
\beeq
z:=z_1=z_2=z_3~,\Dquad \Vang:=\Vang_1=\Vang_2=-\Vang_3~.
\eneq
This is the most squeezed triangular configuration obtainable
in observations at a fixed redshift. The true squeezed limit
for the consistency relation in Eq.~\eqref{consi}
could be reached, only when the observed redshift becomes
infinite. In fact, two observed positions need not be identical 
for the squeezed triangular configuration, but we take this triangular
configuration for simplicity. Finally, the three-point correlation function
of the observed matter density fluctuation in the squeezed triangular
configuration is given in Eq.~\eqref{sqzhere} and derived in Eq.~\eqref{sqz} as
\beeq
\label{sqz2}
\xi_{\rm sqz}=
\int\Dkone\int\Dktwo~ e^{i(\mathbf k_1+\mathbf k_2)\cdot(\xone-\xthree)}
\bigg[B_{112}+B_{211}+B_{121}\bigg]
(\mathbf k_1,\mathbf k_2,-\mathbf k_{12})~,
\eneq
where  the connected bispectra in the integrand are defined 
in Eqs.~\eqref{BBBA}$-$\eqref{BBBB}. Compared to the calculations in
Section~\ref{MVformalism}, the computation of the bispectra is much more
involved, as they depend on the line-of-sight integration and the observed
angle (see Appendix~\ref{thrpt}).

While the expression for~$\xi_\up{sqz}$ is general for the squeezed triangular
configuration, we further simplify the expression by taking the limit,
in which the separation between~$\xone$ and~$\xthree$
goes to infinity, i.e.,  the observed redshift becomes
sufficiently large $z\RA\infty$. In the squeezed limit, a highly oscillating
phase in the exponential factor cancels all the contributions, except 
for the contribution from the modes:
\beeq
\label{cond3}
L:=|\xvec_1-\xvec_3|~,\Dquad
{1\over L}\simeq k_{12}\RA0~,\Dquad \kvec_1\approx-\kvec_2~,
\eneq
and the connected bispectra are further simplified 
in Eqs.~\eqref{B112}$-$\eqref{B121}, where
the connected bispectrum $B_{112}$ is negligible
compared to the other two bispectra.
In this squeezed limit, the integral over~$k_2$ in Eq.~\eqref{sqz2} is 
essentially removed to satisfy the condition in Eq.~\eqref{cond3}
imposed by the exponential factor with $L\RA\infty$, and the three-point
correlation function is simplified as
\beeq
\label{simple}
\xi_\up{sqz}\simeq\int {d^3k_1\over(2\pi)^3}\left({B_{211}+B_{121}\over L^3}
\right)\propto \int d\ln k_1\int d\mu_{k_1}~k_l^3k_1^3
\left(B_{211}+B_{121}\right)~,
\eneq
where we used (mind the dimension)
\beeq
e^{i(\kvec_1+\kvec_2)\cdot \bdv{L}}\sim{1\over L^3}\de^D(\kvec_1+\kvec_2)
\sim k_l^3\de^D(\kvec_1+\kvec_2)~,\Dquad k_l:={1\over L}~.
\eneq
With two dominant bispectra in the squeezed limit given
in Eqs.~\eqref{B211} and~\eqref{B121} 
\bear
B_{211}&=&D_1^2k_1^2k_l^2P_\RR(k_1)P_\RR(k_l)
\bigg[\FF(\mathbf k_1,\mathbf k_l;\Vang)
  +\FF(\mathbf k_l,\mathbf k_1;\Vang)\bigg]~,\\
B_{121}&=&D_1^2k_l^2k_1^2P_\RR(k_l)P_\RR(k_1)
\bigg[\FF(\mathbf k_l,-\mathbf k_1;\Vang)
  +\FF(-\mathbf k_1,\mathbf k_l;\Vang)\bigg]~,
\enar
and the Fourier kernels~$\FF(\kvec_1,\kvec_2;\Vang)$ computed in 
Appendix~\ref{details}, the important quantity in computing~$\xi_\up{sqz}$
is the angle average of the Fourier kernels from $B_{211}$ and $B_{121}$
in Eq.~\eqref{simple},
\beeq
\label{avgF}
\int {d\mu_{k_1}\over2} \bigg[\FF(\mathbf k_1,\mathbf k_l)
  +\FF(\mathbf k_l,\mathbf k_1)+\FF(\mathbf k_l,-\mathbf k_1)
  +\FF(-\mathbf k_1,\mathbf k_l)\bigg]\propto k_l^{n_\FF}~,
\eneq
and its dependence on the long mode~$k_l$, where we suppressed 
the dependence of the Fourier kernels on the observed angle~$\Vang$
and defined the power-law coefficients~$n_\FF$.
The detailed computation of individual Fourier kernels and their
contributions to the sum of two connected bispectra are presented
in Appendix~\ref{details}.

In the limit $k_l\RA0$, most of the individual components
in Eq.~\eqref{mainRRmain} vanish, because they have $n_\FF>0$ or their
angle average in Eq.~\eqref{avgF} vanish.
Adding up all the surviving contributions in Eq.~\eqref{avgF}, we obtain
\beeq
\int{d\mu_{k_1}\over2}\bigg[\cdots\bigg]=\underbrace{0}_{\rm SPT}-
\underbrace{8D_1k_1^2}_{\rm GRE}+\underbrace{{-2D_V\over\Hz}k_1^2}_{\rm src.}
+\underbrace{{2D_V\over\Hz}k_1^2}_{\rm obs.}+\underbrace{0}_{\rm nl.}
+\underbrace{4D_1\frac35\fnl k_1^2}_{\rm ini.}~,
\eneq
where each contribution is labeled according to their origin.
First, the contribution from the standard perturbation theory (SPT) in the
matter density fluctuation naturally vanishes in the squeezed limit.
Second, the general relativistic effects (GRE) in the matter density
fluctuation survive, as their contribution is essentially identical
to that from the primordial non-Gaussianity in the initial condition 
(denoted as ``ini'') 
in proportion to~$\fnl$. The other two surviving contributions
are associated with the light propagation, and they cancel each other.
The first one arises from the coupling of the Sachs-Wolfe effect ($\propto\RR$)
at the source position to the matter density fluctuation ($\propto\Delta\RR$),
and the other contribution arises from the same mechanism, but due to the
coupling of the Sachs-Wolfe effect at the observer position. Finally,
the coupling of the line-of-sight contributions to the matter density
fluctuation is denoted as ``nl,'' and it vanishes. In short, the
contributions associated with the light propagation (or the sum of
those denoted as src, obs, and nl) {\it completely vanish}
by cancellation,\footnote{These effects are often referred to as the
projection effects, but it is a misnomer, as they involve contributions
at the source position and the observer position, as well as those along
the light propagation.} but the relativistic effects intrinsic to the
matter density fluctuation or the primordial non-Gaussian contribution
survive, both of which are at the source position, independent of observations
or light propagation.

Therefore, with these leading bispectra, we show that
the three-point correlation function in Eq.~\eqref{sqz2}
scales with the long-mode in the squeezed limit as 
\beeq
\label{scale}
\xisqzlim\propto D_1^2 k_1^{n_s+1}k_l^{n_s-2+n_\FF}L^{-3}\propto 
D_1^2k_l^{n_s+1+n_\FF}~,
\eneq
where $n_\FF=0$ for the surviving (hence the leading) contributions in 
long modes. In conclusion, the observed
three-point correlation function {\it vanishes}
in the squeezed limit $k_l=1/|\xvec_1-\xvec_3|\RA0$,
even accounting for all the relativistic corrections from the light propagation
and the nonlinearity in the matter density fluctuation due to the Hamiltonian
constraint equation. 
With all the relativistic effects, there exist no contribution that
scale as $P_m(k_l)/k_l^4$, if we phrase it in terms of the power spectrum
in a hypersurface, as in Section~\ref{MVformalism}.

This conclusion is indeed consistent with
the single-field consistency relation in Eq.~\eqref{consi}. In terms of
the matter density fluctuation, the consistency relation in Eq.~\eqref{consi}
can be recast as
\beeq
B_\de\propto k_1^2k_2^2k_3^2B_\RR\propto(n_s-1)k_s^{4+n_s-4}k_l^{2+n_s-4}~,
\eneq
and its contribution to $\xi_\up{sqz}$ in Eq.~\eqref{simple}
corresponds to the leading corrections in~$k_l$ we obtained in Eq.~\eqref{scale}
with $n_\FF=0$. We stress
that in deriving the Fourier kernels~$\FF$ for individual contributions
in Eq.~\eqref{mainRRmain}, we have used the Einstein equation 
in the standard model to relate each component such as $\varphi_v$, $v_\chi$,
and so on to the initial condition~$\RR$. The consistency we obtained is
by no chance a coincidence.

\section{Summary of new findings}
\label{findings}
Assuming the standard $\Lambda$CDM model in a single-field inflationary
scenario, we have analytically
computed the three-point correlation function of the matter density
fluctuation in the squeezed triangular configuration, accounting for 
the intrinsic relativistic effects in the matter density fluctuation 
and the relativistic effects associated with light propagation and 
observations. The squeezed three-point correlation function 
of the matter density fluctuation on large scales is sensitive to the
primordial non-Gaussianity and is expected to be the source of a prominent
feature in the galaxy power spectrum on large scales. The intrinsic
non-Gaussianity in the matter density fluctuation is always present
from the Hamiltonian constraint in general relativity, and it has been
extensively debated in literature whether such non-Gaussianity in a 
single-field inflationary model can give rise to signals similar to the
presence of primordial non-Gaussianity. In contrast, it is generally
accepted in the community that the relativistic effects
associated with the light propagation in observations are expected
contribute to the non-Gaussian signals, if not directly in the galaxy power
spectrum. Our findings are summarized as follows.

\begin{itemize}
\item While the linear-order calculations are independent of spatial gauge
choice, the calculations beyond the linear order in perturbations are
affected by a choice of spatial gauge condition. Consequently, 
{\it the three-point
correlation function depends on a choice of spatial gauge condition}
(see Section~\ref{spatial}). 

\item Since there is no physical argument to prefer one choice to another
spatial gauge, {\it any theoretical descriptions
beyond the linear order in perturbations have ambiguities in spatial gauge
choice, or a choice by hand in such theoretical descriptions would require
further physical explanations}  (see below how these ambiguities are
resolved in the theoretical
descriptions of cosmological observables). For example, the non-Gaussian
correction to the two-point galaxy correlation arises from the three-point
matter density correlation \cite{MALUBO86,MAVE08}. This relation {\it cannot}
be valid for all spatial gauge choices, because the transformation properties
of the two-point and the three-point correlation functions are different,
which leaves us two possibilities: This relation is not valid at all 
in general relativity, or it should be valid for only one specific 
choice of spatial gauge. Even in the latter case, 
it still needs a physical explanation behind its choice of spatial gauge.

\item If valid for one specific choice of spatial gauge condition, it has
to be the spatial C-gauge in Eq.~\eqref{CCC} in conjunction with the
temporal comoving gauge in Eq.~\eqref{gauge}, because only this gauge choice
yields that the matter density fluctuation includes
the correct second-order Newtonian contributions in the standard
perturbation theory.

\item The matter density fluctuation in general relativity exhibits
extra relativistic contributions, originating from the nonlinear Hamiltonian
constraint in general relativity 
\cite{BRHIET14,BRHIWA14,CAMASA15,VIVEMA14,MAPIRO21}.
These intrinsic and  nonlinear contributions exist, 
even if the initial condition is set to be Gaussian
at the linear order in perturbations. According to the relation for
non-Gaussian correction \cite{MALUBO86,MAVE08}, the intrinsic relativistic
effects in the matter density fluctuation generate signals like the primordial
non-Gaussianity. In a single-field inflationary scenario, it was argued 
\cite{PASCZA13,DEDOGR15,YOGO16,DEDOET17,KOUMET18,UMKOET19}
that these contributions can be removed by extra coordinate transformations.
However, we showed \cite{MIYO22b}
 in Section~\ref{extra} that large diffeomorphisms
such as dilatation and special conformal transformation,
which are not part of gauge transformation, do not affect the theoretical
descriptions of cosmological observables, and  coordinate transformations
over a finite range of validity regime such as the conformal Fermi coordinate
(CFC) can be recast as a gauge transformation over the entire manifold,
while matching the CFC coefficients when expanded over the finite 
validity range. Consequently, {\it there exist no residual symmetries or
coordinate transformations
that can affect the gauge-invariant calculations of cosmological observables,
while they can change the functional form of the expression.}

\item The matter density fluctuation expressed at the observed position 
involves extra relativistic
contributions associated with light propagation and observation,
and {\it this observed matter density fluctuation
is independent of spatial gauge choice, which naturally
resolves the ambiguities in choosing a spatial gauge condition.}
However, the theoretical descriptions
that relate the matter density fluctuations to the galaxy number density
fluctuations are independent of observations, and hence the ambiguities
in spatial gauge choice still remain.

\item Gauge-invariant calculations of cosmological observables demonstrated
(see, e.g., \cite{YOFIZA09,YOO10,YOSC16,BIYO17,SCYOBI18,GRSCET20})
that there exist perturbation contributions at the observer position.
Our calculations beyond the linear order in perturbations show that {\it these
contributions at the observer position couple to those at the source position
and thereby obtain a positional dependence, which cannot be ignored
in computing the correlation function or their Fourier counterpart.}

\item Accounting for all the relativistic effects, we computed the three-point
correlation function of the observed matter density fluctuation 
in the squeezed limit and showed that {\it 
the relativistic effects associated with the light propagation 
and observations produce zero non-Gaussian signals 
like the primordial non-Gaussianity by cancellation}, but the intrinsic
relativistic effects in the matter density fluctuation are {\it not} 
countered by any relativistic effects in the light propagation.

\item The squeezed three-point correlation function receives the dominant
contribution from the bispectra in the squeezed triangle in Fourier space,
which admit a contribution that scales like 
the matter-potential cross power spectrum $P_{m\phi}\propto k_l^{n_s-2}$,
but no contribution that scales like the potential power spectrum
$P_\phi\propto k_l^{n_s-4}$. {\it The three-point correlation function vanishes
in the squeezed limit, under the standard $\Lambda$CDM model in a single-field
inflationary scenario.}
\end{itemize}

\section{Discussion}
\label{sec:discuss}
As pointed out in Section~\ref{findings}, the largest uncertainties in the
theoretical description of galaxy clustering reside in galaxy bias, 
or the relation between the
galaxy and the matter distributions. Considered in general relativity,
galaxy bias models need a specific choice of gauge condition. The proper-time
hypersurface is a preferred choice of temporal gauge, as it is the only
hypersurface a local observer can construct without any extra information
(see Section~\ref{sec:gauge}). In contrast, the spatial gauge choice is
completely left arbitrary. Given the correspondence to the Newtonian 
perturbation theory, the spatial C-gauge choice might be preferred, but this
is {\it not} an explanation for the choice. Nonetheless, 
if this is the right gauge choice 
for the bias model \cite{MALUBO86,MAVE08}, 
the galaxy two-point correlation function receives the
non-Gaussian correction from the intrinsic relativistic effects in the
matter density fluctuation, even in the absence of the primordial 
non-Gaussianity. As shown in Section~\ref{extra}, this correction
{\it cannot} be removed by any other subsequent transformations.

On the other hand, it has been argued 
\cite{PASCZA13,DEDOGR15,YOGO16,DEDOET17,KOUMET18,UMKOET19}
 that a long mode in a single-field
inflationary model can be absorbed into a local coordinate transformation,
and its coupling to short modes can be removed. While we showed in 
Section~\ref{extra} that any gauge-invariant calculations remain 
unaffected by extra transformations, it is possible that the theoretical
descriptions of galaxy bias might be valid only in such local coordinates, 
rather than in the entire manifold. Since those models are based on Newtonian
descriptions to begin with, there is no reason to be surprised 
to encounter ambiguities, when they are cast in general relativity.
If that is the case, the coupling of long and short modes such as 
$\RR\Delta\RR$ in Eq.~\eqref{propt} can be removed in a single-field 
inflationary scenario due to the consistency relation 
\cite{PASCZA13,DEDOGR15,DEDOET17,KOUMET18,UMKOET19}, and there
is {\it no} non-Gaussian correction. 
It is evident that we need a better theoretical description of galaxy bias,
as we can only observe galaxies, not matter. Naturally, the theoretical
description of cosmological observables is independent of spatial gauge choice,
as the observed redshift and angle fully specifies the observed position of
the source. However, note that galaxy bias or the relation between the matter
and the galaxy distributions is independent of observations, and any successful
model of galaxy bias should address the ambiguities associated with spatial
gauge choice.

As opposed to the non-Gaussian contributions from the intrinsic relativistic
effects, it has been argued \cite{PASCZA13,KOUMET18,UMKOET19}
in the community that the light propagation
in observations will inevitably generate non-vanishing non-Gaussian signals
in the squeezed triangular configuration, even in the case of a single-field
inflationary scenario. Indeed, this expectation was based on general arguments,
and no accurate calculations have been performed. 
Here we showed that in the observed three-point correlation function
the relativistic contributions 
from the light propagation in fact {\it cancel each other} in a single-field
inflationary model, if all the relativistic effects are accounted for.
This conclusion is independent of galaxy bias, as it only involves the
linear-order matter density fluctuation and the linear-order light propagation.
However, we only computed the observed matter density fluctuation. In galaxy
clustering, the fluctuation in the physical volume compared to the observed
volume also contributes, though this contribution does not mix with the
calculations in this work due to the linear-order galaxy bias factor.
To be definite about no relativistic contribution from the light propagation
in a single-field inflationary scenario, we need to repeat the same computation
here with the second-order description of the volume fluctuation~$\de V$.
With the complete verification of the gauge-transformation properties
of all the second-order expressions in the light propagation and observations
\cite{MAYO22}, this computation will be soon performed in a near future.
Furthermore, the cancellation of the relativistic contributions
to the primordial non-Gaussian signals
takes place in the observed three-point correlation function, 
while the relativistic effects are in general present in other statistics.

In literature, the power spectrum analysis is often performed to predict
the primordial non-Gaussian signals in galaxy clustering and to
quantify the detectability in a given survey. Since the local-type primordial
non-Gaussianity features strong signals in the squeezed triangle,
the power spectrum analysis appears natural. However, a Fourier transformation
in a hypersurface as in Eq.~\eqref{pokmain} or Eq.~\eqref{pokint} yields
a prediction that is different from the power spectrum obtainable by integrating
over a light cone volume (see, e.g., \cite{GRSCET20}), as the correlation
function is {\it not} just a function of separation in a hypersurface.
In particular, on such large scales, where the primordial non-Gaussian signals
are strong, the line-of-sight evolution and the geometry of the survey
drive the observed power spectrum away from the simple prediction in
a hypersurface, which would be accurate enough on small scales.
While the correlation function is free of these issues in the power spectrum
analysis, its signal is close to zero and nearly featureless on such large
scales, hence making it vulnerable to other systematic errors.
The angular power spectrum analysis is also free of the issues associated with
the geometry of surveys.
However, as it is computed by projecting the observed
fluctuation along the line-of-sight, it loses the redshift information in the
statistics, and the non-Gaussian signals on large scales show up only at low
angular multipoles  (see, e.g., \cite{SLHIET08}),
where the cosmic variance is largest and cannot be
further reduced by increasing the survey volume.
Other methods to quantify the observable signals on such large scales
need to be further developed and improved such as the spherical Fourier
analysis (see, e.g., \cite{BIQU91,FILAET95,HETA95,YODE13,RARE12}),
which uses the spherical harmonics for angular
decomposition and the spherical Bessel function for radial Fourier analysis.
Of course, all these two-point statistics are relevant for probing the
local-type
primordial non-Gaussianity, if the squeezed three-point correlation contributes
to the two-point correlation of the galaxy distribution, or if the galaxy bias
model is valid.

\acknowledgments
We acknowledge support by the Swiss National Science Foundation
(SNF CRSII5\_173716). J.Y. is further supported by
a Consolidator Grant of the European Research Council (ERC-2015-CoG grant
680886).
N.G. has received funding through a grant 
from the European Research Council (ERC) 
under the European Union's Horizon
2020 research and innovation program 
(Grant agreement No.~863929: project title 
``Testing the law of gravity with novel large-scale structure observables'').

\appendix

\section{Analytical expressions}
\label{append}
\subsection{Analytic solutions and simplification in the standard cosmology}
\label{setup}
Here we adopt the most general representation of a spatially flat
Friedmann-Robertson-Walker (FRW) metric and choose a rectangular coordinate:
\beeq
\label{metric}
ds^2=-a^2\left(1+2\alpha\right)d\eta^2-2a^2\beta_{,\al}dx^\al d\eta
+a^2\left[(1+2\varphi)\de_{\al\be}+2\ga_{,\al\be}\right]dx^\al dx^\be~,
\eneq
where $\alpha,\beta,\cdots$ represent the spatial indicies, $\eta$ is the
conformal time coordinate, and~$a(\eta)$ is the expansion scale factor.
Four perturbations $\alpha,\beta,\gamma,\varphi$ represent the
full scalar degrees of freedom in metric tensor. We also assumed that there is
no vector or tensor perturbations at the linear order. Though second-order
scalar perturbations generate the vector and tensor perturbations at second
order, they do~not couple to scalar perturbations if there is none at linear
order. Furthermore, we use the superscript to indicate the perturbation order
of each variable, for instance,
\beeq
\alpha(x^\mu)=\alpha^{(1)}+\alpha^{(2)}+\cdots~.
\eneq

While our theoretical descriptions of the observable quantities are 
gauge-invariant as a whole, individual components are gauge-dependent. 
Furthermore, since their expressions take different forms in different gauges,
we introduce two gauge choices convenient for our calculations:
conformal Newtonian gauge ($\chi\equiv0$) and the comoving gauge ($v\equiv0$).
Both gauge conditions fix the spatial gauge condition by setting at all 
perturbation orders
\beeq
\label{CCC}
\gamma\equiv0~ \Dquad\Dquad (\up{spatial~C~gauge})~,
\eneq
or spatial C-gauge \cite{NOHW04,YOZA14}.
The temporal gauge conditions are fixed again at all orders in perturbations by
\bear
\label{gauge}
&&\chi:=a\beta+a\gamma'\equiv0\Dquad\Dquad {\rm (Newtonian)}~,\\
&&u_\al=g_{\al\mu}u^\mu=:-av_{,\al}\equiv0\Dquad ~~~{\rm (comoving)}~,
\enar
where $u^\mu$ is the four velocity. It is clear from the definition of the
comoving gauge that there exist many different choices 
for comoving gauge, depending on which component's $u_\al$ is set zero.
Here we choose $v=0$ for the matter four velocity~$u^\mu$.
Both choices completely fix the gauge symmetry and leave no unphysical degree
of freedom. The other popular choice of gauge condition is the synchronous
gauge, where the temporal gauge is fixed with vanishing fluctuation in the
time component
\beeq
\label{sync}
\alpha\equiv0\Dquad {\rm (synchronous)}~,
\eneq
combined with the vanishing off-diagonal component 
\beeq
\label{BBB}
\beta\equiv0 \Dquad\Dquad (\up{spatial~B~gauge})~.
\eneq
The synchronous gauge conditions leave spatial gauge freedoms
(see, e.g., \cite{MABE95,YOO14b})
\beeq
T={c_1(\xvec)\over a}~,\Dquad L=c_1(\xvec)\int{dt\over a^2}+c_2(\xvec)~,
\eneq
in the temporal and spatial gauge, so that an extra temporal comoving condition
for the matter velocity $(v_m\equiv0)$ is often
imposed to fix gauge freedom, which still leaves 
$c_2(\xvec)$ arbitrary. According to the convention in \cite{YOO14b},
this choice of the synchronous gauge ($\al\equiv\be\equiv v\equiv0$)
is referred to as gauge-II, while the standard comoving gauge ($v=0$)
with spatial C-gauge ($\ga=0$) is called gauge-I.

Here we consider the standard cosmology, in which the initial condition is set
during the single-field inflationary period and the subsequent evolution
leads to a $\Lambda$CDM universe today.
Assuming a pressureless medium, we derive the
linear-order analytical relations among the metric perturbation variables 
by solving the Einstein equation (see \cite{YOGO16,BIYO17,SCYOBI18,GRSCET20}
for derivations),  and they are all related to the
initial condition~$\RR$ characterized by the comoving-gauge curvature
perturbation~$\varphi_v$:
\beeq
\label{cv}
\varphi_v:=\varphi-\HH v~,\Dquad \dot\varphi_v^{(1)}=0~,\Dquad 
\RR(\bm{x}):=\varphi_v(\bm{x},t_i)~,
\eneq
where the comoving-gauge curvature perturbation~$\varphi_v$ is conserved 
in time in a $\Lambda$CDM universe
and we defined the initial condition~$\RR$ in a hypersurface 
at some early time~$t_i$. Note that the comoving-gauge curvature perturbation
beyond the linear order evolves in time and this second-order growing solution
vanishes in the limit $t_i\RA0$. So, the initial condition~$\RR(\bm{x})$
includes non-vanishing time-independent solution beyond the linear order
in perturbations.

In the conformal Newtonian gauge, three perturbation
variables are relevant for our calculations: two gravitational potentials
and the scalar velocity potential of the matter four velocity:
\beeq
\label{Newtg}
\alpha_\chi:=\alpha-\dot \chi~,\Dquad \varphi_\chi:=\varphi-H\chi~,\Dquad
v_\chi:=v-\frac1a\chi~.
\eneq
In a $\Lambda$CDM universe, two gravitational potentials are identical with
different sign:
\beeq
\Psi:=\alpha_\chi^{(1)}=-\varphi_\chi^{(1)}\equiv D_\Psi\RR~,\Dquad 
D_\Psi(t):=\frac1\Sigma-1~,
\Dquad \Sigma(t):=1+\frac32{\Omega_m(t)\over f(t)}~,
\eneq
where we defined the time-dependent growth factor~$D_\Psi$ for the Newtonian
gauge potential~$\Psi$ and~$f(t)$ is the standard logarithmic growth rate 
of structure. The analytic relation to the initial condition is derived
 \cite{YOGO16,BIYO17,SCYOBI18,GRSCET20} from the Einstein equation.
Similarly, the scalar velocity potential is then
\beeq
 v_\chi^{(1)}=-D_V\RR\,,\Dquad D_V(t):={1\over\HH\Sigma}~,\Dquad
u^\al:=-\frac1av_\chi{}^{,\al}~.
\eneq
In the comoving gauge, the spatial velocity of the matter fluid is zero
($v=0$), and
the comoving-gauge curvature perturbation is conserved at the linear order.
The density fluctuation in the comoving gauge describes the growth of 
structure in the rest frame:
\beeq
\label{growth}
\delta_v:=\delta+3\HH v~,\Dquad \delta_v^{(1)}=-D_1\Delta\RR~,\Dquad
 D_1(t):=H\int_0^t{dt'\over \HH^2(t')}={1\over\HH^2f\Sigma}~,
\eneq
corresponding to the standard matter density fluctuation in literature.
Note that the growth factor~$D_1$ is not normalized to unity at the present 
time~$t_o$ and its relation to the logarithmic growth rate is
\beeq
f(t)={d\ln D_1\over d\ln a}~,\Dquad D_1'=\HH fD_1~.
\eneq
The second-order growth functions in a $\Lambda$CDM universe were derived
in \cite{YOGO16}, and they are explicitly
\beeq
\label{growth2}
D_A(t):={7\over10}H\int_0^tdt'~D_1^2f\left(\Sigma+\frac12f+2\right)~,\Dquad
D_B(t):={7\over4}H\int_0^tdt'~D_1^2f\left(\Sigma-\frac12f\right)~,
\eneq
in relation to the second-order solution for the matter density in
Eq.~\eqref{newt}. The growth factors are dimensionful, s.t.,
the density fluctuation~$\de_v$ is dimensionless:
$[D_1]=L^2$ and $[D_A]=[D_B]=L^4$.

Now we use these linear-order relations to derive the analytic expression
of the observed matter density fluctuation in Eq.~\eqref{main}. We compute
the individual components of the analytic expression in the temporal comoving
gauge $(v\equiv0)$. Apart from the second-order matter density fluctuation in 
Eq.~\eqref{propt}, the individual components need to be computed only up to
the linear order in perturbations (see, e.g., \cite{SCYOBI18,GRSCET20}
for the linear-order
expressions for the individual components associated with the distortion
of the source position compared to the observed position).
First, the distortion in the observed redshift is 
\beeq
\delta z_v:=- H_z\chi_v+H_o\chi_v(x_o)+V_\parallel-V_\parallel(x_o)
-\alpha_\chi+\alpha_\chi(x_o)-\int_0^{\bar r_z}\dr \left(\alpha_\chi
-\varphi_\chi\right)' ~,
\eneq
where $V_\parallel=-v_{\chi,\al}n^\al$ is the line-of-sight peculiar velocity
set by the observed direction~$n^\al$, the integration is along the 
line-of-sight direction, and~$x_o$ (or just subscript~$o$) 
represents that quantities are evaluated
at the observer position. We also used the subscript~$v$ to indicate that 
the gauge-dependent terms~$\dz$ and~$\chi$
are evaluated in the comoving gauge ($v=0$):
\beeq
\chi_v=a\beta_v=-av_\chi=aD_V\RR~.
\eneq
The coordinate time lapse~$\delta\eta_o$ in~$\delta z$ vanishes in the comoving
gauge, while non-vanishing in the conformal Newtonian gauge (see, e.g., 
\cite{GRSCET20}).
The line-of-sight integral can be simplified using integration by part
\beeq
-2\int_0^\rz\dr~ \alpha_\chi'=2\int_0^\rz\dr\frac d{d\rbar}
\left[D_\Psi(\rbar)\right]\RR(\bar r\hat n) 
=2\rbr{D_\Psi R-D_{\Psi o} R_o}-2\int_0^\rz\dr D_\Psi(\bar r)
 \partial_r R(\bar r\hat n)\,,
\eneq
and we arrive at the simplified expression of the distortion in the observed
redshift
\beeq
\dz_v=\RR_o-\RR+n^\alpha\left(D_V\RR_{,\alpha}-D_V\RR_{,\alpha}\big|_o\right)
-2\int_0^\rz\dr~ D_\Psi(\bar r) ~\partial_r R(\bar r\hat n)\,,
\eneq
where we used the relation
\beeq
D_\Psi(t)=\HH(t)D_V(t)-1~.
\eneq
Since the distortion in the time coordinate of the source position
from the observed redshift is
\beeq
\label{nota}
\delta\eta_s={\dz_v\over\Hz}~,
\eneq
and the time derivative of the linear-order matter density fluctuation is
\beeq
\delta'_v=-D_1'\Delta\RR=-\HH fD_1\Delta\RR=-D_V\Delta\RR\,,
\eneq
we can evaluate the first coupling term  in Eq.~\eqref{main} from
the temporal distortion
\beeq
\label{delta2temporal}
\delta\eta_s~ \delta_v'=\frac{1}{\Hz}D_V \Delta \RR\left[\RR-\RR_o
-n^\alpha\left(D_V\RR_{,\alpha}-D_V\RR_{,\alpha}\big|_o\right)
+2\int_0^\rz\dr D_\Psi(\bar r)~ \partial_r \RR(\bar r\hat n)\right]\,. 
\eneq
Mind that we used the notation $\Delta\eta_s$ for~$\de\eta_s$ in
Eq.~\eqref{nota} in previous work \cite{SCYOBI18,GRSCET20}.

Next we move to the remaining coupling terms in Eq.~\eqref{main} and
compute the spatial distortion in the source position. Compared to the observed
position, the spatial distortion of the source can be decomposed along and
perpendicular to the line-of-sight direction. The radial distortion
along the line-of-sight direction is
\beeq
\delta r_v:=\chi_v(x_o)-\frac{\delta z_\chi}{\HH_z}+\int_0^{\bar r_z}
\mathrm d\bar r\, (\alpha_\chi-\varphi_\chi)+n_\alpha \delta x^\alpha_v(x_o)~,
\eneq
where $\delta z_\chi=\delta z+H\chi$ is a gauge-invariant combination and
the spatial shift of the observer position is obtained by integrating the
shift until the present time~$\bar\eta_o$ as
\beeq
\de x^\al_v=\int_0^{\bar\eta_o}d\eta~\beta_v{}^{,\al}=
\int_0^{\bar\eta_o}d\eta~D_V\nabla^\al\RR\bigg|_{\bm{x}=0}~.
\eneq
Using the analytical relations, we simplify the radial distortion as
\bear
\delta r_v&=&D_V(x_o)\RR_o+\frac{1}{\Hz}\left[-D_\Psi\RR-\RR_o-n^\alpha\left(
D_V\RR_{,\alpha}-D_V\RR_{,\alpha}\bigg|_o\right)
+2\int_0^\rz\dr~ D_\Psi(\bar r)~ \partial_r \RR(\bar r\hat n)\right] \nnn
&&
+2\int_0^\rz\dr ~D_\Psi(\bar r)\RR(\bar r\hat n)+\int_0^{\bar\eta_o}
\mathrm d\eta\,D_V~\partial_r\RR\bigg|_{\bm{x}=0}\,,
\enar
where the integration along the time coordinate in~$\delta x^\al_v$ is
not to be confused with the line-of-sight integration. 
The angular distortion of the source position along the polar direction is
\bear
\bar r_z~\delta\theta_v&=&-\bar r_z\theta_\alpha V^\alpha_o
-\int_0^\rz\dr (\bar r_z-\bar r)\theta_\alpha (\alpha_\chi
-\varphi_\chi)^{,\alpha}+\theta_\alpha \delta x^\alpha_v~\\
&=&
-\rz \theta^\alpha (D_V\RR_{,\alpha})\big|_o
-2\int_0^\rz\dr (\rz-\bar r) D_\Psi(\bar r)\theta^\alpha \RR_{,\alpha}
+\int_0^{\bar\eta_o}\mathrm d\eta~D_V(\bar\eta)\theta^\alpha \RR_{,\alpha}
\bigg|_{\bm{x}=0} ~,\nn
\enar
where we ignored the orientation $\Omega^i$ of the observer frame in the
full expression of~$\dtt$, 
as it is not correlated
with~$\RR$. Similarly, the angular distortion along the azimuthal direction
is identical with $\rz\sin\theta\de\phi_v$ in the left-hand side and
with $\theta_\alpha$ replaced by~$\phi_\alpha$ in the right-hand side.

Now that we have expressed all the individual components of the matter 
density fluctuation in Eq.~\eqref{main} in terms of the initial 
condition~$\RR$, we put them together here to show our main analytic equation
for the matter density fluctuation in the standard cosmology,  and we will
use this equation to compute the bispectrum in the squeezed limit:
\bear
\label{mainRR}
\deobs&=&D_1\left(-\Delta \RR+\frac 32 \RR^{,\al}\RR_{,\al}
+4\RR\Delta \RR
  \right)+\frac 57D_{A}\nabla_\al\left(\RR^{,\al}\Delta \RR
\right)+\frac 17D_{B}\Delta\left(\RR^{,\al}\RR_{,\al}\right) \\
&&
+\frac{1}{\Hz}D_V \Delta \RR\left[\RR-\RR_o
-n^\alpha\left(D_V\RR_{,\alpha}-D_V\RR_{,\alpha}\big|_o\right)
+2\int_0^\rz\dr D_\Psi(\bar r)~ \partial_r \RR(\bar r\hat n)\right] \nnn
&&
-D_1{\pa\over\partial r}\Delta\RR
\bigg\{D_V(x_o)\RR_o-\frac{1}{\Hz}\bigg[D_\Psi\RR+\RR_o+n^\alpha
\left(D_V\RR_{,\alpha}-D_V\RR_{,\alpha}\big|_o\right)   \nnn
&&
\Dquad
-2\int_0^\rz\dr D_\Psi(\bar r) ~\partial_r \RR(\bar r\hat n)\bigg]
+2\int_0^\rz\dr D_\Psi(\bar r)\RR(\bar r\hat n)\bigg\}
+D_1 \left(D_V\nabla^\alpha\RR\right)_o\habla_\alpha\left(\Delta \RR\right) \nnn
&&
+2D_1\habla_\alpha\left(\Delta\RR\right)\int_0^\rz\dr
\left(\frac{\rz-\bar r}{\rz\bar r}\right)D_\Psi(\bar r)\habla^\alpha \RR 
-D_1\nabla^\al\left(\Delta\RR\right)
\int_0^{\bar\eta_o}\mathrm d\eta\,D_V\nabla_\al \RR\bigg|_{\bm{x}=0}
  \nn~,
\enar
where $\hat\nabla$ is the angular gradient, 
the first line is the second-order matter density in Eq.~\eqref{propt},
and the remaining terms are from $\Delta x^\mu\partial_\mu\delta_v$ in
 sequential
order. Three components from the spatial shift of the observer position
are combined into one in the last line, and the product of two
angular gradients in the last line represents
\beeq
2D_1\int_0^\rz\dr\rbr{\frac{\rz-\bar r}{\rz\bar r}} D_\Psi(\bar r)
\left[\left({\partial\over\partial\theta}\Delta\RR\right)_{\rz}
\left({\partial\over\partial\theta}\RR\right)_{\rbar}+
{1\over\sin^2\theta}\left({\partial\over\partial\phi}\Delta\RR\right)_{\rz}
\left({\partial\over\partial\phi}\RR\right)_{\rbar}\right]~.
\eneq
Equation~\eqref{mainRR} is the main equation shown in Eq.~\eqref{mainRRmain}
for our computation.

Further simplification can be made in the limit that the Universe is 
approximated as the Einstein-de~Sitter universe, or matter-dominated universe,
in which $\Lambda=0$ and the matter density parameter~$\Omega_m=1$ ($f=1$).
Due to the simplicity, we can derive the analytic solutions, regarding
the Hubble parameter and the angular diameter distance
\bear
a&=&\left({t\over t_o}\right)^{2/3}=\left({\eta\over\eta_o}\right)^2~,
\Dquad {t\over t_o}=\left({\eta\over\eta_o}\right)^3~,\Dquad \eta_o=3t_o~,\\
H&=&{2\over3t}~,\Dquad \HH={2\over\eta}~,\Dquad \rbar_z=\bar\eta_o-\bar\eta=
{2\over H_0}\left(1-{1\over\sqrt{1+z}}\right)~,
\enar
and the relation among the perturbation variables
\beeq
\Sigma=\frac 52\,,\Dquad D_\Psi=-\frac 35\,,
\Dquad D_1=\frac{\eta^2}{10}\,,\Dquad D_V=\frac \eta5\,,\Dquad
D_A=D_B=D_1^2={\eta^4\over100}~.
\eneq

\subsection{Fourier decomposition and one-point ensemble average}
\label{FDEV}
To facilitate the subsequent calculations, we introduce a set of second-order
Fourier kernels~$F(\kone,\ktwo)$ 
that capture various contributions to the matter density fluctuation 
in Eq.~\eqref{mainRR}
at the observed redshift and describe their time evolution from the initial 
condition~$\RR$. Schematically, the matter density fluctuation at the second
order will be expressed as
\beeq
\delta(\mathbf x) \propto
\int\Dkone\int\Dktwo ~e^{i\kone\cdot\mathbf x}e^{i\ktwo\cdot\mathbf x}
F(\mathbf k_1,\mathbf k_2;\Vang,\etaz)\RR(\kone)\RR(\ktwo)\,,
\eneq
where the Fourier kernel is dimensionless.
As shown in Eqs.~\eqref{main} and~\eqref{mainRR}, 
the observed matter density fluctuation at the second order is coupled with 
the contributions at the source position, at the observer position, or
along the line-of-sight direction. So we discuss three different 
types of Fourier
kernels in turn. The derivation of Fourier kernels for the individual 
contributions is presented in Section~\ref{details}.

Furthermore, it proves useful for the computational convenience and also 
conceptual clarity to define Fourier counter parts, given the 
individual contributions to the matter density fluctuation. These Fourier
counter parts in configuration space and Fourier space are defined in 
a hypersurface set by the observed redshift~$z$ in the usual way:
\beeq
\label{deltatheory}
\delta(\mathbf x)=\int\Dkk\,e^{i\mathbf k\cdot \mathbf x}
~\delta(\mathbf k;\Vang, \etaz)\,,\Dquad
\delta(\mathbf k;\Vang, \etaz)=\int d^3x\,e^{-i\mathbf k\cdot \mathbf x}
~\delta(\mathbf x)~,
\eneq
where the volume integral over the position~$\mathbf x$
is all over the infinite hypersurface, not over the light cone volume.
Note that this hypersurface encompasses a volume outside the observed light 
cone volume and hence these Fourier counter parts are not directly observable,
except at the intersection with the light cone volume.
However, we can imagine that
fictitious observers at different spatial position in 
the Universe perform the same observations. In other words,
for the same observed redshift
and angle, these fictitious observers can construct the
fluctuation field~$\delta(\mathbf x)$ with the observed position
$\mathbf x=\rbar_z\Vang+\mathbf x_o$,
where $\mathbf x_o$ is the spatial position of the fictitious observer
and can be set zero for the real observer (us). Noting that our position
in the Universe is not special, it is conceptually useful to think of such
fictitious observations and to construct the observed matter density field 
and its Fourier counter part outside our own light cone volume. However, it
is noted that since the observed matter density fluctuation depends not only
on the redshift, but also on the angle, the observed angle should be specified,
when the fictitious observations are considered, and hence the angular
dependence in
Eq.~\eqref{deltatheory}. We refer the reader to the work \cite{GRSCET20}
for more detailed discussion and computation.

The Fourier kernels are also useful in computing the one-point ensemble 
average. The ensemble average~$\langle\delta(\mathbf x)\rangle$ is an 
average of the fluctuation~$\delta(\mathbf x)$ over many realizations of
the Universe, as discussed in Section~\ref{ensavg}.
With the ergodic theorem, the ensemble average is equivalent
to the Euclidean average over the hypersurface, which can be readily
computed in our formalism
with the observations by the fictitious observers. Note that
the cosmic variance in practice arises due to our limitation to one 
observer position or the lack of average over translation in the Euclidean
average (see, e.g., \cite{MIYOET20}). The average over the hypersurface
yields
\beeq
\label{ensemble}
\langle\delta(\mathbf x)\rangle \underset{\up{Erg.}}{=}\lim_{V\RA\infty}
{1\over V}\int d^3x\int\Dkk\,
e^{i\mathbf k\cdot \mathbf x}~\delta(\mathbf k;\Vang, \etaz)
={1\over V}\big\langle\delta(\mathbf k\equiv0;\Vang, \etaz)\big\rangle~.
\eneq
In general, the ensemble average of a fluctuation is zero for a Gaussian
distribution. However, at the second order, and in particular with the
relativistic contributions, the ensemble average of the matter density
fluctuation is {\it non-vanishing}, as we show in the following.
More importantly, it is apparent by the definition
of the average over a hypersurface that the ensemble average depends on
a choice of hypersurface.\\

\Ietc{Coupling terms with contributions at the source position}
First, we consider the contributions~$\de_s(\xvec)$
at the source position only, and
these terms in Eq.~\eqref{mainRR} can be expressed as
\beeq
\delta_s(\mathbf x)=:
\int\Dkone\int\Dktwo ~e^{i\kone\cdot\mathbf x}e^{i\ktwo\cdot\mathbf x}
F_s(\mathbf k_1,\mathbf k_2;\Vang,\etaz)\RR(\kone)\RR(\ktwo)\,, 
\eneq
and this yields the Fourier mode according to Eq.~\eqref{deltatheory}
\beeq
\label{deltasth}
\delta_s(\mathbf k;\Vang,\etaz)=\int\frac{\mathrm d^3q}{(2\pi)^3}~
F_s(\mathbf q,\mathbf k-\mathbf q;\Vang,\etaz)~
\RR(\mathbf q)\RR(\mathbf k-\mathbf q)\,,
\eneq
where the Fourier kernel~$F_s$ is subject to
\beeq
F^\ast_s(\mathbf k_1,\mathbf k_2;\n,\etaz)
=F^\ast_s(-\mathbf k_1,-\mathbf k_2;\n,\etaz)\,.
\eneq
Note that the Fourier kernels include the time-dependence. For example,
the linear-order  matter density contribution to the Fourier kernel is
\beeq
\delta^{(1)}_m(\mathbf x,t)=-D_1(t)\Delta\RR(\mathbf x)~,\Dquad
F_s\ni~D_1(t)k^2~.
\eneq
For later convenience, we introduce another notation for Fourier
kernels~$\FF$ that are directly related to the bispectrum, and for the
contributions~$\de_s(\xvec)$ at the source position we simply have
\beeq
\FF_s(\kone,\ktwo;\Vang,\etaz):=F_s(\kone,\ktwo,\Vang,\etaz)~.
\eneq

Finally, the ensemble average of the coupling terms with contributions 
at the source position is obtained by using Eq.~\eqref{ensemble} as
\beeq
\label{avgs}
\ens:=\langle\delta_s(\mathbf x)\rangle=
\int\frac{\mathrm d^3q}{(2\pi)^3}F_s(\mathbf q,-\mathbf q;\Vang,\etaz)P_\RR(q)
=\int d\ln q~\Delta_\RR^2(q)\int{d\mu_q\over2}~
F_s(\mathbf q,-\mathbf q;\Vang,\etaz)~,
\eneq
where $\mu_q=\mathbf q\cdot\Vang/|\mathbf q|$ 
is the cosine angle with respect to the observed direction.\\

\Ietc{Coupling terms with contributions at the observer position}
Next we consider the contributions~$\de_o(\xvec)$
to the matter density fluctuation
coupled with those at the observer position. To compute the field in a 
hypersurface outside the light cone, we consider fictitious observers
at different spatial positions ($\mathbf x_o\neq0$). Those coupling terms
in Eq.~\eqref{mainRR} can be expressed in terms of Fourier kernel~$F_o$ as
\beeq
\label{deltao}
\delta_o(\mathbf x)=:\int\Dkone\int\Dktwo~ e^{i\kone\cdot\mathbf x_o}
e^{i\ktwo\cdot \mathbf x}F_o(\kone,\ktwo;\Vang,\etao,\etaz)
\RR(\kone)\RR(\ktwo)\,,
\eneq
where the Fourier kernel also has the time-dependence on~$\etao$
set at the observer position~$\bdv{x}_o$. 
Note that since $\rbar_z$ and $\Vang$ are fixed 
in a hypersurface, the position~$\bdv{x}_o$ for the fictitious observer
is a function of~$\mathbf x$ in consideration. Also mind that the Fourier
kernel~$F_o$ is {\it not symmetric} over the arguments~$\kone$ and~$\ktwo$
in our convention, where the wave vector~$\kone$ belongs to the Fourier 
transform of the contribution evaluated at the observer position
and $\ktwo$ belongs to the contribution at the source position.
The Fourier counter part is then
\beeq
\label{deltaoth}
\delta_o(\mathbf k;\Vang,\etao,\etaz)=\int\Dq~e^{-i\mathbf q\cdot \rbar_z\n} 
~F_o(\q,\mathbf k-\q;\Vang,\etao,\etaz)~\RR(\q)\RR(\mathbf k-\q)\,, 
\eneq
and its ensemble average is
\bear
\ens&=&\langle\delta_o(\mathbf x)\rangle=
\int\Dq ~e^{-i\rz\q\cdot\n}F_o(\q,-\q;\n,\etao,\etaz)P_\RR(q)\\
&=&\int d\ln q~\Delta_\RR^2(q)\int{d\mu_q\over2}~e^{-iq\rbar_z\mu_q}
F_o(\mathbf q,-\mathbf q;\Vang,\etao,\etaz)~. \nn
\enar
We define the Fourier kernel  
\beeq
\label{FFO}
\FF_o(\kone,\ktwo;\Vang,\etao,\etaz):
=e^{-i\kone\cdot\rbar_z\n}F_o(\kone,\ktwo;\Vang,\etao,\etaz)~,
\eneq
and note that it includes exponential factor.\\

\Ietc{Coupling terms with contributions along the line-of-sight direction}
Finally, we consider the contributions~$\de_{nl}(\xvec)$
to the matter density fluctuation
coupled with those along the line-of-sight direction:
\beeq
\label{deltanl}
\delta_{nl}(\mathbf x)=:\int\Dkone\int\Dktwo\int_0^\rz\dr ~
e^{i\bar r \kone\cdot\Vang}e^{i\rz\ktwo\cdot\Vang}
~F_{nl}(\kone,\ktwo;\Vang,\rbar,\etaz)\RR(\kone)\RR(\ktwo)\,, 
\eneq
where the time-dependence of the Fourier kernel is specified by~$\rbar$
in the line-of-sight integration. We use the subscript~$nl$ to refer to this
as non-local Fourier kernel. The kernel is also {\it not symmetric} in 
the arguments, and the wave vector $\kone$ belongs to the contribution 
evaluated along the line-of-sight, while $\ktwo$ to the one evaluated 
at the source position. The Fourier counter part is then
\beeq
 \label{deltanlth}
\delta_{nl}(\mathbf k;\Vang,\etaz)=\int\Dq\int_0^\rz\dr 
e^{-i\Delta r \q\cdot\Vang}~F_{nl}(\q,\mathbf k-\q;\rbar,\etaz)
~\RR(\q)\RR(\mathbf k-\q)\,,
\eneq
and its ensemble average is
\bear
\ens&=&\langle\delta_{nl}(\mathbf x)\rangle=
\int \Dq\int_0^\rz\dr e^{-i\Delta r\mathbf q\cdot\Vang}~
F_{nl}(\q,-\q;\Vang,\rbar,\etaz)P_\RR(q) \\
&=&\int d\ln q~\Delta_\RR^2(q)\int_0^\rz\dr 
\int{d\mu_q\over2}~e^{-iq\Delta r\mu_q}
F_{nl}(\mathbf q,-\mathbf q;\Vang,\rbar,\etaz)~. \nn
\enar
where $\Delta r:=\rbar_z-\rbar$. The Fourier kernel for the non-local
contributions is then defined as 
\beeq
\label{FFnl}
\FF_{nl}(\kone,\ktwo;\Vang,\etaz):=\int_0^{\rbar_z}\dr~
e^{-i\kone\cdot\Delta r\n}F_{nl}(\kone,\ktwo;\Vang,\rbar,\etaz)~,
\eneq
and it includes the line-of-sight integration and the exponential factor
in the kernel.
While $F_{nl}$ is dimensionful, the Fourier kernel~$\FF_{nl}$ is dimensionless
due to the line-of-sight integral.

\subsection{Three-point correlation and the bispectrum}
\label{thrpt}
Here we compute the three-point correlation function and its bispectrum.
The leading-order contribution to the three-point statistics
arises from the contraction
of two linear-order contributions and a second-order contribution of
the matter density fluctuation.
Since the Fourier mode of the observed matter density fluctuation
at the second order in perturbations is collectively expressed in terms of
Fourier kernels as
\beeq
\delta(\mathbf k;\Vang,\etaz)=\int\frac{\mathrm d^3q}{(2\pi)^3}~
\FF(\mathbf q,\mathbf k-\mathbf q;\Vang,\etaz)~
\RR(\mathbf q)\RR(\mathbf k-\mathbf q)\,,
\eneq
and the linear-order matter density fluctuation is
\beeq
\delta^{(1)}(\mathbf k)=D_1(z)k^2\RR(k)~,
\eneq
the leading-order contribution to the bispectrum is then
\bear
\label{bispec}
\Big\langle\delta^{(1)}(\mathbf k_1)\delta^{(1)}
(\mathbf k_2)\delta^{(2)}(\mathbf k_3)\Big\rangle
&=&D_1^2k_1^2k_2^2P_\RR(k_1)P_\RR(k_2)(2\pi)^3\delta^D(\kone+\ktwo+\mathbf k_3)
\\
&&
\times\left[\FF(-\mathbf k_1,-\mathbf k_2;\Vang_3)
+\FF(-\mathbf k_2,-\mathbf k_1;\Vang_3)\right]
+\Big\langle\delta^{(1)}(\mathbf k_1)\delta^{(1)}(\mathbf k_2)\Big\rangle
\Big\langle\delta^{(2)}(\mathbf k_3)\Big\rangle~,\nn
\enar
where we omitted the time dependence in the growth function~$D_1$
and the Fourier kernels~$\FF$. Note that the tadpole term exists because
the one-point ensemble average $\AVE{\de^{(2)}(\kthree)}$
is {\it not}
vanishing. Accounting for the permutation of three leading-order
contributions and for three different observed positions at the same observed
redshift, we derive the observed three-point correlation function 
\bear
\label{ddd}
&&
\Big\langle \delta(\mathbf x_1)\delta(\mathbf x_2)\delta(\mathbf x_3)
\Big\rangle=\int\Dkone\int\Dktwo\int\Dkthr e^{i\mathbf k_1\cdot\mathbf x_1}
e^{i\mathbf k_2\cdot\mathbf x_2}e^{i\mathbf k_3\cdot\mathbf x_3}
\Big\langle\delta(\mathbf k_1)\delta(\mathbf k_2)\delta(\mathbf k_3)
\Big\rangle   \\
&&\Dquad
=\int\Dkone\int\Dktwo e^{i\mathbf k_1\cdot(\mathbf x_1-\mathbf x_3)}
e^{i\mathbf k_2\cdot(\mathbf x_2-\mathbf x_3)}\bigg\{
D_1^2k_1^2k_2^2P_\RR(k_1)P_\RR(k_2)\bigg[\FF(-\mathbf k_1,-\mathbf k_2;\Vang_3)
~~~~  \nnn
&&\Dquad\Dquad
+\FF(-\mathbf k_2,-\mathbf k_1;\Vang_3)\bigg]
+\up{perm.}\bigg\}
 +\Big[\xi_m(\mathbf x_1,\mathbf x_2)+\xi_m(\mathbf x_2,\mathbf x_3)
+\xi_m(\mathbf x_3,\mathbf x_1)\Big]\ens~,\nn
\enar
where the permutation terms contain the Fourier kernels~$\FF$ with two
other angular positions~$\Vang_1$ and~$\Vang_2$ and
the two-point correlation function is the just the linear-order
matter correlation function
\beeq
\xi_m(\mathbf x_1,\mathbf x_2)=\left\langle\de^{(1)}(\mathbf x_1)
\de^{(1)}(\mathbf x_2)\right\rangle~.
\eneq
The three-point correlation is non-vanishing, even
for a configuration $\xone=\xtwo$ and $\xthree\RA\infty$ due to the
non-vanishing constant contribution
$\ens=\langle\de(\mathbf x)\rangle\neq0$,
or the difference between the ensemble average and
the background at a given time.

To address this subtlety, we first define a dimensionless fluctuation
\beeq
\DD(\mathbf x):=\delta(\mathbf x)-\ens~,\Dquad
\Big\langle \DD(\mathbf x)\Big\rangle=0~,
\eneq
though the correct fluctuation should be with extra constant $(1+\ens)$.
For instance, the matter density can be split as
\beeq
\rho(\mathbf x)=\bar\rho(t)\left(1+\ens\right)\left(1+{\DD(\mathbf x)
\over 1+\ens}\right)~,\Dquad \AVE{\rho(\xvec)}=\bar \rho(t)(1+\ens)~.
\eneq
The three-point correlation function of~$\DD(\mathbf x)$ is then
\beeq
\Big\langle\DD(\mathbf x_1)\DD(\mathbf x_2)\DD(\mathbf x_3)\Big\rangle=
\Big\langle \delta(\mathbf x_1)\delta(\mathbf x_2)\delta(\mathbf x_3)
\Big\rangle-\bigg[\Big\langle \delta(\mathbf x_1)\delta
(\mathbf x_2)\Big\rangle~\Big\langle\delta(\mathbf x_3)\Big\rangle +\up{perm.}
\bigg]+2\big\langle\delta\big\rangle^3 ~,
\eneq
where the extra terms in the three-point correlation including
the constant contribution ensure that the tadpole terms do not contribute
and only the connected contribution to the three-point correlation function
remains. Computing to the second-order in the matter density fluctuation,
we derive the three-point correlation function
\bear
\label{DDD}
\Big\langle\DD(\mathbf x_1)\DD(\mathbf x_2)\DD(\mathbf x_3)\Big\rangle
&=&2~\ens^3+
\int\Dkone\int\Dktwo ~e^{i\mathbf k_1\cdot(\mathbf x_1-\mathbf x_3)}
e^{i\mathbf k_2\cdot(\mathbf x_2-\mathbf x_3)} \\
&&
\times\left\{
D_1^2k_1^2k_2^2P_\RR(k_1)P_\RR(k_2)
\bigg[\FF(-\mathbf k_1,-\mathbf k_2;\Vang_3)+\FF(-\mathbf k_2,
-\mathbf k_1;\Vang_3)\bigg]+\up{perm.}\right\}~,\nn
\enar
where the tadpoles in Eq.~\eqref{ddd} are all removed.
Furthermore, since we only considered the field
up to the second order in perturbations or keep terms up to~$P_\RR^2$,
the constant term~$\ens^3\sim P_\RR^3$ will be ignored for consistency,
as a proper treatment of such term would need one-loop
contributions~$P_\RR^3$ in the bispectrum.

Having derived the general expression for the three-point statistics,
we now consider a special triangular configuration, in which two observed
positions are identical
and the third position points in the opposite
side of the sky\footnote{While these two identical points are then subject
to large non-linear corrections, these two points for the squeezed 
triangular configuration only need to be close to each other, compared
to the third position, avoiding any extra complication due to nonlinearity
on small scales.}
\beeq
\label{sqzcond}
z:=z_1=z_2=z_3~,\Dquad \Vang:=\Vang_1=\Vang_2=-\Vang_3~,
\eneq
i.e., one observed redshift~$z$
and one angular vector~$\Vang$ for the three points in the sky.
In terms of three-dimensional position vectors, the squeezed
triangular configuration is represented by
\beeq
\label{sqzcond2}
\mathbf x_2=\mathbf x_1~,\Dquad \mathbf L:=\xone-\xthree~,
\eneq
and our primary interest lies in the squeezed triangular configuration
in the limit $L\RA\infty$.
The squeezed correlation function can then be further simplified as
\beeq
\label{sqz}
\xi_{\rm sqz}:=
\Big\langle \DD(\mathbf x_1)\DD(\mathbf x_2)\DD(\mathbf x_3)\Big\rangle=
\int\Dkone\int\Dktwo~ e^{i(\mathbf k_1+\mathbf k_2)\cdot\mathbf L}
\bigg[B_{112}+B_{211}+B_{121}\bigg]
(\mathbf k_1,\mathbf k_2,-\mathbf k_{12})~,
\eneq
where $\mathbf k_{12}:=\kone+\ktwo$. The connected bispectra with 
dimension~$L^6$ in the integrand are defined as
\bear
\label{BBBA}
B_{112}&:=&D_1^2k_1^2k_2^2P_\RR(k_1)P_\RR(k_2)
\bigg[\FF(-\mathbf k_1,-\mathbf k_2;-\Vang)+\FF(-\mathbf k_2,-\mathbf k_1;-\Vang)\bigg]~,\\
B_{211}&:=&D_1^2k_2^2k_3^2P_\RR(k_2)P_\RR(k_3)
\bigg[\FF(-\mathbf k_2,-\mathbf k_3;\Vang)+\FF(-\mathbf k_3,-\mathbf k_2;\Vang)\bigg]~,\\
\label{BBBB}
B_{121}&:=&D_1^2k_3^2k_1^2P_\RR(k_3)P_\RR(k_1)
\bigg[\FF(-\mathbf k_3,-\mathbf k_1;\Vang)+\FF(-\mathbf k_1,-\mathbf k_3;\Vang)\bigg]~,
\enar
where the subscripts represent the order of perturbations, e.g.,
\beeq
B_{112}:=\left\langle\DD^{(1)}_{\kvec_1}\DD^{(1)}_{\kvec_2}\DD^{(2)}_{\kvec_3}
\right\rangle~,
\eneq
and the dimension of the bispectra is~$L^6$.
The task for computing the squeezed correlation boils down to computing the
Fourier kernels for Eqs.~\eqref{BBBA}$-$\eqref{BBBB}, which are presented
in Section~\ref{details}. Despite the simplicity in the
notation, it is noted that the kernels~$\FF_o$ and~$\FF_{nl}$
for the observer and the non-local contributions
include the exponential factors or the line-of-sight integral.
Moreover, though those two types of Fourier kernels are not symmetric
over the arguments, their contributions to the squeezed correlation function
are made symmetric in the bispectra.
However, mind the dependence of~$\Vang$ and its sign.

Treating the squeezed three-point correlation function as a two-point
correlation or just a function of separation~$\bdv{L}$,
the Fourier counter part or the power spectrum can be defined as 
\beeq
\label{pokint}
P(k):=\int d^3L~e^{-i\mathbf k\cdot\mathbf L}~\xi_{\rm sqz}\simeq
\int\Dkone \bigg[B_{112}+B_{211}+B_{121}\bigg]
(\mathbf k_1,\mathbf k-\mathbf k_1,-\mathbf k)~,
\eneq
where $\mathbf k=\mathbf k_1+\mathbf k_2$. It is shown \cite{MAVE08} that
this power spectrum is a useful
quantity to measure the correction to the standard galaxy power spectrum,
arising from the non-Gaussian nature of the matter density fluctuation, or 
 the squeezed three-point correlation~$\xi_\up{sqz}$. However, it is important
to note that $\xi_\up{sqz}$ here is 
the observed three-point correlation function,
which is not just a function of its separation~$\bm{L}$ alone.
Furthermore, $\xi_\up{sqz}$ is defined on the past light cone, not in an
infinite hypersurface of simultaneity at the observed redshift, 
so that any integration over~$\bm{L}$
should involves the variation in time or the observed redshift.
Therefore, the final equality for the power spectrum in terms of three 
bispectrum should be taken as a simple theoretical measure of non-Gaussianity,
rather than a real observed power spectrum that can be obtained in the light
cone volume. While one can still define the observed power spectrum 
as in Eq.~\eqref{pokint} with the volume integration over the light-cone
volume, it is {\it not} equal to the expression in the RHS of Eq.~\eqref{pokint}
in terms of the bispectra.

Before we proceed to derive the individual Fourier kernels,
we discuss the general scaling of the squeezed correlation
function (see also \cite{PEEBL80}).
Though we have computed the three-point correlation
function $\xi_{\rm sqz}$, this squeezed correlation function
contributes to the galaxy two-point correlation function as a
non-Gaussian correction discussed in Section~\ref{MVformalism}.
For such two-point correlation function, 
in the absence of non-vanishing ensemble average,
we expect it to vanish in the limit $L\RA\infty$. Assuming a
power-law relation, we {\it expect} the power-law index to be at least positive:
\beeq
\label{condition}
\xi_{\rm sqz}\propto L^{-n}~,\Dquad
\lim_{L\RA\infty}\xi_{\rm sqz}=0~,\Dquad n>0~.
\eneq
This expectation allows the possibility for the power spectrum
$P(k)\propto k^{n-3}$ to scale as 
$P(k)\propto  P_m/k^2\propto P_{\delta\phi}$
or the primordial non-Gaussianity signature discussed in 
Section~\ref{MVformalism}.

In the limit $L\RA\infty$, the squeezed triangular configuration takes the
form
\beeq
\label{sqzlimit}
\mathbf k_l:=\mathbf k_{12}=\mathbf k_1+\mathbf k_2\RA0~,\Dquad
\mathbf k_1\approx -\mathbf k_2~,\Dquad \mathbf k_3=-\mathbf k_{12}=-\mathbf
k_l~,
\eneq
and the connected bispectra {\it in this limit} are
\bear
\label{B112}
B_{112}&=&D_1^2k_1^4P_\RR(k_1)P_\RR(k_1)
\bigg[\FF(-\mathbf k_1,\mathbf k_1;-\Vang)+\FF(\mathbf k_1,-\mathbf k_1;-\Vang)\bigg]~,\\
\label{B211}
B_{211}&=&D_1^2k_1^2k_l^2P_\RR(k_1)P_\RR(k_l)
\bigg[\FF(\mathbf k_1,\mathbf k_l;\Vang)+\FF(\mathbf k_l,\mathbf k_1;\Vang)\bigg]~,\\
\label{B121}
B_{121}&=&D_1^2k_l^2k_1^2P_\RR(k_l)P_\RR(k_1)
\bigg[\FF(\mathbf k_l,-\mathbf k_1;\Vang)+\FF(-\mathbf k_1,\mathbf k_l;\Vang)\bigg]~.
\enar
While $k_l\RA0$, the other wave vector~$\kone$ can take any arbitrary value.
As the exponential factor in Eq.~\eqref{sqz} vanishes, the squeezed correlation
function is determined by the integration of these connected bispectra.
Note that the contribution of~$B_{112}$ is a constant in this limit or
independent of~$k_l$, and its ratio to the other contributions is
\beeq
{B_{112}\over B_{211}}\sim{B_{112}\over B_{121}}\sim{k_1^2P_\RR(k_1)\over k_l^2
P_\RR(k_l)}\sim\left({k_l\over k_1}\right)^{2-n_s}~,
\eneq
suppressed at least by $k_l/k_1$ before considering extra suppression factors
from the ratio of
the Fourier kernels. The calculations in Appendix~\ref{details} show that
the ratio of two Fourier kernels vanish due to the symmetry of the kernels
in $B_{112}$ or scales with some power of~$k_l$.
Therefore, we will ignore the contribution from~$B_{112}$.
The long-mode contributions in $B_{211}$ and $B_{121}$ to the power spectrum
are schematically
\beeq
\Delta P(k_l)\propto\int d\ln k_1~k_1P_m(k_1)P_m(k_l)\int d\mu_1~
{[\FF+\FF]\over k_l^2}~,\Dquad P_m(k)\propto k^4P_\RR(k)~,
\eneq
suggesting that the scaling with~$k_l$ for the sum of two kernels
in $B_{211}$ or $B_{121}$ after the angular integration should be
\beeq
\Delta P\propto k_l^{n_s-2+n_{\FF}}~,\Dquad n_{\FF}>-1-n_s\simeq-2~,
\eneq
to satisfy the condition in Eq.~\eqref{condition},
where we approximated
\beeq
\int d\mu \bigg[\FF+\FF\bigg]\propto k_l^{n_{\FF}}~.
\eneq
As demonstrated in Appendix~\ref{details}, the power-law slope
is non-negative: $n_\FF\geq0$.

\section{Detailed calculations of Fourier kernels}
\label{details}

Here we present the detailed calculations of the Fourier kernels for three
different types of contributions to the observed matter density fluctuation.
With the kernels in hands, the one-point ensemble average and three
connected bispectra can be readily computed. Each subsection specifies
the calculations according to three different types of Fourier kernels.

The squeezed triangular configuration, which is our primary interest,
is specified by only two angular
directions~$\Vang_1$ and~$\Vang_2=-\Vang_1$, opposite directions in the sky.
To facilitate the calculation, we set this direction to be aligned with
$z$-direction ($\Vang=\Vang_1=\Zang=-\Vang_2$), in which
the azimuthal integration is trivial. We will use the following relations 
for the polar integration in terms of spherical
Bessel functions~$j_n(x)$
\bear
&&
\int\mathrm {d\mu\over2}\cos(\mu a)=j_0(a)\,,\Dquad
\int{\mathrm d\mu\over2}\mu\sin(\mu a)=j_1(a)\,,\\
&&
\int{\mathrm d\mu\over2}\mu^2\cos(\mu a)={j_0(a)-2j_2(a)\over3}~.\nn
\enar

\subsection{Contributions at the source position}
\label{srccont}
The contributions at the source position arise from three different origins.
The first is the second-order matter density contributions in the
standard perturbation theory shown Eq.~\eqref{newt}, and the second is
the second-order relativistic contribution to the matter density fluctuation
in Eq.~\eqref{propt}. Finally, the last arises from the coupling terms of the
linear-order matter density fluctuation and the relativistic contribution
at the source position in Eq.~\eqref{main}. We present the detailed
calculations according to this classification.\\

\IIetc{Contributions from the standard perturbation theory} 
The second-order contributions in the standard perturbation
theory consist of two terms in Eq.~\eqref{newt}. The Fourier kernel
for the first term is
\beeq
\label{spt1}
\bullet~~
\frac 57 D_A(z)\nabla_\al\rbr{\RR^\al\Delta \RR}~:\Dquad F_s(\kone,\ktwo)
=\frac{5}{7} D_A(z) \left(\kone+\ktwo\right)\cdot \kone~k_2^2 ~,
\eneq
and given the structure of the connected bispectra we symmetrize the Fourier
kernel
\beeq
\label{sptone}
\FF_s(\kone,\ktwo)=\frac{5}{14} D_A(z) \left(\kone+\ktwo\right)\cdot
\left(k_2^2\kone+k_1^2\ktwo\right)~.
\eneq
Having derived the Fourier kernel, the one-point ensemble average in
Eq.~\eqref{avgs} is readily shown to be vanishing
\beeq
\ens=0~,
\eneq
as the sum of two wave vectors is zero.
The contribution to the connected bispectra can be derived by
using Eqs.~\eqref{B112}$-$\eqref{B121}. 
The first connected bispectra~$B_{112}$ in
Eq.~\eqref{B112} is independent of~$k_l$ and is simply zero
\beeq
B_{112}=0
\eneq
due to symmetry of the wave vectors. 
Since the pre-factors of three
bispectra are the same for different Fourier kernels,
we present calculations of the sum of Fourier kernels
in the square bracket in Eqs.~\eqref{B112}$-$\eqref{B121}.
The other two connected bispectra are non-zero, 
\bear
&&
B_{211}\ni \frac57D_A(z)\left[2k_l^2k_1^2+(k_l^2+k_1^2)\kone\cdot\klong
  \right]\RA0~,\nnn
&&
B_{121}\ni \frac57D_A(z)\left[2k_l^2k_1^2-(k_l^2+k_1^2)\kone\cdot\klong
\right]\RA0~
\enar
but they vanish in the limit $k_l\RA0$. 

The Fourier kernel for the second term of the standard perturbation theory
is already symmetric:
\beeq
\label{spttwo}
\bullet~~
\frac 17 D_B(z) \Delta\rbr{\RR^\al\RR_\al}:\Dquad
\FF_s=F_s(\kone,\ktwo)=\frac 17 D_B(z) \left(\kone+\ktwo\right)^2
\kone\cdot \ktwo ~,
\eneq
and the one-point ensemble average and the first connected bispectra
vanish due to symmetry of the wave vectors
\beeq
\ens=0~,\Dquad B_{112}=0~.
\eneq
The two connected bispectra are obtained by Eqs.~\eqref{B211} and~\eqref{B121}
\beeq
B_{211}\ni\frac27D_B(z)(\kone+\klong)^2\kone\cdot\klong\RA0~,\Dquad
B_{121}\ni-\frac27D_B(z)(\kone-\klong)^2\kone\cdot\klong\RA0~,
\eneq
and they also vanish in the limit $k_l\RA0$.

In the standard perturbation theory, two growth functions~$D_A(t)$ and~$D_B(t)$
are assumed to be equal to~$D_1^2(t)$, and the equality is only valid 
in the Einstein-de~Sitter universe, while the approximation yields only
small errors in the late universe \cite{BECOET02}.
The first kernel in Eq.~\eqref{sptone} can then be re-arranged as
\beeq
{5\over7}D_1^2(t)\left[1+{\kone\cdot\ktwo\over2k_1k_2}\left({k_1\over k_2}
+{k_2\over k_1}\right)\right]~,
\eneq
and the second kernel in Eq.~\eqref{spttwo} is
\beeq
\frac17D_1^2(t)\left[{\kone\cdot\ktwo\over k_1k_2}\left({k_1\over k_2}
+{k_2\over k_1}\right)+2\left({\kone\cdot\ktwo\over k_1k_2}\right)^2\right]~,
\eneq
where we scaled each Fourier kernel by $k_1^2k_2^2$ in the denominator,
as the Fourier kernels in the standard perturbation theory are expressed
in terms of the matter density fluctuation $\delta(\kvec)=-k^2\RR(\kvec)$.
They add up to yield the standard kernel
\beeq
\label{stdf2}
F_2(\kone,\ktwo)=D_1^2(t)\left[\frac57+\frac12{\kone\cdot\ktwo\over k_1k_2}
\left({k_1\over k_2}+{k_2\over k_1}\right)+\frac27
\left({\kone\cdot\ktwo\over k_1k_2}\right)^2\right]~,
\eneq
but note that the shift term in proportion to $\kone\cdot\ktwo$
originates from two terms with different time dependence.\\

\IIetc{Relativistic contributions in the matter density fluctuation}
At the second order in perturbations, there exist two extra relativistic
contributions to the matter density fluctuation in addition to the standard
contributions discussed in the context of the standard perturbation theory.
They originate from the constraint equation of general relativity.
The Fourier kernels for these two terms in Eq.~\eqref{propt} are computed
here. The Fourier kernel for the first relativistic contribution is
\beeq
\label{GR1}
\bullet~~
\frac 32 D_1(z)\RR^{,\al} \RR_{,\al}:\Dquad
\FF_s=F_s(\kone,\ktwo)=-\frac 32 D_1(z)\kone\cdot \ktwo ~,
\eneq
and its one-point ensemble average is non-vanishing
\beeq
\label{grd1}
\ens=\frac 32 D_1(z)\avg_2\,,
\eneq
where the variance~$\avg_n$ is defined in Eq.~\eqref{Deltan}.
The contributions to the connected bispectra are rather simple
\beeq
B_{112}\ni3D_1k_1^2~,\Dquad B_{211}=-B_{121}\ni-3D_1\kone\cdot \klong\RA0~.
\eneq

The Fourier kernel for the second relativistic contribution
in Eq.~\eqref{propt} is
\beeq
\label{GR2}
\bullet~~
4D_1(z)\RR\Delta \RR:\Dquad F_s(\kone,\ktwo)=-4D_1(z)k^2_2~,
\eneq 
and its symmetrized Fourier kernel is
\beeq
\FF_s(\kone,\ktwo)=-2D_1 \left(k_1^2+k_2^2\right) ~.
\eneq
For the bispectrum contributions, the kernels~$F_s$ get symmetrized, so we
can use the symmetric kernel~$\FF_s$, but for the one-point ensemble
average in Eq.~\eqref{avgs}, the kernel~$F_s$ should be used.
The one-point ensemble average of the second relativistic contribution is
\beeq
\label{grd2}
\ens=-4D_1(z)\avg_2~.
\eneq
Finally, the contributions to the connected bispectra are then obtained by
using Eqs.~\eqref{B112}$-$\eqref{B121}
\beeq
B_{112}\ni-8D_1k_1^2~,\Dquad B_{211}=B_{121}\ni-4D_1(k_1^2+k_l^2)\RA
-4D_1k_1^2~,
\eneq
and all of them are non-zero in the limit $k_l\RA0$.\\

\IIetc{Non-Gaussian contributions in the presence of~$\fnl$}
The primordial non-Gaussianity in the initial condition with 
non-vanishing~$\fnl$ in Eq.~\eqref{ini} gives rise two extra terms 
in Eq.~\eqref{fnleq}, and their Fourier kernels can be 
readily read-off from Eqs.~\eqref{GR1} and~\eqref{GR2} as
\bear
\label{fnlker}
&&\bullet~~-\frac65D_1(z)\fnl~\RR^{,\al}\RR_{\,a}:\qquad \quad
\FF_s=F_s(\kone,\ktwo)=\frac 65 D_1(z)\fnl~\kone\cdot \ktwo ~,\\
&&\bullet~~-\frac65D_1(z)\fnl~\RR\Delta\RR:\Dquad 
\FF_s(\kone,\ktwo)=\frac35D_1(z)\fnl\left(k_1^2+k_2^2\right)~.
\enar
As opposed to the two relativistic contributions in the matter density
fluctuations, the coefficients for these two contributions arrange
in a way that the sum of these two contributions in proportion to~$\fnl$
is 
\beeq
\FF_s(\kone,\ktwo)=\frac35D_1(z)\fnl|\kone+\ktwo|^2~,
\eneq
and the one-point ensemble average vanishes by cancellation of the two
terms. The contribution to the first bispectrum vanishes, and the other
two bispectra
\beeq
B_{211}\ni \frac65D_1f_{\rm nl}|\kone+\kvec_l|^2\quad
\RA~~\frac65D_1f_{\rm nl}k_1^2 ~,\Dquad
B_{121}\ni \frac65D_1f_{\rm nl}|\kone-\kvec_l|^2\quad
\RA~~\frac65D_1f_{\rm nl}k_1^2 ~,
\eneq
survive in the squeezed limit.\\

\IIetc{Coupling contributions at the source position} The linear-order
matter density fluctuation is coupled with the relativistic effects in
the light propagation, which appears as $\Delta x^\mu~\partial_\mu\delta$
in Eq.~\eqref{main}. This coupling results in numerous extra terms 
detailed in Eq.~\eqref{mainRR}, as there are many contribution terms
in~$\Delta x^\mu$. Among those, there are four terms, contributing
at the source position, and we compute the Fourier kernels for those four
terms.

The first two terms arise from the coupling $\delta\eta~\pa_\eta\delta$,
where $\delta\eta=\dz/\HH$ and $\delta'\propto D_1'=D_V$. The first of
the two contributions is the Sachs-Wolfe contribution at the source, and
its Fourier kernel is 
\beeq
\bullet~~
\frac{D_V}{\HH}\RR\Delta \RR: \Dquad
F_s(\kone,\ktwo)=-{D_V(z)\over\Hz}k_2^2~,\Dquad
\FF_s(\kone,\ktwo)=-\frac{D_V(z)}{2\Hz}\rbr{k_1^2+k_2^2}~.
\eneq
The one-point ensemble average of this coupling term is
\beeq
\ens=-\frac{D_V(z)}{\Hz}\avg_2~,
\eneq
and the connected bispectra are
\beeq
B_{112}\ni-{2D_V(z)\over\Hz}k_1^2~,\Dquad B_{211}=B_{121}\ni
-{D_V(z)\over\Hz}(k_1^2+k_l^2)\RA{D_V(z)\over\Hz}k_1^2~.
\eneq
None of the one-point ensemble average or the connected bispectra vanish
in the limit $k_l\RA0$.

The second contribution in the coupling $\delta\eta~\pa_\eta\delta$ arises
from the line-of-sight velocity contribution at the source position. The
Fourier kernel for the contribution is
\beeq
\bullet~~
-\frac{D_V^2}{\Hz}\partial_r \RR\Delta \RR:\Dquad
F_s(\kone,\ktwo)={D_V^2(z)\over\Hz}i\mu_1k_1k_2^2~,
\eneq
and its symmetrized Fourier kernel is
\beeq
\FF_s(\kone,\ktwo)=\frac{D_V^2(z)}{2\Hz}ik_1k_2\rbr{\mu_1\, k_2+\mu_2k_1} ~.
\eneq
The one-point ensemble average and the first connected bispectrum then
trivially vanish due to the symmetry of the wave vectors
\beeq
\ens=0~,\Dquad B_{112}=0~,
\eneq
and the remaining two connected bispectra also vanish
\beeq
B_{211}\ni {D_V^2\over\Hz}ik_1k_l\left(\mu_1 k_l+\mu_l
k_1\right)\RA0~,\Dquad
B_{121}\ni{D_V^2\over\Hz}ik_1k_l\left(\mu_l k_1-\mu_1k_l\right)\RA0~,
\eneq
upon angle average in the limit.

The remaining two contributions at the source position come from the coupling
term $\delta r~\pa_r\delta$ (no contributions at the source position from
the angular distortion). The first term is the coupling of the matter
density fluctuation and the gravitational potential at the source position.
The Fourier kernel for this contribution is
\beeq
\bullet~~
\frac{D_1D_\Psi}{\HH}\RR\partial_r\Delta \RR:\Dquad
F_s(\kone,\ktwo)=-{D_1(z)D_\Psi(z)\over\Hz}i\mu_2k_2^3~,
\eneq
and its symmetrized kernel is
\beeq
\FF_s(\kone,\ktwo)=-\frac{D_1(z)D_\Psi(z)}{2\Hz}i
\rbr{\mu_1\,k_1^3+\mu_2\, k_2^3}~.
\eneq
The one-point ensemble average and the first connected bispectrum vanish
\beeq
\ens=0~,\Dquad B_{112}=0~,
\eneq
upon angle average. Mind that the one-point ensemble average is computed by
using~$F_s$ in Eq.~\eqref{avgs}, while the connected bispectra are computed
by using~$\FF_s$ in Eq.~\eqref{B112}. The remaining two connected bispectra 
\beeq
B_{211}\ni
-{D_1D_\Psi\over\Hz}i\left(\mu_1k_1^3+\mu_lk_l^3\right)\RA0~,\Dquad
 B_{121}\ni-{D_1D_\Psi\over\Hz}i\left(-\mu_1k_1^3+\mu_lk_l^3\right)
  \RA0~,
\eneq
also vanish upon angle average in the limit $k_l\RA0$. 

The last contribution is the coupling with the line-of-sight velocity at
the source position. The Fourier kernel for the fourth term is
\beeq
\bullet~~
\frac{D_1D_V}{\Hz}\partial_rR\partial_r\rbr{\Delta R}:\Dquad
F_s(\kone,\ktwo)={D_1(z)D_V(z)\over\Hz}\mu_1k_1\mu_2k_2^3~,
\eneq
and its symmetrized kernel is
\beeq
\FF_s(\kone,\ktwo)=\frac{D_1(z)D_V(z)}{2\Hz}
\mu_1\mu_2\,k_1k_2\rbr{k_1^2+k_2^2} ~.
\eneq
The one-point ensemble average and the first connected bispectrum are
non-zero
\beeq
\ens=-\frac{D_1(z)D_V(z)}{3\Hz}\avg_4 \,,\Dquad
B_{112}=-{2D_1(z)D_V(z)\over\Hz}\mu_1^2k_1^4\RA-{2D_1(z)D_V(z)\over3\Hz}k_1^4~,
\eneq
where the angular dependence is removed upon angular
integration. The remaining two connected bispectra, however, 
\beeq
B_{211}=-B_{121}
\ni{D_1(z)D_V(z)\over\Hz}\mu_1\mu_lk_1k_l(k_1^2+k_l^2)\RA0~,
\eneq
vanish, if we take the limit $k_l\RA0$.

\subsection{Coupling contributions from the source and the observer positions}
Since cosmological observables are measured by the observer,
these observables depend not only on the physical properties of the
source, but also on the state of the observer. As the observers perform 
measurements in the rest frame different from the FRW frame, there exist
various contributions in the cosmological observables from the gravitational
potential and the line-of-sight velocity at the observer position, as shown
in Eqs.~\eqref{drr}$-$\eqref{dtt}. These contributions contribute only
through the coupling terms in $\Delta x^\mu~\pa_\mu\delta$ in Eq.~\eqref{main}.
As discussed in  Appendix~\ref{FDEV}, the Fourier kernels~$\FF_o$ 
for such coupling contributions involve the exponential factor in 
Eq.~\eqref{FFO}, in addition to the ordinary Fourier kernels~$F_o$
from the coupling terms.

There exist six different contributions in Eq.~\eqref{mainRR}, involving
those at the observer position. The first two terms come from the coupling
$\delta\eta~\pa_\eta\delta$. The Sachs-Wolfe contribution at the observer
position makes the first of such terms, and its Fourier kernel is
\beeq
\bullet~~
-\frac{D_V}{\HH}\RR_o\Delta \RR:\Dquad 
F_o(\kone,\ktwo)=\frac{D_V(z)}{\Hz}k_2^2~,
\eneq
and its symmetrized kernel with the exponential factor is
\beeq
\FF_o={D_V(z)\over2\Hz}
\left(e^{-i\kone\cdot\rbar_z\Vang}k_2^2+e^{-i\ktwo\cdot\rbar_z\Vang}
k_1^2\right)~,
\eneq
where the subscript~$o$ indicates that~$\RR$ is evaluated at the observer
position and there is no time dependence of~$\RR_o$ on~$\etao$,
as the initial condition is constant in time. Furthermore, we keep the
convention that the first wave vector~$\kone$ describes the quantity
at the observer position, while the second wave vector~$\ktwo$ describes
one at the source position.
The one-point ensemble average and the first connected bispectrum are
\beeq
\ens(z)={D_V(z)\over{\cal H}_z}\avg_{2,0}(z)~, \Dquad
B_{112}\ni{2D_V(z)\over\Hz}k_1^2j_0(k_1\rbar_z)~,
\eneq
where the time-dependent variance~$\avg_{n,m}(z)$ is defined in 
Eq.~\eqref{avgnm}.
Furthermore, in the derivation we used 
\beeq
\int_{-1}^1\mathrm {d\mu\over2}~e^{\pm ix\mu}=j_0(x)~,
\eneq
and also for the following derivations we will use the useful relation
involving the integration of the exponential factor:
\beeq
\int_{-1}^1\mathrm {d\mu\over2}~\mu e^{\pm ix\mu}=\pm ij_1(x)~,\Dquad
\int_{-1}^1\mathrm {d\mu\over2}~\mu^2 e^{\pm ix\mu}={j_0(x)-2j_2(x)\over3}
~.\nonumber
\eneq
Two connected bispectra are non-vanishing
\bear
B_{211}&\ni&{D_V(z)\over\Hz}\left(e^{-i\kone\cdot\rz\Vang}k_l^2
+e^{-i\mathbf k_l\cdot\rz\Vang}k_1^2\right)\RA{D_V(z)\over\Hz}k_1^2~,\\
B_{121}&\ni&{D_V(z)\over\Hz}\left(e^{i\kone\cdot\rz\Vang}k_l^2+
e^{-i\mathbf k_l\cdot\rz\Vang}k_1^2\right)\RA{D_V(z)\over\Hz}k_1^2~,
\enar
even in the limit $k_l\RA0$.

The second term arises from the line-of-sight velocity
at the observer position, contributing to~$\delta\eta$. The Fourier kernel
for this term is
\beeq
\bullet~~
\frac{D_VD_V^o}{\HH}(\partial_r\RR)_o\Delta \RR:\Dquad
F_o(\kone,\ktwo)=-\frac{D_V(z)D_V(\etao)}{\Hz}i\mu_1k_1k_2^2\,,
\eneq
and its symmetrized kernel is
\beeq
\FF_o=-\frac{D_V(z)D_V(\etao)}{2\Hz}ik_1k_2\left(e^{-i\kone\cdot\rbar_z\Vang}
\mu_1k_2+e^{-i\ktwo\cdot\rbar_z\Vang}\mu_2k_1\right)~.
\eneq
The one-point ensemble average and the first connected bispectrum are
\beeq
\ens=-{D_V(z)D_V(\etao)\over{\cal H}_z}\avg_{3,1}(z)~,\Dquad
B_{112}\ni-\frac{2D_V(z)D_V(\etao)}{\Hz}k_1^3j_1(k_1\rbar_z)~,
\eneq
non-vanishing, while two remaining connected bispectra 
\bear
 B_{211}&\ni&-{D_V(z)D_V(\etao)\over\Hz}ik_1k_l\left(e^{-i\kone\cdot\rbar_z
\Vang} \mu_1k_l+e^{-i\mathbf k_l\cdot\rbar_z\Vang}\mu_lk_1\right)\RA0~,\\
 B_{121}&\ni&-{D_V(z)D_V(\etao)\over\Hz}ik_1k_l\left(-e^{i\kone\cdot\rbar_z
\Vang}
 \mu_1k_l+e^{-i\mathbf k_l\cdot\rbar_z\Vang}\mu_lk_1\right)\RA0~,
 \enar
vanish in the limit $k_l\RA0$, upon angular integration.

The next two terms come from the coupling term $\delta r~\pa_r\delta$,
in which the gravitational potential and the line-of-sight velocity terms
couple to the derivative of the matter density fluctuation. The Fourier
kernel for the gravitational potential contribution at the observer position
is 
\beeq
\bullet~~
-D_1\rbr{D_V^o-\frac{1}{\Hz}}\RR_o\partial_r\rbr{\Delta \RR}:\Dquad
F_o(\kone,\ktwo)=D_1(z)\left[D_V(\etao)-\frac{1}{\Hz}\right]i\mu_2 k_2^3\,,
\eneq
and its symmetrized kernel is
\beeq
\FF_o={D_1(z)\over2}\left[D_V(\etao)-{1\over\Hz}\right]i\left(
e^{-i\kone\cdot\rbar_z\n}\mu_2k_2^3+e^{-i\ktwo\cdot\rbar_z\n}\mu_1k_1^3
\right)~.
\eneq
Evident in the time-dependent pre-factors, there are multiple contributions
from the gravitational potential at the observer position 
with different pre-factors.
The one-point ensemble average and the first connected bispectrum are
\beeq
\ens(z)=-D_1(z)\left[D_V(\etao)-{1\over\Hz}\right]\avg_{3,1}(z)~,\Dquad
B_{112}\ni-2D_1(z)\left[D_V(\etao)-{1\over\Hz}\right]k_1^3j_1(k_1\rbar_z)~,
\eneq
and the remaining connected bispectra are
\bear
 B_{211}&\ni&D_1(z)\left[D_V(\etao)-{1\over\Hz}\right]i
\left(e^{-i\kone\cdot\rbar_z\n}
\mu_lk_l^3+e^{-i\mathbf k_l\cdot\rbar_z\n}\mu_1k_1^3\right)\RA0~,\\
 B_{121}&\ni&D_1(z)\left[D_V(\etao)-{1\over\Hz}\right]
i\left(e^{i\kone\cdot\rbar_z\n}
\mu_lk_l^3-e^{-i\mathbf k_l\cdot\rbar_z\n}\mu_1k_1^3\right)\RA0~,
 \enar
vanishing in the limit $k_l\RA0$ upon angular integration.

The next term is the line-of-sight velocity contribution to the coupling term
at the observer position, and its Fourier kernel is
\beeq
\bullet~~
-\frac{D_1D_V^o}{\Hz}(\partial_r \RR_o)\partial_r\rbr{\Delta \RR}:\Dquad
F_o(\kone,\ktwo)=-\frac{D_1(z)D_V(\etao)}{\Hz} \mu_1\mu_2k_1k_2^3\,,
\eneq
and its symmetrized kernel is
\beeq
\FF_o=-{D_1(z)D_V(\etao)\over2\Hz}\mu_1\mu_2k_1k_2\left(
e^{-i\kone\cdot\rbar_z\n}k_2^2+e^{-i\ktwo\cdot\rbar_z\n}k_1^2\right)~,
\eneq
The one-point ensemble average and the first connected bispectrum are
\beeq
\ens(z)={D_1(z)D_V(\etao)\over3\Hz}\left[\avg_{4,0}(z)
-2\avg_{4,2}(z)\right]~,\Dquad
B_{112}\ni{2D_1(z)D_V(\etao)\over3\Hz}k_1^4
\left[j_0(k_1\rbar_z)-2j_2(k_1\rbar_z)\right]~,
\eneq
non-vanishing, while the remaining connected bispectra vanish
\bear
 B_{211}&\ni&-{D_1(z)D_V(\etao)\over\Hz}\mu_1\mu_lk_1k_l
\left(e^{-i\kone\cdot\rbar_z\n}
k_l^2+e^{-i\mathbf k_l\cdot\rbar_z\n}k_1^2\right)\RA0~,\\
 B_{121}&\ni&{D_1(z)D_V(\etao)\over\Hz}
\mu_1\mu_lk_1k_l\left(e^{i\kone\cdot\rbar_z\n}
k_l^2+e^{-i\mathbf k_l\cdot\rbar_z\n}k_1^2\right)\RA0~.
 \enar

The other coupling contribution 
$(\delta\theta~\pa_\theta+\delta\phi~\pa_\phi)\delta_m$
comes from the angular distortion, which includes only the line-of-sight
velocity term at the observer position. The Fourier kernel for this
term is
\beeq
\bullet~~
D_1D_V^o(\partial^\alpha \RR)_o\habla_\alpha(\Delta \RR):\Dquad
F_o(\kone,\ktwo)=D_1(z)D_V(\etao)\rz\rbr{-\mu_1\mu_2k_1k_2
+\kone\cdot\ktwo}k_2^2\,,
\eneq
and its symmetrized kernel is
\beeq
\FF_o=\frac12D_1(z)D_V(\etao)\rbar_z
\left(-\mu_1\mu_2k_1k_2+\kone\cdot\ktwo\right)
\left(e^{-i\kone\cdot\rbar_z\n}k_2^2+e^{-i\ktwo\cdot\rbar_z\n}k_1^2\right)~.
\eneq
Note that there exist two contributions in total, 
one from~$\delta\theta$ and one from~$\delta\phi$, which are combined as
the angular gradient~$\hat\nabla$.
The one-point ensemble average and the first connected bispectrum are
\beeq
\ens=-2D_1(z)D_V(\etao)\avg_{3,1}(z)~,\Dquad
B_{112}\ni-4D_1(z)D_V(\etao)k_1^3j_1(k_1\rbar_z)~,
\eneq
non-zero, while the remaining connected bispectra are
\bear
 B_{211}&\ni&D_1D_V^o\rbar_z\left(-\mu_1\mu_lk_1k_l+\kone\cdot\klong\right)
\left(e^{-i\kone\cdot\rbar_z\n}k_l^2+e^{-i\mathbf k_l\cdot\rbar_z\n}k_1^2
\right)\RA0 ~,\\
 B_{121}&\ni&D_1D_V^o\rbar_z\left(\mu_1\mu_lk_1k_l-\kone \cdot\klong\right)
\left(e^{i\kone\cdot\rbar_z\n}k_l^2+e^{-i\mathbf k_l\cdot\rbar_z\n}k_1^2
\right)\RA0 ~,
\enar
vanishing in the limit $k_l\RA0$, where we used $j_1(x)=x(j_0+j_2)/3$.

One last contribution is from the spatial shift~$\delta x^\alpha_o$
of the observer position, contained in spatial distortions~$\delta r$,
$\delta\theta$, and~$\delta\phi$. While it contains an integration over
time, its spatial position is fixed at the observer position at the 
perturbation order of our interest, rendering it essentially the same
as any other contributions at the observer position in this section.
The Fourier kernel for the spatial shift contribution is
\beeq
\bullet~~
-D_1\rbr{\int_0^{\etao}\mathrm d\bar\eta\, D_V(\bar\eta)\nabla_\alpha \RR}
\nabla^\alpha(\Delta \RR): \Dquad
F_o(\kone,\ktwo)=-D_1(z)\rbr{\int_0^{\etao}\mathrm d\bar\eta\,
  D_V(\bar\eta)}\rbr{\kone\cdot\ktwo} k_2^2~,
\eneq
and its symmetrized kernel is
\beeq
\FF_o=-{D_1(z)\over2}\rbr{\int_0^{\etao}\mathrm d\bar\eta\, D_V(\bar\eta)}
\rbr{\kone\cdot\ktwo}
\left(e^{-i\kone\cdot\rbar_z\n}k_2^2+e^{-i\ktwo\cdot\rbar_z\n}k_1^2\right)~,
\eneq
where the integration over~$\bar\eta$ is along the time coordinate with
spatial position fixed (not the line-of-sight integration).
The one-point ensemble average and the first connected bispectrum are
\beeq
\ens(z)=D_1(z)\avg_{4,0}(z)
\int_0^{\etao}\mathrm d\bar\eta~D_V(\bar\eta)~, \Dquad
B_{112}\ni2D_1(z)\rbr{\int_0^{\etao}\mathrm d\bar\eta\, D_V(\bar\eta)}k_1^4
j_0(k_1\rbar_z)~,
\eneq
and the remaining connected bispectra are
\bear
B_{211}&\ni&
-{D_1(z)}\rbr{\int_0^{\etao}\mathrm d\bar\eta\, D_V(\bar\eta)}\kone\cdot
\klong\left(e^{-i\kone\cdot\rbar_z\n}k_l^2+e^{-i\mathbf k_l\cdot\rbar_z\n}k_1^2
\right)\RA0~,\\
B_{121}&\ni&
{D_1(z)}\rbr{\int_0^{\etao}\mathrm d\bar\eta\, D_V(\bar\eta)}\kone\cdot
\klong\left(e^{i\kone\cdot\rbar_z\n}k_l^2+e^{-i\mathbf k_l\cdot\rbar_z\n}k_1^2
\right)\RA0~.
 \enar
In the Einstein-de~Sitter universe, the pre-factors are further simplified
as
\beeq
-D_1(z)\left[D_V(\etao)-\frac{1}{\Hz}\right]={\etaz^2\over10}\left({\etao\over
5}-{\etaz\over2}\right)~,\Dquad
-D_1\rbr{\int_0^{\etao}\mathrm d\bar\eta\, D_V(\bar\eta)}=
-{\etaz^2\etao^2\over100}~.
\eneq

\subsection{Coupling contributions involving the line-of-sight integral}
The last type of contributions to the observed matter density fluctuation
involves the coupling with contributions from the line-of-sight direction.
The well-known components in this category are the contributions of the 
gravitational lensing and the integrated Sachs-Wolfe effects.
The Fourier kernels for such contributions in Eq.~\eqref{FFnl} 
include not only the line-of-sight integration, but also the exponential
factor.
There exist four coupling terms arising from the line-of-sight integration.

The first term arises from the coupling term in $\delta\eta~\pa_\eta\delta$,
and the integrated Sachs-Wolfe contribution in~$\delta\eta=\dz/\HH$ 
results in the coupling contribution involving the line-of-sight integral.
Its Fourier kernel is
\beeq
\bullet~~
\frac{2D_V}{\HH}\rbr{\int_0^\rz d\rbar~D_\Psi(\bar r)\partial_r\RR}\Delta\RR:
\Dquad
F_{nl}(\bar r,\kone,\ktwo)=-\frac{2D_V(z)}{\Hz} 
D_\Psi(\bar r)i\mu_1k_1k_2^2\,, 
\eneq
where the first wave vector~$\kone$ belongs to the contribution along the
line-of-sight direction and the second wave vector~$\ktwo$ describes the
matter fluctuation at the source position. The full and symmetrized
Fourier kernel is
\beeq
\FF_{nl}=-\frac{D_V(z)}{\Hz} \int_0^{\rz}d\rbar~
D_\Psi(\bar r)ik_1k_2\left(
e^{-i\kone\Delta r\n}\mu_1k_2+e^{-i\ktwo\Delta r\n}\mu_2k_1\right)\,,
\eneq
and its one-point ensemble average is
\beeq
\ens(z)=-{2D_V(z)\over{\cal H}_z}\int\Dkk~k^3P_\RR(k)
\int_0^{\rz}d\rbar~D_\Psi(\rbar)j_1(k\rbar)~,
\eneq
where $\Delta r:=\rz -\bar r$.
The connected bispectra are then obtained as
\bear
B_{112}&\ni&-\frac{4D_V(z)}{\Hz} \int_0^{\rz}d\rbar~
D_\Psi(\bar r)k_1^3j_1(k_1\Delta r)~,\\
 B_{211}&\ni&-{2D_V(z)\over\Hz}\int_0^{\rz}d\rbar~D_\psi(\rbar)i
k_1k_l\left(e^{-i\kone\cdot\Delta r\n}\mu_1k_l+e^{-i\mathbf k_l\cdot\Delta r\n}
\mu_lk_1\right)\RA0~,\\
 B_{121}&\ni&-{2D_V(z)\over\Hz}\int_0^{\rz}d\rbar~D_\psi(\rbar)i
k_1k_l\left(-e^{i\kone\cdot\Delta r\n}\mu_1k_l+e^{-i\mathbf k_l\cdot\Delta r\n}
\mu_lk_1\right)\RA0~.
 \enar
In the Einstein-de~Sitter universe, the gravitational potential is
constant in time:
\beeq
D_\Psi=-\frac35~,
\eneq
allowing us to analytically perform the line-of-sight integration by using
\beeq
k\int_0^\rz\dr j_1(k\Delta r)=1-j_0(k\rz)\,,\Dquad
\frac{k}{3}\int_0^\rz\dr \Big[j_0(k\Delta r)-2j_2(k\Delta r)\Big]
=j_1(k\rz)\,.
\eneq
The one-point ensemble average and the first connected bispectrum becomes
\beeq
\ens(z)=\frac{3\etaz^2}{25}\left[\avg_2-\avg_{2,0}(z)\right]\,,
\Dquad
B_{112}\ni{24\over25}{k_1^2\left[1-j_0(k_1\rz)\right]\over H_0^2(1+z)}~.
\eneq

The second term comes again from the integrated Sachs-Wolfe contribution 
in~$\dz$, but through the coupling term $\delta r~\pa_r\delta$. Hence
the line-of-sight integration is identical to the first term, but with
different derivative coupling to the matter density fluctuation.
The Fourier kernel for this contribution is
\beeq
\bullet~~
-\frac{2D_1}{\HH}\rbr{\int_0^\rz\dr D_\Psi(\bar r)\partial_r \RR}
\partial_r(\Delta \RR): \Dquad
F_{nl}(\bar r,\kone,\ktwo)
=-\frac{2D_1(z)}{\Hz}D_\Psi(\bar r)\mu_1\mu_2 k_1k_2^3\,,
\eneq
and its symmetrized kernel is
\beeq
\FF_{nl}=-\frac{D_1(z)}{\Hz}\int_0^{\rz}d\rbar~
D_\Psi(\bar r)\mu_1\mu_2 k_1k_2\left(
e^{-i\kone\Delta r\n}k_2^2+e^{-i\ktwo\Delta r\n}k_1^2\right)\,.
\eneq
The one-point ensemble average is
\beeq
\ens(z)={2D_1(z)\over3{\cal H}_z}\int\Dkk~k^4P_\RR(k)\int_0^{\rz}d
\rbar~D_\Psi(\rbar)\left[j_0(k\rbar)-2j_2(k\rbar)\right]~, 
\eneq
 and the first connected bispectrum is
\beeq
B_{112}\ni{4D_1(z)\over3\Hz}\int_0^{\rz}d\rbar~
D_\Psi(\rbar)k_1^4\left[j_0(k_1\Delta r)-2j_2(k_1\Delta r)\right]~.
\eneq
Two remaining connected bispectra 
\bear
 B_{211}&\ni& -{2D_1(z)\over\Hz}\int_0^{\rz}d\rbar~
D_\Psi(\rbar)\mu_1\mu_lk_1k_l
\left(e^{-i\kone\cdot\Delta r\n}
k_l^2+e^{-i\mathbf k_l\cdot\Delta r\n}
k_1^2\right)\RA0~,\\
 B_{121}&\ni&{2D_1(z)\over\Hz}\int_0^{\rz}d\rbar~
D_\Psi(\rbar)\mu_1\mu_lk_1k_l
\left(e^{i\kone\cdot\Delta r\n}k_l^2+e^{-i\mathbf k_l\cdot\Delta r\n}
k_1^2\right)\RA0~,
 \enar
vanish in the limit $k_l\RA0$ upon angular integration.
In the Einstein-de~Sitter universe, the one-point ensemble average and
the first connected bispectrum are further simplified as
\beeq
\ens=-\frac{3\etaz^3}{50}\avg_{3,1}(z)~,\Dquad B_{112}\ni
-{24\over25}{k_1^3j_1(k_1\rz)\over H_0^3(1+z)^{3/2}}~.
\eneq

The third term arises from the line-of-sight integration of the gravitational
potential contribution in the radial distortion~$\delta r$. Its Fourier kernel
is
\beeq
\bullet~~
-2D_1\rbr{\int_0^\rz\dr D_\Psi(\bar r)\RR}\partial_r(\Delta \RR):\Dquad
F_{nl}(\bar r,\kone,\ktwo)=2D_1(z)D_\Psi(\bar r)i\mu_2k_2^3\,, 
\eneq
and its symmetrized kernel is
\beeq
\FF_{nl}=D_1(z)\int_0^{\rz}d\rbar~
D_\Psi(\bar r)i\left(e^{-i\kone\Delta r\n}\mu_2k_2^3
+e^{-i\ktwo\Delta r\n}\mu_1k_1^3\right)\,.
\eneq
The one-point ensemble average  is
\beeq
\ens(z)=-2D_1(z)\int\Dkk~k^3P_\RR(k)\int_0^{\rz}d
\rbar~D_\Psi(\rbar)j_1(k\rbar)~,
\eneq
and the first connected bispectrum is
\beeq
B_{112}\ni-4D_1(z)\int_0^{\rz}d\rbar~D_\Psi(\rbar)k_1^3j_1(k_1\Delta r)~.
\eneq
The remaining connected bispectra are
\bear
 B_{211}&\ni&2D_1(z)\int_0^{\rz}d\rbar~D_\Psi(\rbar)i\left(
e^{-i\kone\cdot\Delta r\n}\mu_lk_l^3
+e^{-i\mathbf k_l\cdot\Delta r\n}\mu_1k_1^3\right)\RA0~,\\
 B_{121}&\ni&2D_1(z)\int_0^{\rz}d\rbar~D_\Psi(\rbar)i\left(
e^{i\kone\cdot\Delta r\n}\mu_lk_l^3
-e^{-i\mathbf k_l\cdot\Delta r\n}\mu_1k_1^3\right)\RA0~.
 \enar
In the Einstein-de~Sitter universe, we have
\bear
\ens(z)=\frac{3\etaz^2}{25}\left[\avg_2-\avg_{2,0}(z)\right]\,, 
\Dquad B_{112}\ni{24\over25}{k_1^2\left[1-j_0(k_1\rz)\right]\over H_0^2(1+z)}~.
\enar

The last term comes from the gravitational lensing contribution, which 
couples to the matter density fluctuation through the angular distortion
$(\delta\theta~\pa_\theta+\delta\phi~\pa_\phi)\delta_m$. The Fourier kernel
for this contribution takes the form
\bear
&&
\bullet~~
2D_1\rbr{\int_0^\rz\dr\rbr{\frac{\rz-\bar r}{\rz \bar r}}D_\Psi(\bar r)
  \habla^\alpha \RR}\habla_\alpha\rbr{\Delta \RR}:\\
&&\Dquad
F_{nl}(\bar r,\kone,\ktwo)
=2D_1(z)\Delta r D_\Psi(\bar r)k_2^2\rbr{-\mu_1\mu_2k_1k_2+\kone\cdot\ktwo}\,,
\nn
\enar
and its symmetrized kernel is
\beeq
\FF_{nl}=D_1(z)\int_0^{\rz}d\rbar~
 D_\Psi(\bar r)\Delta r\left(e^{-i\kone\Delta r\n}k_2^2
+e^{-i\ktwo\Delta r\n}k_1^2\right)
\rbr{-\mu_1\mu_2k_1k_2+\kone\cdot\ktwo}~.
\eneq
The one-point ensemble average is 
\beeq
\ens(z)=-4D_1(z)\int\Dkk~k^3P_\RR(k)\int_0^{\rz}d
\rbar~D_\Psi(\rbar)j_1(k\rbar)~, 
\eneq
and the first connected bispectrum is
\beeq
B_{112}\ni-8D_1(z)\int_0^{\rz}d\rbar~D_\Psi(\rbar)k_1^3j_1(k_1\Delta r)~,
\eneq
and the remaining connected bispectra are
\bear
 B_{211}&\ni&2D_1(z)\int_0^{\rz}d\rbar~\Delta rD_\Psi(\rbar)
\left(-\mu_1\mu_lk_1k_l+\kone\cdot\klong\right)
\left(e^{-i\kone\cdot\Delta r\n}k_l^2
+e^{-i\mathbf k_l\cdot\Delta r\n}k_1^2\right)\RA0~,\\
 B_{121}&\ni&2D_1(z)\int_0^{\rz}d\rbar~\Delta rD_\Psi(\rbar)
\left(\mu_1\mu_lk_1k_l-\kone\cdot\klong\right)
\left(e^{i\kone\cdot\Delta r\n}k_l^2
+e^{-i\mathbf k_l\cdot\Delta r\n}k_1^2\right)\RA0~.
 \enar
In the Einstein-de~Sitter universe, the non-vanishing components are
further simplified as
\beeq
\ens={6\etaz^2\over25}\left[\avg_2-\avg_{2,0}(z)\right]~,\Dquad
B_{112}\ni{48\over25}{k_1^2\left[1-j_0(k_1\rz)\right]\over H_0^2(1+z)}~.
\eneq

\bigskip\bigskip\bigskip

\bibliography{ms.bbl}

\end{document}